\documentclass[twoside,12pt]{PhDthesisPSnPDF}

\usepackage{color}
\usepackage{slashed}
\usepackage{mathrsfs}
\usepackage{verbatim}
\usepackage{bbold}

\DeclareMathAlphabet{\mathcal}{OMS}{cmsy}{m}{n}

\newcommand{\figuremacroW}[4]{
	\begin{figure}[htbp]
		\centering
		\includegraphics[width=#4\textwidth]{#1}
		\caption[#2]{\textbf{#2} - #3}
		\label{#1}
	\end{figure}
}

\ifpdf  
    \pdfcatalog { /PageMode (/UseOutlines)
                  /OpenAction (fitbh)  }
\fi

\title{Dynamical Yukawa Couplings}

\ifpdf
  \author{\normalsize Memoria de Tesis Doctoral realizada por\\[2mm]
{\bf \large Rodrigo Alonso de Pablo}}
  \collegeordept{\normalsize y  dirigida por\\{\bf Bel\'en Gavela Legazpi} \\
  Catedr\'atica del Departamento de F\'isica Te\'orica\\ Universidad Aut\'onoma de Madrid
}
  \university{\,}

    \crest{    \includegraphics[width=4cm]{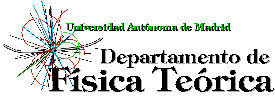}\,\,\,\,\,\,\,
    \includegraphics[width=4cm]{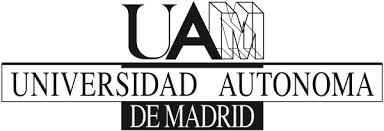}\,\,\,\,\,\,\,\,
        \includegraphics[width=4cm]{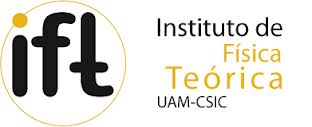}
}
  
\else
  \author{YourName}
  \collegeordept{CollegeOrDept}
  \university{University}
  \crest{\includegraphics[width=4cm]{logo}}
\fi

\renewcommand{\submittedtext}{ \normalsize Presentada ante el Departamento de F\'isica Te\'orica\\ de la Universidad Aut\'onoma de Madrid para la obtenci\'on del T\'itulo}
\degree{Philosophi\ae Doctor (PhD)}
\degreedate{\normalsize Madrid Junio de 2013}

\begin{document}


\renewcommand\baselinestretch{1.2}
\baselineskip=18pt plus1pt


\maketitle  


\cleardoublepage
\pagenumbering{gobble}

\begin{center}
{\bf Tribunal}\\[4mm]

{\it Presidente}\\[2mm]
Luciano Maiani\\[3mm]
{\it Secretaria}\\[2mm]
Maria Jos\'e Herrero\\[3mm]
{\it Vocales}\\[2mm]
Alberto Casas\\[2mm]
Christophe Grojean\\[2mm]
Alejandro Pomarol\\[3mm]
{\it Suplentes}\\[2mm]
Pilar Hern\'andez\\[2mm]
Stefano Rigolin\\[2mm]

\vspace{10mm}
\emph {Expertos Doctores}\\[2mm]  Gino Isidori\\[2mm]
 Silvia Pascoli
\end{center}
\cleardoublepage






\frontmatter


\setcounter{secnumdepth}{3} 
\setcounter{tocdepth}{3}    
\tableofcontents            


\listoffigures	

\listoftables  
\cleardoublepage


\include{glossary} 
\thispagestyle{empty}
\vspace{4cm}
\begin{multicols}{2} 

\begin{footnotesize} 
{\huge \bf Glossary}
\vspace{2cm}
\\*
{\bf  CKM -} Cabibbo-Kobayashi-Maskawa [quark mixing matrix]\\
{\bf CP -} Parity and Charge conjugation\\
{\bf EFT -} Effective Field Theory\\
{\bf EWSB -} Electroweak Symmetry Breaking\\
{\bf  FCNC -} Flavour Changing Neutral Current\\
{\bf FSB -} Flavour Symmetry Breaking\\
{\bf $\mathcal G$ -}  Gauge Group of the Standard Model: $SU(3)_c\times SU(2)_L\times U(1)_Y$\\
{\bf $\mathcal G_\mathcal F$ - } Flavour Group: $U(n_g)^5\times O(n_g)$\\
{\bf $\mathcal G_\mathcal F^{\slashed A}$ -} Explicitly Axial Symmetry Breaking case Flavour Group: $SU(n_g)^5\times SO(n_g)$\\
{\bf  GIM - } Glashow-Iliopoulos-Maiani [mechanism]\\
{\bf HP -} Hierarchy Problem\\
{\bf IH -} Inverted Hierarchy\\
{\bf $J$ -} Jacobian of the change of coordinates from the physical parameters to the invariants\\
{\bf LHC -} Large Hadron Collider\\
{\bf ME -} Modelo Est\'andar\\
{\bf $n_g$ - } Number of fermion generations \\
{\bf NH -} Normal Hierarchy\\
{\bf PJ -} Problema de la Jerarqu\'ia\\
{\bf  PMNS - } Pontecorvo-Maki-Nakagawa-Sakata [lepton mixing matrix]\\
{\bf QFT -} Quantum Field Theory\\
{\bf QCD -} Quantum Chromodynamics\\
{\bf RESE -} Rotura Espont\'anea de Simetr\'ia Electrod\'ebil\\
{\bf SM -} Standard Model\\
{\bf $\mathcal Y$ - } Flavour scalar fields in the bi-fundamental
representation \\
{\bf $\chi$ -}  Flavour scalar fields in the fundamental
representation\\
\end{footnotesize}
\end{multicols}


\mainmatter

\renewcommand{\chaptername}{} 



\chapter{Objetivo y motivaci\'on} 

El campo de f\'isica de part\'iculas se encuentra actualmente en
un punto crucial.
La exploraci\'on del mecanismo de rotura espont\'anea de simetr\'ia
electrod\'ebil (RESE) en el gran colisionador de hadrones (LHC) ha desvelado
la presencia de un bos\'on que se asemeja al escalar de Higgs \cite{Aad:2012tfa,Chatrchyan:2012ufa}
dada la precisi\'on de los datos experimentales disponibles
\cite{Corbett:2012dm,Giardino:2013bma}. La descripci\'on del Modelo 
Est\'andar (ME) de la generaci\'on de masas \cite{Englert:1964et,Higgs:1964ia,Higgs:1964pj}
ha demostrado ser acertada y la auto-interacci\'on del bos\'on de Higgs
que desencadena la RESE es ahora la quinta fuerza de la naturaleza, junto con la gravedad, el eletromagnetismo la interacci\'on d\'ebil y la fuerte. 

Esta nueva fuerza, como el resto
de las fuerzas cuantizadas, var\'ia en intensidad dependiendo de la escala a la que se
la examine, pero al contrario que la fuerza d\'ebil o fuerte, esto plantea  un problema 
\cite{Callaway:1988ya} ya que a una escala de alta energ\'ia o corta distancia del orden de
$10^{-12}fm$ el mecanismo de RESE se desestabilizar\'ia, pues el acoplo cu\'artico se
cancelar\'ia  \cite{Degrassi:2012ry,EliasMiro:2011aa}. Dicho problema podr\'ia ser
resuelto por la introducci\'on de nueva f\'isica, lo cual conduce a otra cuesti\'on te\'orica,
el Problema de la Jerarqu\'ia (PJ). Cualquier tipo de nueva f\'isica que se acople a la
part\'icula de Higgs produce gen\'ericamente una contribuci\'on radiativa al t\'ermino de masa de dicho bos\'on
del orden de la nueva escala, lo que significar\'ia que la escala electrod\'ebil
es naturalmente cercana a la escala de f\'isica m\'as alta que interacciona con los campos
del ME. Las propuestas para solucionar este problema pueden ser clasificadas
en soluciones de f\'isica perturbativa, siendo el paradigma la supersimetr\'ia,
y ansazts de din\'amica fuerte.  Supersimetr\'ia es una  elegante simetr\'ia  entre
bosones y fermiones que implica cancelaciones sistem\'aticas entre las contribuciones
radiativas bos\'onicas y fermi\'onicas al t\'ermino de masa del Higgs. Por otro lado la hip\'otesis de que
el bos\'on de Higgs sea un estado ligado producido por nueva din\'amica fuerte implica que
el mecanismo de RESE del ME es simplemente una descripci\'on efectiva que debe ser completada
por una teor\'ia m\'as fundamental. Todas estas hip\'otesis suponen naturalmente nueva f\'isica a la
escala del TeV y est\'an siendo testeadas de manera decisiva en el LHC.

En el frente cosmol\'ogico la interacci\'on gravitatoria ha sido la
fuente de  nuevos desaf\'ios en f\'isica de part\'iculas.
El universo est\'a expandi\'endose aceleradamente, algo que 
en cosmolog\'ia est\'andar requiere la presencia de ``energ\'ia oscura", 
una energ\'ia de vac\'io cuya presi\'on negativa provoca que el universo
se ensanche con velocidad creciente. Estimaciones naif en la teor\'ia est\'andar de la contribuci\'on
a este tipo de energ\'ia difieren del valor observado 120 \'ordenes de magnitud, un hecho que
muestra enf\'aticamente nuestra ignorancia sobre la naturaleza de la energ\'ia oscura.
 Cosmolog\'ia y astrof\'isica proporcionaron
 la s\'olida evidencia de materia extra no bari\'onica en el universo, llamada
`materia oscura", como otra muestra experimental no explicable en el ME.
Hay un activo programa experimental para la b\'usqueda de materia oscura en 
este din\'amico sector de f\'isica de part\'iculas. La tercera evidencia de nueva f\'isica
en cosmolog\'ia proviene de un hecho muy familar del mundo visible: est\'a constituido
de mucha mas materia que antimateria, y aunque el ME proporciona una fuente
de exceso de part\'iculas sobre antipart\'iculas el resultado no es suficiente para
explicar la proporci\'on observada.

La parte de nueva f\'isica que concierne m\'as de cerca al ME es el hecho de
que los neutrinos han demostrado ser masivos. La evidencia de masa de neutrinos
proveniente de los datos de oscilaci\'on es una de las selectas evidencias de nueva
f\'isica mas all\'a del ME. En este sector la b\'usqueda de violaci\'on lept\'onica de conjugaci\'on
de carga y paridad (CP),
transiciones de sabor de leptones cargados y la relaci\'on fundamental entre neutrinos y
antineutrinos; su car\'acter Majorana o Dirac, tienen ambiciosos programas experimentales
que producir\'an resultados en los pr\'oximos a\~nos.

Para completar la lista de desaf\'ios en f\'isica de part\'iculas, 
deben ser mencionados la tarea pendiente de la 
cuantizaci\'on de gravedad y el presente pobre entendimiento
del vac\'io de QCD representado en el problema-$\theta$.

El tema de esta tesis es un problema horizontal: el puzle de sabor.
La estructura de sabor del espectro de part\'iculas est\'a conectada en la teor\'ia
est\'andar a la RESE, y las masas de los neutrinos son parte esencial de este
puzle. \'Estos son temas que han sido tratados en el trabajo del estudiante de doctorado en otro contexto:
la fenomenolog\'ia de sabor en el caso de din\'amica fuerte de RESE \cite{Alonso:2012jc,Alonso:2012pz},
la determinaci\'on del Lagrangiano bos\'onico general en el mismo contexto \cite{Alonso:2012px}  y la fenomenolog\'ia de
sabor de
 un modelo para masas de neutrinos \cite{Alonso:2012ji} han formado parte del programa de doctorado del candidato.
El tema central de esta tesis es sin embargo la exploraci\'on de una posible explicaci\'on a la
estructura de sabor \cite{Alonso:2011yg,Alonso:2012fy,onpreparation,onpreparation2}.

El principio gauge puede ser se\~nalado como la fuente creadora de progreso
en f\'isica de part\'iculas, bien entendido y elegantemente implementado en el ME. Por
el contrario el sector de sabor permanece durante d\'ecadas como una de las partes
peor entendidas del ME. El ME muestra la estructura de sabor de una manera param\'etrica,
dejando sin respuesta preguntas como el origen de la fuerte jerarqu\'ia en masas de fermiones
o la presencia de grandes angulos de mezcla de sabor para leptones en constraste con la
peque\~na mezcla del sector de quarks; \'estas preguntas conforman 
el conocido como puzle de sabor. Dicho puzle permanece por lo tanto como una
cuesti\'on fundamental sin respuesta en f\'isica de part\'iculas.

La principal gu\'ia en este trabajo es el uso de simetr\'ia para explicar el puzle de sabor.
La simetr\'ia, que juega un papel central en nuestro entendimiento en f\'isica de part\'iculas,
es empleada en esta tesis para entender la estructura de sabor. Un n\'umero variado
de simetr\'ias han sido postuladas con respecto a este problema \cite{Maiani:2011zz,Froggatt:1978nt,Georgi:1979md,Berezhiani:1983hm,Chivukula:1987py,Barbieri:1995uv,Berezhiani:2001mh,D'Ambrosio:2002ex,Cirigliano:2005ck,Davidson:2006bd,Alonso:2011jd,Barbieri:2011ci}. En este estudio
la simetr\'ia ser\'a seleccionada como la mayor simetr\'ia  continua global posible en la teor\'ia libre \footnote{Alternativamente se puede definir en t\'erminos mas t\'ecnicos como la mayor simetr\'ia posible en el l\'imite
de acoplos de Yukawa ausentes  \cite{Chivukula:1987py,D'Ambrosio:2002ex,Cirigliano:2005ck}.}. 
La elecci\'on est\'a motivada por las exitosas consequencias fenomenol\'ogicas de selectionar
la susodicha simetr\'ia en el caso de la hip\'otesis de Violaci\'on M\'inima de Sabor~\cite{Chivukula:1987py,D'Ambrosio:2002ex,Cirigliano:2005ck,Davidson:2006bd,Alonso:2011jd}, un campo en el
que el autor tambi\'en a trabajado \cite{Alonso:2011jd}. Debe ser destacado que los diferentes
or\'igenes posibles para la masa de los neutrinos resultan en distintas simetr\'ias de sabor en el
sector lept\'onico; de especial relevancia es la elecci\'on del car\'acter Dirac o Majorana. En cualquiera
de los casos la simetr\'ia de sabor no es evidente en el espectro, luego debe estar escondida.
En este trabajo el estudio de rotura espont\'anea de la simetr\'ia de sabor para leptones y quarks
ser\'a desarrollado con \'enfasis en el resultado natural contrastado con la estructura observada
en la naturaleza. Se mostrar\'a como la diferencia entre quarks y leptones en la estructura de sabor
resultante, en particular los \'angulos de mezcla, se origina en la naturaleza Majorana o Dirac de los fermiones.

En el presente an\'alisis, el criterio de naturalidad ser\'a la regla para
decidir si la soluci\'on propuesta es aceptable o introduce
puzles mas complicados  que los que resuelve. Es relevante por lo tanto la acepci\'on de 
naturalidad, siguiendo el criterio de t'Hooft, todos los par\'ametros adimensionales
no restringidos por una simetr\'ia deben ser de orden uno, mientras que todos
los par\'ametros con dimesiones deben ser del orden de la escala de la teor\'ia.
Exploraremos por lo tanto en qu\'e casos este criterio permite la explicaci\'on de la estructura
de masas y angulos de mezcla.

Respecto a las diferentes partes de nueva f\'isica involucradas conviene distinguir tres escalas
distintas i) la escala de RESE establecida por la masa del bos\'on W, ii) un escala posiblemente
distinta de sabor, denotada $\Lambda_f$ y caracter\'istica de la nueva f\'isica responsable de la estructura
de sabor, iii) la escala efectiva de violaci\'on de numero lept\'onico $M$ responsable de las masas de los
neutrinos, en el caso de que \'estas sean de Majorana.

\chapter{Aim and Motivation} 

The field of particle physics is presently at a turning point. 
The exploration
of the mechanism of electroweak symmetry breaking (EWSB) at the LHC
has unveiled the presence of a boson with the characteristics of the Higgs scalar \cite{Aad:2012tfa,Chatrchyan:2012ufa}
given the precision of presently available data \cite{Corbett:2012dm,Giardino:2013bma}. The Standard Model (SM) description of 
mass generation \cite{Englert:1964et,Higgs:1964ia,Higgs:1964pj} has proven successful, and the Higgs self-interaction
that triggers EWSB stands now as the fifth force in nature, after gravity, electromagnetism,
weak and strong interactions.

This new force, as every other quantized force
in nature, varies in strength depending on the scale at which it is probed
but, unlike for strong or weak forces, this poses a problem \cite{Callaway:1988ya} as at
a high energy or short distance scale of order $10^{-12}fm$ the
mechanism of electroweak symmetry breaking would be destabilized 
since the coupling of this force vanishes \cite{Degrassi:2012ry,EliasMiro:2011aa}.
This problem could be solved by the introduction of new physics
which brings the discussion to another theoretical issue, 
the Hierarchy Problem. The point is that any new physics 
that couples to the Higgs particle produces generically a radiative contribution
to the Higgs mass term of order of the new mass scale,
which would mean that the electroweak scale is naturally close to the highest new physics
scale that couples to the SM fields. Proposals to address this
problem can be classified in perturbative physics solutions, the paradigm
being supersymmetry, and strong dynamics ansatzs.
Supersymmetry is an elegant symmetry between bosons and fermions
that implies systematic cancellations among the contributions to the Higgs mass term of 
these two types of particles.
On the other hand the hypothesis of the Higgs boson being a bounded state
produced by new strong dynamics implies that the
mechanism of electroweak symmetry breaking of the SM is just an effective description
to be completed by a more fundamental theory. All these hypothesis involve
new physics at the TeV scale and are being crucially tested at the LHC.

In the cosmology front, the gravitational interaction has been the source of 
new challenges in particle physics. The universe is accelerating, something that in 
standard cosmology requires  the presence of ``Dark Energy": a vacuum energy
 whose negative pressure makes the universe expand with increasing
rate. Naive estimates of the contribution to this type of energy from the standard theory
are as far off from the observed value as $120$ orders of magnitude, a fact that emphatically
reflects our ignorance of the nature of Dark Energy.
Furthermore cosmology together with astrophysics brought the solid piece of evidence of extra matter in the
universe which is not baryonic, the so called ``Dark Matter" as another
experimental evidence not explainable within the Standard Model. 
There is an active experimental program for the search of Dark Matter in this
lively sector of particle physics.
The third piece of evidence of new physics in cosmology
stems from one very familiar fact of the visible
universe: it is made out of much more matter than antimatter, and even 
if the SM provides a source for particle over antiparticle abundance in cosmology,
this is not enough to explain the ratio observed today.

The evidence of new physics that concerns more closely the Standard Model is the
fact that neutrinos have shown to be
massive. The data from oscillation experiments revealed that neutrinos have mass, 
a discovery that stands as one of the selected few sound pieces of evidence 
of physics beyond the SM. In this sector the search for 
leptonic CP violation, charged lepton generation transitions and most of all the 
fundamental relation among neutrino particles and antiparticles; their Majorana
or Dirac nature, are exciting and fundamental quests pursued by
ambitious experimental programs.

To complete the list of challenges in particle physics, it shall be
mentioned that there is the pending task of the quantization of gravity and the present
 poor understanding
of the vacuum of QCD embodied in the $\theta$ problem. 

The focus of  this project is a somehow horizontal problem: the flavour puzzle, which is
constituted by the mass and mixing pattern of the known elementary fermions. 
The flavour structure of the particle spectrum is connected in the standard theory to EWSB,
and the masses of neutrinos are an essential part of the flavour puzzle.  EWSB and neutrino masses have been
subject of study in a different context for the PhD candidate: the flavour phenomenology in a
strong EWSB realization \cite{Alonso:2012jc,Alonso:2012pz}, the determination of the general bosonic Lagrangian in 
the same scheme \cite{Alonso:2012px} and the flavour phenomenology of a neutrino mass model~\cite{Alonso:2012ji}
are part of the author's work. The focus of this write-up is nonetheless on the exploration
of a possible explanation of the flavour pattern developed in Refs. \cite{Alonso:2011yg,Alonso:2012fy,onpreparation,onpreparation2}.

The gauge principle can be singled out
as the driving engine of progress in particle physics, well understood and
elegantly realized in the SM. In contrast the flavour sector stands since decades 
as the less understood part of the SM.
The SM
displays the flavour pattern merely parametrically, leaving unanswered
questions like the origin of the strong hierarchy in fermion masses or
the presence of large flavour mixing in the lepton sector versus the
little overlap in the quark sector. The flavour 
puzzle stays therefore a fundamental open question in particle physics.

The main guideline behind this work is the use of symmetry to address the
flavour puzzle. Symmetry, that plays a central role in our understanding of particle physics,
is called here to explain the structure of the flavour sector. A number of
different symmetries have been postulated with respect to this problem
\cite{Maiani:2011zz,Froggatt:1978nt,Georgi:1979md,Berezhiani:1983hm,Chivukula:1987py,Barbieri:1995uv,Berezhiani:2001mh,D'Ambrosio:2002ex,Cirigliano:2005ck,Davidson:2006bd,Joshipura:2009gi,Alonso:2011jd,Barbieri:2011ci}.
Here the symmetry will be selected as the largest possible continuous global symmetry arising
in the free theory \footnote{Alternatively defined as the largest possible symmetry in the limit of vanishing Yukawa couplings
 \cite{Chivukula:1987py,D'Ambrosio:2002ex,Cirigliano:2005ck}, to be introduced later.}.
This choice is motivated by the successful phenomenological consequences of selecting this symmetry, 
as in the case of the Minimal Flavour Violation (MFV) ansatz~\cite{Chivukula:1987py,D'Ambrosio:2002ex,Cirigliano:2005ck,Davidson:2006bd, Joshipura:2009gi, Alonso:2011jd},
 a field to which the author has also contributed \cite{Alonso:2011jd}.
It must be underlined that the different possible origins of neutrino masses result in different
flavour symmetries in the lepton sector; of  special relevance is the
choice of Majorana or Dirac nature.
The flavour symmetry in any case is
 not evident in the spectrum, ergo it must be somehow hidden.
 
In this dissertation the study of the mechanism of flavour symmetry breaking
 for both quark and leptons will be carried out with emphasis on its
natural outcome in comparison with the observed flavour pattern.
It will be shown how the difference between quark and leptons in the resulting flavour structure, in particular mixing, 
stems from the Majorana or Dirac  nature of fermions.

In the analysis presented here, naturalness criteria shall be the guide to tell whether
the implementation is acceptable or introduces worse puzzles than
those it solves.
A relevant issue is what will be meant by natural; following 't Hooft's naturalness 
 criteria, all dimensionless  free parameters not 
constrained by a symmetry should be of order one, and all dimensionful ones should be 
of the order of the scale of the  theory. We will thus explore in which cases 
 those criteria allow for an explanation of
the pattern of mixings and mass hierarchies. 

As for the different physics involved in this dissertation, there will be three relevant scales; i) the EWSB scale set by the $W$ mass 
and which in the SM corresponds to the vacuum expectation value (vev) $v$ of the Higgs field; ii) a possible distinct flavour scale $\Lambda_f$ 
characteristic of the new physics underlying  the flavour puzzle; iii) the effective lepton number violation scale $M$ responsible for 
light neutrinos masses, if neutrinos happen to be Majorana particles.






\chapter{Introduction}

\ifpdf
    \graphicspath{{1_introduction/figures/PNG/}{1_introduction/figures/PDF/}{1_introduction/figures/}}
\else
    \graphicspath{{1_introduction/figures/EPS/}{1_introduction/figures/}}
\fi



As all pieces of the Standard Model fall into place when confronted with
experiment, the last one being the discovery of a Higgs-like boson at the LHC 
\cite{Aad:2012tfa,Chatrchyan:2012ufa}, 
one cannot help but stop and wonder at the theory that the scientific community
has carved to describe the majority of phenomena we have tested in the
laboratory.
This theory comprises both the forces we have been able to understand at
the quantum level and the matter sector. The former shall be briefly reviewed first.

\section{Forces of the Standard Model}
Symmetries have shed light in numerous occasions in particle physics, in particular
the understanding of local space-time or gauge symmetries stands as the deepest 
insight in particle physics.
The gauge principle, at the heart of the SM, is as beautifully formulated
as powerful and predictive for describing how particles interact through forces. 
The SM gauge group, \begin{equation}\mathcal{G}=SU(3)_{c}\times SU(2)_L\times U(1)_Y\,,\end{equation} encodes the 
strong, weak and electromagnetic interactions and describes the spin 1 (referred to as vector-boson) elementary 
particle content that mediate these forces. The strong interactions concern those 
particles that transform under $SU(3)_c$ with $c$ standing for color, and are the 
subject of study of quantum chromodynamics (QCD). The electroweak sector
 $SU(2)_L\times U(1)_Y$ comprises the weak isospin group $SU(2)_L$ and the abelian 
 hypercharge group $U(1)_{Y}$ which reduce to the familiar electromagnetic gauge group and
 Fermi interaction
 below the symmetry breaking scale. This part of the theory is specified, in the unbroken phase, given the group and
 the coupling constants of each subgroup, here $g_s$ for $SU(3)_c$, $g$ for $SU(2)_L$
 and $g'$ for $U(1)_Y$ at an energy scale $\mu$. This information is enough to
 know that 8 vector-boson mediate the strong interaction, the so-called gluons, and that
 4 vector bosons enter the electroweak sector: the $Z, W^{\pm}$ and the photon.

The implementation of the gauge principle in a theory that allows the prediction
of observable magnitudes as cross sections, decay rates etc. makes use of Quantum
Field Theory (QFT). In the canonical fashion we write down the Lagrangian density
denoted $\mathscr{L}$, that for the pure gauge sector of the Standard Model reads:
\begin{equation}
\mathscr{L}_{gauge}= -\frac{1}{2}\mbox{Tr}\left\{G^{\mu\nu} G_{\mu\nu}\right\}-\frac{1}{2}\mbox{Tr}\left\{W^{\mu\nu} W_{\mu\nu}\right\}\,
-\frac{1}{4}B^{\mu\nu} B_{\mu\nu},
\end{equation}
where and $\mu$ and $\nu$ are Lorentz indexes and $G_{\mu\nu},W_{\mu\nu}$ and $B_{\mu\nu}$ 
stand for the field strengths of $SU(3)_c\,,\,SU(2)_L$ and $U(1)_Y$ respectively.
This part of the Lagrangian describes forces mediators 
and these mediators self-interaction. 
The field strengths are defined through the covariant derivatives
:
\begin{equation}
D_\mu=\partial_\mu +i g_s \frac{\lambda_i}{2} G_\mu^i+i g \frac{\sigma_i}{2}W^i_\mu+ig' Q_Y B_\mu\,,
\end{equation}
with Gell-Mann matrices $\lambda_i$ as generators of color transformations, Pauli matrices $\sigma_i$ as
weak isospin generators, and $Q_Y$ is the hypercharge of the field that the covariant derivative acts on.
$G_\mu^i$ denote the 8 gluons, $W^i_\mu$ the three weak isospin bosons and $B_\mu$ the hypercharge
mediator. The photon ($A_\mu$) and $Z$ are the usual combination of neutral electroweak bosons: $Z_\mu=\cos\theta_W W_\mu^3-\sin\theta_W B_\mu$, $A_\mu=\sin\theta_W W_\mu^3+\cos\theta_W B_\mu $  and the weak angle, $\tan\theta_W=g'/g$. In terms of the
covariant derivatives the field strengths are defined as:
\begin{equation}
G_{\mu\nu}=-\frac{i}{g_s}\,\left[D_\mu,D_\nu \right]\,,\quad
W_{\mu\nu}=-\frac{i}{g}\,\left[D_\mu,D_\nu \right]\,,\quad 
B_{\mu\nu}=-\frac{i}{g'}\,\left[D_\mu,D_\nu \right]\,,
\end{equation}
where the covariant derivative acts on a fundamental or unit-charge implicit object of the corresponding
gauge subgroup.
However the fact that the $W$ and $Z$ spin-1 bosons are massive requires of the introduction of
further bosonic fields in the theory. This brings the discussion to the electroweak breaking sector.
Masses are not directly implementable in the theory as bare or ``hard" mass
terms are not allowed by the gauge symmetry. The way the
SM describes acquisition of masses is the celebrated Brout-Englert-Higgs mechanism \cite{Englert:1964et,Higgs:1964ia,Higgs:1964pj},
a particularly economic description requiring the addition of a $SU(2)_L$ doublet spin-0 boson (scalar), denoted $H$.
This bosonic field takes a vev and its interactions  with the rest of fields when
expanding around the true vacuum produce mass terms for the gauge bosons.
The interaction of this field with the gauge fields is given by its transformation properties or 
charges, reported in table \ref{HiggsRep}, the masses produced for the $W$ and $Z$ boson
being in turn specified by the vev of the field $\left\langle H \right\rangle\equiv (0,v/\sqrt{2})^T$ together with the coupling constants
$g$ and $g'$.
\begin{table}
\centering
\begin{tabular}{cccc} 
& $ SU(3)_c $ & $SU(2)_L $ & $U(1)_Y$ \\ 
\hline 
$H$   &   1 & 2 & 1/2\\
\end{tabular}
\caption{The Higgs field charges under $\mathcal{G}$}
\label{HiggsRep}
\end{table}
This vev is acquired via the presence of the quartic coupling of the Higgs, the fifth force,
and the negative mass term. These two pieces conform the potential
that triggers EWSB and imply the addition of two new parameters to the theory, explicitly:
\begin{equation}
\mathscr{L}_H=\left(D_\mu H\right)^\dagger D^\mu H - \lambda\left(H^\dagger H-\frac{v^2}{2}\right)^2\,.
\label{HiggsLag}
\end{equation}
where the $v$ is the electroweak scale $v\simeq 246$GeV and $\lambda$ 
the quartic coupling of the Higgs, which can be extracted from the measured Higgs mass
 $\lambda= m_h^2/(2v^2) \simeq 0.13$.
Note that the potential, the second term above, has the minimum
at $\left\langle H^\dagger H \right\rangle =v^2/2$. 

As outlined in the previous section, the Higgs could be elementary
or composite; the paradigm of composite bosons are pions, understood through the
 Goldstone theorem. In the pions
chiral Lagrangian the relevant scale is the pion decay constant $f_\pi$
associated to the strong dynamics, in the analogy with a composite Higgs
the scale is denoted $f$ which, unlike in technicolor \cite{Susskind:1978ms,Dimopoulos:1979es,Dimopoulos:1981xc},
 in Composite Higgs Models \cite{Kaplan:1983fs,Kaplan:1983sm,Agashe:2004rs,Agashe:2006at,Gripaios:2009pe}
 is taken different from the electroweak vev $v$. In the limit in which these two scales are close, a more suitable
parametrization of the Higgs is, alike to the exponential parametrization of the $\sigma$-model,
\begin{equation}
\left(\tilde H\,,\,H\right)=U\, \frac{\left\langle h \right\rangle +h}{\sqrt{2}}\,,
\qquad\qquad U^\dagger U=U U^\dagger =1\, ,
\end{equation}
where $\tilde H=i\sigma_2 H^*$ with $\sigma_2$ the second Pauli matrix in weak isospin space.
 $U$ is a $2 \times 2$ unitary matrix which can be thought of as a space-time dependent
element of the electroweak group and consequently absorbable in a gauge 
transformation while $\left\langle h \right\rangle +h$ is the constant ``radial" component plus
the physical bosonic degree of freedom, both invariant under a gauge transformation. The value
of $\left\langle h \right\rangle$ is fixed by $v$ and $f$.

In this way gauge invariance of the corrections to Eq.~\ref{HiggsLag} concerns the dimensionless $U$ matrix
and its covariant derivatives whereas the series in $H/f$ can be encapsulated in general dimensionless functions
$\mathcal{F}[(\left\langle h \right\rangle +h)/f]$ different for each particular model.

Since both $U$ and $\mathcal{F}$ are dimensionless, the expansion is in powers of momentum (derivatives) over the analogous of
the chiral symmetry breaking scale \cite{Manohar:1983md,GeorgiWeakInt}. The Lagrangian up to chiral dimension 4 in this scheme 
for the bosonic sector was given in \cite{Alonso:2012px} and the flavour phenomenology in this scenario was studied in \cite{Alonso:2012jc,Alonso:2012pz} as part of the authors work
that however does not concern the discussion that follows.

 \section{Matter Content}
The course of the discussion leads now to the matter content of the Standard Model.
Completing the sequence of intrinsic angular momentum, between the spin 1
vector bosons and the spin 0 scalars, the spin $1/2$ ultimate constituents of matter, the elementary
fermions, are placed. These fermions constitute what we are made of and surrounded by. 
Their interactions follow from their transformation properties under the gauge group.
Quarks are those fermions that sense the strong interactions and are classified in three
types according of their electroweak interactions; a weak-isospin doublet $Q_L$ and two singlets $U_R, D_R$ . Leptons do not feel the strong 
but only the electroweak interaction and come in two shapes; a doublet $\ell_L$, and a singlet $E_R$ of $SU(2)_L$. The explicit transformation
properties of the fermions are reported in table \ref{SMReps}.
\begin{table}[htdp]
\centering
\begin{tabular}{cccc} 
& $ SU(3)_c $ & $SU(2)_L $ & $U(1)_Y$ \\ 
\hline 
$Q_L$   &   3  & 2 & 1/6\\
$U_R $   &  3  & 1 & 2/3\\
$D_R$    &  3  & 1  & -1/3\\
$\ell_L$  & 1  & 2 & -1/2\\
$E_R$     &  1 &1  & -1
\end{tabular}
\caption[Fermion content of the SM]{\centering \textbf{Fermion content of the SM} - Transformation properties under the gauge group $\mathcal{G}$.}
\label{SMReps} 
\end{table}

The subscripts $L$ and $R$ refer to the two irreducible components of any fermion; left and right-handed.
Right-handed fermions, in the limit of vanishing mass, have a spin projection on the direction of motion
of $1/2\hbar$ whereas left-handed fermions have the opposite projection, $-1/2\hbar$. These two components
are irreducible in the sense that they are the smallest pieces that transform in a closed form under the
Lorentz group with a spin $1/2$.  
The explicit description of the interaction of fermions with gauge fields is read from the Lagrangian;
\begin{equation}
\mathscr{L}_{matter}=i \sum_{\psi = Q_L}^{E_R} \overline \psi \slashed D \psi\,,
\label{kin}
\end{equation}
where $\slashed D =\gamma_\mu D^\mu$ and $\gamma_\mu$ are the Dirac matrices.

There is a discreet set of representations for the non-abelian groups ($SU(3)_c$ and $SU(2)_L$): the
fundamental representation, the adjoint representation etc.
All fermions transform in the simplest non-trivial of them\footnote{The trivial representation is just not to transform, a case denoted
by ``$1$" in the first to columns of table \ref{SMReps}}: the
fundamental representation, hereby denoted $N$ for $SU(N)$. 
For the abelian part, the representation (charge) assignation can be a priori any real number
normalized to one of the fermion's charges, e.g. $E_R$. There is however yet another predictive feature
in the SM connected to the gauge principle: the extra requirement for the consistency of the theory of the cancellation of
anomalies or the conservation of the symmetry at the quantum level imposes a number of constraints. These constraints,
for one generation, are just enough to fix \emph{all} relative $U(1)_Y$ charges, leaving no arbitrariness in this sector of the SM.

Let us  summarize the simpleness of the Standard Model up to this point;
 we have specified a consistent theory based on local symmetry
described by 4 coupling constants for the 4 quantized forces of nature,
a doublet scalar field acquiring a vev $v$ and a matter content of 5 types of
particles whose transformation properties or ``charges" are chosen from a discreet set.

There is nonetheless an extra direction perpendicular to the previous which 
displays the full spectrum of fermions explicitly, that is, the flavour structure. 
Each of the fermion fields in table \ref{SMReps} appears replicated three times
in the spectrum with wildly varying masses and a connection with the rest of
the replicas given by a unitary mixing matrix. Explicitly:
\begin{align}
Q_L^{\alpha}\,=&\left \{ \left(\begin{array}{c} u_L \\ d_L\end{array}\right)\,,\,\left(\begin{array}{c} c_L \\ s_L\end{array}\right)\,,\,\left(\begin{array}{c} t_L \\ b_L\end{array}\right)\right\},  &U_R^{\alpha}=&\left\{u_R, c_R, t_R\right\},\\[2mm]
   \ell_L^{\alpha}\,=& \left \{ \left(\begin{array}{c} \nu^e_L \\ e_L\end{array}\right)\,,\,\left(\begin{array}{c} \nu^\mu_L \\ \mu_L\end{array}\right)\,,\,\left(\begin{array}{c} \nu^\tau_L \\ \tau_L\end{array}\right)\right\}, & D_R^{\alpha}=&\left\{d_R, s_R, b_R\right\},\\[2mm]
E_R^{\alpha}=&\left\{e_R, \mu_R, \tau_R\right\}\,, && 
\end{align}
where e stands for the electron, $\mu$ for the muon, $\tau$ for the $\tau$-lepton, $u$ for
the up quark, $d$ for the down quark, $c$ for charm, $s$ for strange, $b$ for bottom
and $t$ for the top quark. 
The flavour structure is encoded in the Lagrangian,
\begin{equation}
\mathscr{L}_{fermion-mass}=-\overline{Q}_L Y_U\tilde H U_R -\overline{Q}_L Y_D H D_R -\overline{\ell}_L Y_E E_R H+h.c.+
\mathcal{L}_{\nu-mass}\,,
\end{equation}

 where the $3\times 3$ matrices $Y_U , Y_D, Y_E$ have indices in flavour space.
\subsection{Neutrino Masses}
The character of neutrino masses is not yet known, however if we restrict to the matter content we have 
observed so far, the effective field theory approach displays a suggestive first correction to the SM.
Effective field theory, implicit when discussing the Higgs sector, is a model independent
description of new physics implementing the symmetries and particle content present in the 
known low energy theory. Corrections appear in an
expansion of inverse powers of the new physics scale $M$.
This generic scheme yields a remarkably strong result,
at the first order in the expansion the \emph{only} possible term
produces neutrino Majorana masses after EWSB:
\begin{equation}
\mathscr{L}^{d=5}=\frac{1}{M}\mathcal{O}^W+h.c.\equiv\frac{1}{M}\overline \ell_L^{\alpha} \tilde H\, c_{\alpha\beta}  \,\tilde H^T \ell_L^{c,\beta}+h.c.\,,
\label{WeinOp}
\end{equation}
where $c$ is a matrix of constants in flavour space.
This operator, known as Weinberg's Operator \cite{Weinberg:1979sa}, violates lepton number
 however this does not represent a problem since lepton number is an accidental symmetry of the SM, the fundamental symmetries are
the gauge symmetries.
As to what is the theory that produces this operator, there are three possibilities corresponding
to three different fields as mediators of this interaction: the type I~\cite{Minkowski:1977sc,GellMann:1980vs,Mohapatra:1979ia}, II~\cite{Magg:1980ut,Schechter:1980gr,Wetterich:1981bx,Lazarides:1980nt,Mohapatra:1980yp} and III~\cite{Foot:1988aq,Ma:1998dn} seesaw models.
The mediator could transform as a fermionic singlet of the Standard Model (type I), a scalar triplet
of $SU(2)_L$ (type II) and a fermionic triplet of $SU(2)_L$ (type III) diagrammatically depicted in Fig.~\ref{nj438447f1_online}. 
\figuremacroW{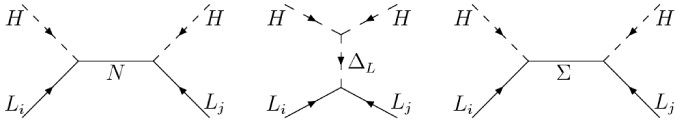}{The three types of seesaw model}{ The three mediator fields that can produce the Weinberg Operator at low energies are the fermion singlet ($N$), the scalar triplet ($\Delta_L$) and the fermion triplet ($\Sigma$). In this figure $L$ stands for the
lepton doublets $\ell_L$.}{1}
\emph{Here we will select the type I seesaw model}
which introduces right-handed neutrinos in analogy with the rest of fermions. These
particles are perfect singlets under the Standard Model, see table \ref{NR}, something that
allows for their Majorana character which is
 transmitted to the left-handed neutrinos detected in experiment through the Yukawa 
couplings.
\begin{table}[htdp]
\centering
\begin{tabular}{cccc} 
& $ SU(3)_c $ & $SU(2)_L $ & $U(1)_Y$ \\ 
\hline 
$N_R$   &   1 & 1 & 0\\
\end{tabular}
\caption{Right-handed neutrino charges under the SM group}
\label{NR}
\end{table}
The complete Lagrangian for the fermion masses is therefore:
\begin{align}\label{fer-mass}
&\mathscr{L}_{fermion-mass}=\mathscr{L}_{Yukawa}+\mathscr{L}_{Majorana}\,,\\ \label{Yukawa}
\mathscr{L}_{Yukawa}=&-\overline{Q}_L Y_U\tilde H U_R -\overline{Q}_L Y_D H D_R -\overline{\ell}_L Y_E E_R H-\overline{\ell}_L Y_\nu\tilde H N_R+h.c.\,,\\ \label{Maj}
\mathscr{L}_{Majorana}=&- \overline {N}_R^c \frac{M}{2} N_R+h.c.\,,
\end{align}
where $M$ is a symmetric $3\times3$ matrix and $N_R$ stands for the right-handed neutrinos which now also enter the
sum of kinetic terms of Eq. \ref{kin}. 
The limit in which the right-handed neutrino scale $M$ is much larger than the Dirac scale $Y_\nu v$ yields as first correction after
integration of the heavy degrees of freedom the Weinberg Operator with the constants $c_{\alpha\beta}$ in Eq.~\ref{WeinOp} being $c_{\alpha\beta}=(Y_\nu Y_\nu^T)_{\alpha\beta}/2$.
\subsection{The Flavour Symmetry}
 If the gauge
part was described around the gauge group one can do the same,
if only formally a priori, for the flavour side. A way to characterize
it is then choosing the largest symmetry that the free theory could
present given the particle content and orthogonal to the gauge group,
 this symmetry is that of the group \cite{Chivukula:1987py,D'Ambrosio:2002ex,Cirigliano:2005ck}:
\begin{align}\nonumber
&\qquad\qquad\qquad\qquad\qquad\mathcal{G_F}=\mathcal{G}_\mathcal F ^q\times \mathcal{G}_\mathcal F^l\,, \\
\mathcal{G}_\mathcal F^q=&SU(3)_{Q_L}\times SU(3)_{U_R}\times SU(3)_{D_R}\times U(1)_B \times U(1)_{A^U}\times U(1)_{A^D} \,,\\
\mathcal{G}_\mathcal F^l=& SU(3)_{\ell_L}\times SU(3)_{E_R}\times O(3)_{N}\times U(1)_L \times  U(1)_{A^l} \,,
\end{align}
It is clear that each $SU(3)$ factor corresponds to the different gauge representation fields which do not
acquire mass in the absence of interactions.
Right-handed neutrinos have however a mass not arising from interactions, but present already in the
free Hamiltonian. Given this fact the largest symmetry possible in this sector is $O(3)$ for the degenerate case: \begin{equation} M=|M|  I_{3\times 3}\,,\end{equation}which is imposed here. The symmetry selected here can alternatively be defined as that arising, for the right-handed neutrino mass matrix of the above form, in the limit $\mathscr{L}_{Yukawa}\rightarrow 0$.
 
 There is an ambiguity in the definition of the lepton sector symmetry and indeed other definitions are present
 in the literature \cite{Davidson:2006bd,Alonso:2011jd}, in particular for the $N_R$ fields a $U(3)_{N_R}$ 
 symmetry is selected if the symmetry is identified with the kinetic term of the matter fields. This option
 leads to a complete parallelism from the symmetry point of view for leptons and quarks and would 
 consequently lead to similar outcomes in an unsuccessful scenario \cite{onpreparation2}.  

Under the non-abelian part of $\mathcal{G_F}$ the matter fields transform as detailed in table \ref{FLRepsNAB}
and the abelian charges are given in table \ref{FLRepsAB}. In the non-abelian
side one can identify $U(1)_B$ as the symmetry that preserves baryon number and $U(1)_L$ as lepton number
which is broken in the full theory here considered. The remaining $U(1)_A$ symmetries are axial rotations in the quark and
lepton sectors.
\begin{table}[h]
\centering
\begin{tabular}{ccccccc}
& $SU(3)_{Q_L}$ & $SU(3)_{U_R}$ & $SU(3)_{D_R}$ & $SU(3)_{\ell_L}$ & $SU(3)_{E_R}$& $O(3)_{N_R}$\\
\hline
$Q_L$ & 3 &1&1&1&1&1\\
$U_R$ & 1 &3&1&1&1&1\\
$D_R$ & 1 &1&3&1&1&1\\
$\ell_L$ & 1 &1&1&3&1&1\\
$E_R$ & 1 &1&1&1&3&1\\
$N_R$ & 1 &1&1&1&1&3
\end{tabular}
\caption[Representations of the fermion fields under the non-abelian part of $\mathcal{G_F}$]{\centering Representation of the fermion fields under the non-abelian part of $\mathcal{G_F}$}
\label{FLRepsNAB}
\end{table}

\begin{table}[h]
\centering
\begin{tabular}{cccccc}
& $U(1)_B$ & $U(1)_{A^U}$ &$U(1)_{A^D}$ &  $U(1)_L$ &  $U(1)_{A^l}$\\
\hline
$Q_L$ & 1/3 &1&1&0&0\\
$U_R$ & 1/3 &-1&0&0&0\\
$D_R$ & 1/3 &0&-1&0&0\\
$\ell_L$ & 0 &0&0&1&1\\
$E_R$ & 0 &0&0&1&-1\\
$N_R$ & 0 &0&0&0&0
\end{tabular}
\caption[Representations of the fermion fields under the abelian part of $\mathcal{G_F}$]{Representation of the fermion fields under the abelian part of $\mathcal{G_F}$}
\label{FLRepsAB}
\end{table}

$\mathscr{L}_{Yukawa}$ is however non vanishing and encodes the flavour structure, our present knowledge about it being displayed in Eqs. \ref{flavpar1}-\ref{flavpartn}.
The masses for fermions range at least 12 orders of magnitude and the
neutrinos are a factor $10^{6}$ lightest than the lightest charged fermion,
something perhaps connected to their possible Majorana nature. Neutrino masses
are not fully determined, only the two mass squared differences and
and upper bound on the overall scale are known. The fact that one of the mass differences
is only known in absolute value implies that not even the hierarchy is known, 
the possibilities being Normal Hierarchy (NH) $m_{\nu_1}<m_{\nu_2}<m_{\nu_3}$ and Inverted Hierarchy (IH) $m_{\nu_3}<m_{\nu_1}<m_{\nu_2}$.

The mixing shape for quarks is close to an identity matrix, with
deviations given by the Cabibbo angle $\lambda_c$, whereas mixing angles
are large in the lepton sector corresponding to all entries of the same order of magnitude in the mixing matrix.
In the lepton sector the CP phase $\delta$ and the Majorana phases, if present,
are yet undetermined. Altogether, our present knowledge of the flavour structure is encoded in the following data,
\begin{align}\label{flavpar1}
m_d&=4.8^{+0.7}_{-0.3}\,\mbox{MeV}\, , &  m_s&=95\pm 5\,\mbox{MeV}\, , & m_b&=4.18\pm0.03\,\mbox{GeV}\, ,\\
m_u&=2.3^{+0.7}_{-0.5} \,\mbox{MeV} \, ,& m_c&=1.275\pm0.025 \,\mbox{GeV}\, ,& m_t&=173.5\pm0.8 \,\mbox{GeV} \, ,
\end{align}
\begin{align}
m_e&=0.510998928\pm0.000000011 \,\mbox{MeV} \, ,\\
 m_\mu&=105.6583715\pm0.0000035 \,\mbox{MeV}\, ,\\
  m_\tau&=1.776.82\pm 0.16\,\mbox{GeV}\, ,
\end{align}
\begin{align}\label{numassdata}
\sum_i m_{\nu_i}&\leq 0.28 \,\mbox{eV}\, , & \Delta m_{\nu_{12}}^2&= 7.5^{+0.2}_{-0.2}10^{-5} \,\mbox{eV}^2\, , &  |\Delta m_{\nu_{23}}^2|&=2.42^{+0.04}_{-0.07}10^{-3} \,\mbox{eV}^2\, ,
\end{align}
\begin{gather}
V_{CKM}=\left(\begin{array}{ccc}\nonumber
1-\lambda_c^2/2&\lambda_c& A\lambda_c^3\left(\rho-i\eta \right)\\
-\lambda_c&1-\lambda_c^2/2& A\lambda_c^2\\
A\lambda_c^3\left(1-\rho-i\eta\right)&-A\lambda_c^2&1\\
\end{array}\right)+\mathcal{O}(\lambda_c^4)\\[2mm]
A\lambda_c^3\left(\rho+i\eta \right)\equiv\frac{A\lambda_c^3\left(\bar\rho+i\bar\eta \right)\sqrt{1-A^2\lambda_c^4}}{\sqrt{1-\lambda_c^2}\left(1-A^2\lambda_c^4(\bar \rho+i\bar \eta)\right)}\, ,\qquad \lambda_c=0.22535\pm0.00065\, ,\\ A=0.811^{+0.022}_{-0.012}\, , \qquad \bar\rho=0.131^{+0.026}_{-0.013}\, , \qquad \bar \eta =0.345^{+0.013}_{-0.014}\, ,
\label{CKMDef}
\end{gather}
\begin{gather}\nonumber \label{pmns}
U_{PMNS}=\left(
\begin{array}{ccc}
c_{12}c_{13}&s_{12}c_{13}&s_{13}e^{-i\delta}\\
-s_{12}c_{23}-c_{12}s_{23}s_{13}e^{i\delta}&c_{12}c_{23}-s_{12}s_{23}s_{13}e^{i\delta}
&s_{23}c_{13}\\
s_{12}s_{23}-c_{12}c_{23}s_{13}e^{i\delta}&-c_{12}s_{23}-s_{12}c_{23}s_{13}e^{i\delta}
&c_{23}c_{13}\\
\end{array}\right)e^{i\alpha_1 \lambda_3+i\alpha_2 \lambda_8}\,,\\ \label{flavpartn}
\theta_{12}=33^{+0.88}_{-0.78}\,^{\circ}\qquad \theta_{23}=40-50\,^{\circ}\qquad \theta_{13}=8,66^{+0.44}_{-0.46}\,^{\circ}\,,
\end{gather}
where the quark data is taken from \cite{Beringer:1900zz}, the neutrino parameters from \cite{GonzalezGarcia:2010er,PhysRevLett.105.031301}, Majorana phases $\alpha_{1}$ and $\alpha_{2}$, are encoded in the exponentials of
the Gell-Mann matrices of Eq.~\ref{flavpartn}, and $\Delta m_{\nu_{ij}}^2=m_{\nu_j}^2-m_{\nu_i}^2$.

The question arises of what becomes of the anomaly cancellation conditions now that
the flavour structure has been made explicit. The conditions are still fixing the relative hypercharges
of all generations provided all masses are different, all mixing angles nontrivial and
Majorana masses for the right-handed neutrinos.

Comparison of the flavour and gauge sector will be useful for the introduction of
the research subject of this thesis.
First, the ratio of certain parameters of the gauge sector, namely hypercharges, 
cannot take arbitrary values but are fixed due to constraints for the  consistency of the theory, 
while the values for the flavour parameters seem all to be equally valid, at least
from the point of view of  consistency and stability. This brings to a second point, 
the inputs that are arbitrary in the gauge sector, $g_s, g, g',\lambda$ 
are smaller but of $\mathcal{O}(1)$ at the typical scale of the theory $\sim M_Z$, whereas
masses span over $6$ orders of magnitude for charged leptons and including
neutrinos the orders of magnitude escalate to $12$.

Because of gauge invariance particles are fitted into representations of the group,
such that the dimension of the representation dictates the number of particles.
There are left-handed charged leptons and left-handed neutrinos to fit a fundamental
representation of $SU(2)_L$, could  it be that something alike happens in the flavour sector? 
That is, is there a symmetry behind the flavour structure?

If this is the case, the symmetry that dictates the representation is not evident at the scale we are familiar with,
 so it should somehow be hidden; we can tell an electron from a muon because they have different masses. 
But the very same thing happens for $SU(2)_L$,  we can tell the neutrino from the electron as
 the electroweak symmetry is broken.

This comparison led neatly to the study carried out.
We shall assume that there is an exact symmetry behind the flavour structure, and
 if so necessarily broken at low energies; a breaking that we will effectively describe
 via a flavour Higgs mechanism. It is the purpose of this dissertation to study the 
 mechanism responsible for the breaking of such flavour symmetry in the search 
 for a deeper explanation of
 the flavour structure of elementary particles.

\chapter{Flavour Physics} 


\ifpdf
    \graphicspath{{X/figures/PNG/}{X/figures/PDF/}{X/figures/}{3/figures/}}
\else
    \graphicspath{{X/figures/EPS/}{X/figures/}{3/figures/}}
\fi

\section{Flavour in the Standard Model  $+$ type I Seesaw Model}

The model that serves as starting point in our discussion is the Standard 
Model with the addition of the type I seesaw model to account for
neutrino masses, the widely accepted as simplest and most natural extension with lepton
number violation.
This chapter will be concerned with flavour phenomenology and the
way it shapes the flavour structure of new physics at the TeV scale, aiming at the understanding
from a bottom up approach of the sources of flavour violation.
The way in which the flavour symmetry is violated in the theory  here considered
is indeed quite specific and yields sharp experimental predictions that we shall examine next.

The energies considered in this chapter are below the electroweak scale, such that the
Lagrangian of Eq.\,\,\ref{fer-mass}, assuming $M\gg v$, after integrating out the heavy
right-handed neutrinos reads
\begin{align}\nonumber
\mathscr{L}_{fermion-mass}=&-\overline{Q}_L Y_U\tilde H U_R -\overline{Q}_L Y_D H D_R+h.c.\\ &-\overline{\ell}_L Y_E E_R H-\overline{\ell}_L \tilde H\frac{Y_\nu Y_\nu^T}{2M}\tilde H^T \ell^c_L +h.c.+\mathcal{O}\left(\frac{1}{M^2}\right)
\label{fer-mass2}
\end{align}
where we recall that the flavour symmetry here considered sets $M_{ij}=M\delta_{ij}$,
a case that shall not obscure the general low energy characteristics of a type I seesaw model
whereas it simplifies the discussion.
The flavour symmetry in this model is only broken by the above Lagrangian, including $1/M^n$ corrections.
In full generality the Yukawa matrices can be written as the product of a unitary matrix,
a diagonal matrix of eigenvalues and a different unitary matrix on the right end. In the case of
the light neutrino mass term, it is more useful to consider the whole product $Y_\nu Y_\nu^T$
which is a transpose general matrix and therefore decomposable in a unitary matrix
and a diagonal matrix. Explicitly, this parametrization for the Yukawa couplings reads:
\begin{align} \label{BIU1}
Y_U&=\mathcal{U}^U_L \mathbf{y}_U\, \mathcal U^U_R\,, & Y_D&=\mathcal U^D_L\mathbf{y}_D\, \mathcal U^D_R\,,  \\ \label{BIU2}
Y_E&=\mathcal U^E_L \mathbf{y}_E \, \mathcal U^E_R\,, & Y_\nu Y_\nu^T&=\mathcal U^\nu_L\hat{\mathbf{y}}_\nu^2\, \mathcal U^{\nu T}_L\,,
\end{align}
where $\mathcal U^{U,D,E,\nu}_{L,R}$ are the unitary matrices and $\mathbf{y}_{U,D,E}$ and $\hat{\mathbf{y}}_\nu$ the diagonal
matrices containing the eigenvalues of charged fermion Yukawa matrices and $Y_\nu Y_\nu^T$ respectively. 
Even if the symmetry is broken, the rest of the SM and type I seesaw Lagrangian stays invariant under
a transformation under the group $\mathcal{G_F}$ of the fermion fields. In particular the rotations:
\begin{align} 
Q_L& \rightarrow \mathcal U_L^D Q_L\,,  & D_R& \rightarrow \mathcal U_R^{D\dagger} D_R\,,  & U_R&=\mathcal U^{R\dagger}_RU_R\,, \\
\ell_L& \rightarrow \mathcal U_L^{E}\ell_L\,, & E_R& \rightarrow \mathcal U_R^{E\dagger} E_R  \,,
\end{align}
simplify the Yukawa matrices in Eqs. \ref{BIU1},\ref{BIU2} after substitution in Eq.~\ref{fer-mass2} to,
\begin{align}\label{YukParQ}
Y_U&=\mathcal U^{D\dagger}_L\,\mathcal U^U_L \mathbf{y}_U\,,  & Y_D&=\mathbf{y}_D\,,  \\ \label{YukParL}
Y_E&= \mathbf{y}_E\,, & Y_\nu Y_\nu^T&=\mathcal U^{E\dagger}_L\mathcal U^\nu_L\, \hat{\mathbf{y}}_\nu^2\, \mathcal U^{\nu T}_L\mathcal U^{E*}_L\,,
\end{align}
which allows to define:
\begin{align}\label{MixMat}
 V_{CKM}^\dagger& \equiv \mathcal U^{D\dagger}_L\,\mathcal U^U_L\,, & U_{PMNS}& \equiv \mathcal U^{E\dagger}_L \mathcal U^\nu_L\,,\\ \label{EigenYukQ}
\mathbf{y}_{U}&=\mbox{Diag}\left(y_u, y_c, y_t\right)\,, &
\mathbf{y}_{D}&=\mbox{Diag}\left(y_d, y_s, y_b\right)\,,\\ \label{EigenYukL}
\hat{\mathbf{y}}_{\nu}&=\mbox{Diag}\left(\hat y_{\nu_1}, \hat y_{\nu_2}, \hat y_{\nu_3}\right)\,,&
\mathbf{y}_{E}&=\mbox{Diag}\left(y_e, y_\mu, y_\tau\right)\,,
\end{align}
 with $V_{CKM}$ being the usual quark mixing matrix and $U_{PMNS}$ the analogous in the
lepton side; the first encodes three angles and one CP-odd phase and the second 
two extra complex Majorana phases on top the the equivalent of the previous 4 parameters. 
The connection of the eigenvalues with masses will be made clear below.

There are a few things to note here. The right handed unitary matrices $\mathcal U_R^{U,D,E}$ are irrelevant,
the appearance of the irreducible mixing matrix in both sectors 
is due to the simultaneous presence of a Yukawa term for both up and down-type quarks involving the 
same quark doublet $Q_L$,
and the neutrino mass term and charged lepton Yukawa where the lepton doublet $\ell_L$ appears. 
Were the mass terms to commute
there would be no mixing matrix. Were the weak isospin group not present to bind together $u_L$ with
$d_L$ and $\nu_L$ with $e_L$ there would not either be mixing matrix. Weak interactions in conjunction
with mass terms violate flavour. Although mixing matrices
are there and nontrivial it is useful to have in mind these considerations
to remember how they arise.

 After EWSB the independent rotation of the two upper components of the weak isospin doublets,
\begin{align} 
U_L& \rightarrow V_{CKM}^\dagger\, U_L \,, & \nu_L& \rightarrow U_{PMNS}\, \nu_L\,,
\label{rotLH}
\end{align}
takes to the mass basis yielding the Yukawa couplings diagonal,
\begin{align}
\mathscr{L}_{fermion-mass}=& -\frac{y_{\alpha}\,\left(v+h\right)}{\sqrt{2}}\,\overline{U}_L ^{\alpha } U_R^{\alpha} -\frac{y_{i}\,\left(v+h\right)}{\sqrt{2}}\,\overline{D}_L ^{i } D_R^{i}\\&
-\frac{y_{\beta}\left(v+h\right)}{\sqrt{2}}\,\overline{E}_L ^{\beta } E_R^{\beta}-\frac{\hat{y}_{\nu_j}^2\left(v+h\right)^2}{4M}\,\overline{\nu}_L ^{j} \nu_L^{c\,,j}+h.c.\,,
\end{align}
were $h$ is the physical Higgs boson, the unitary gauge has been chosen and hereby greek indices run over up-type quark and charged lepton
mass states and latin indixes over down-type quark and neutrino mass states, see Eqs. \ref{EigenYukQ},\ref{EigenYukL}.

We read from the above that the masses for the charged fermions are $m_\alpha =y_\alpha v/\sqrt{2}=y_\alpha \times 174$GeV
whereas for neutrinos $m_{\nu_i}=\hat y_{\nu_i}^2v^2/(2M)$. The values of masses then fix the Yukawa eigenvalues for the
charged fermions to be:
\begin{align}
\left\{y_t\,,\,y_c\,,\,y_u\right\}=& \,\left\{1.0\,,\,7.3\times 10^{-3}\,,\,1.3\times 10^{-5}\right\}\,,\\
\left\{y_b\,,\,y_s\,,\,y_d\right\}=& \,\left\{2.4\times 10^{-2}\,,\,5.5\times 10^{-4}\,,\,2.7\times 10^{-5}\right\} \,,\\
\left\{y_\tau\,,\,y_\mu\,,\,y_e\right\}=& \,\left\{1.0\times 10^{-2}\,,6.0\times 10^{-4}\,,\,2.9\times 10^{-6}\right\} \,,
\label{YukHKS}
\end{align}
whereas for neutrinos only the mass squared differences are know and the upper bound of Eq. \ref{numassdata} sets $\hat y_{\nu}^2v^2/M\lesssim$\,eV. The values for the Yukawa eigenvalues of the charged fermions display quantitatively the hierarchies in the flavour sector since, as dimensionless couplings of the theory,
they are naturally expected of $\mathcal{O}(1)$, something only satisfied by the top Yukawa\footnote{If the neutrino Yukawa couplings are taken
to be order one the upper bound on masses points towards a GUT \cite{Georgi:1974sy} scale $\sim 10^{15}$GeV for M.}. The smallness
of the eigenvalues is nonetheless stable under corrections since in the limit of vanishing Yukawa eigenvalue a
chiral symmetry arises, which differentiates this fine-tuning from the Hierarchy Problem.

The rest of the Lagrangian does not notice the rotation in Eq.~\ref{rotLH} except for 
the couplings of weak isospin $+1/2$ and $-1/2$ particles:
\begin{equation}
\mathscr{L}_{CC}= i \frac{g}{\sqrt{2}} \overline U_L V_{CKM } \slashed W^+ D_L+i \frac{g}{\sqrt{2}} \overline \nu_L U_{PMNS}^\dagger \slashed W^+ E_L+h.c.\,.
\end{equation}
The rest of couplings, which involve neutral gauge bosons, are diagonal in flavour, to order $1/M^2$. The flavour
changing source has shifted therefore in the mass basis to the couplings of fermions to the $W^{\pm}$ gauge 
bosons. This is in accordance with the statement of the need of both weak isospin and mass terms
for flavour violation.

This process allows to give a physical definition of the unitary matrices
entering the Yukawa couplings: \emph{mixing matrices parametrize the change of basis from the interaction to the mass basis}. This is a 
more general statement than the explicit writing of Yukawa terms or the specification of the character of 
neutrino masses.

The absence of flavour violation in neutral currents
implies the well known and elegant explanation of the smallness of flavour changing neutral 
currents (FCNC) of the Glashow-Iliopoulos-Maiani (GIM) mechanism \cite{Glashow:1970gm}. All neutral current flavour processes are
loop level induced and suppressed by unitarity relations to be proportional to mass differences
and mixing parameters, an achievement of the standard theory that helped greatly to its consolidation.
At the same time this smallness of flavour changing neutral currents stands as a fire proof for
theories that intend to extend the Standard Model, as we shall see next.


\section{Flavour Beyond the Standard Model}

The flavour pattern of elementary particles has been approached in a number 
of theoretical frameworks aiming at its explanation. Shedding light in a problem
as involved as the flavour puzzle has proven not an easy task. Proposed 
explanations are in general partial, in particular reconciling neutrino flavour data
with quark and charged lepton hierarchies in a convincing common framework
is a pending task in the authors view.

In the following a number of the proposed answers to explain flavour are listed,

\begin{itemize}
\item \emph{Froggat -Nielsen theories}. The introduction of an abelian symmetry $R$ under which the different generation
fermions with different chirality have different charges and that is broken by the vev of a field $\left\langle\phi_0\right\rangle$ can explain the hierarchies in the flavour pattern \cite{Froggatt:1978nt}. In this set-up there are extra chiral fermions at a high scale with a
typical mass $\Lambda_f$ such that the magnitude $\epsilon=\left\langle\phi_0\right\rangle/\Lambda_f$ controls the breaking of the abelian symmetry $R$. Interactions among the different
fermions are mediated by the field $\phi_0$ at the high scale and its acquisition of a vev at the low scale implies
 factors of $\epsilon^{a_i+b_j}$ for the coupling of different flavour and chirality fermions $\Psi_{L_i}$, $\Psi_{R_j}$ with  charges $R_{L_i}=b_i$ and $R_{R_j}=-a_j$  normalized to the charge of $\phi_0$ ($R_{\phi_0}=1$). The mass matrix produced in this way contains hierarchies among masses controlled by 
 $m_i/m_j\sim\epsilon^{a_i-a_j+b_i-b_j}$ whereas angles are given by $U_{ij}\sim (m_i/m_j)^{C_{ij}}\gtrsim (m_i/m_j) $.
This symmetry based argument stands as one of the simplest and most illuminating approaches to the flavour puzzle.
\item \emph{Discrete symmetries} discrete symmetries were studied as possible explanations for the flavour pattern in the quark
sector, e.g. \cite{Pakvasa:1977in}, but the main focus today is on the lepton mixing pattern.
 The values of the atmospheric and solar angles motivated
 proposals of values for the angles given by simple integer ratios like the tri-bimaximal mixing pattern \cite{Harrison:2002er} ($\theta_{23}=\pi/4\,,\,\theta_{12}=$arcsin$(1/\sqrt{3}) \,,\,\theta_{13}=0$) . These patterns were later shown to
 be obtainable with breaking patterns of relatively natural discrete symmetries like $A_4$\cite{Altarelli:2005yp,Altarelli:2005yx,Altarelli:2006kg} or $S_4$ \cite{Bazzocchi:2009pv,Altarelli:2009gn}. A discrete flavour treatment of 
 both quarks and leptons requires generally
 of extra assumptions like distinct breaking patterns in distinct fermion sectors
 which have to be kept separate, see e.g. \cite{Kubo:2003iw,Feruglio:2007uu}.
These models though are now in tension with the relatively large reactor angle and new
approaches are being pursued \cite{Feruglio:2013hia,Holthausen:2012dk}. This approach
has the advantage of avoiding Goldstone bosons when breaking the discrete symmetry but the drawback of the ambiguity
in choosing the group.
\item \emph{Extra Dimensions} The case of extra dimension offers a different explanation for the hierarchy in
masses. In Randall-Sundrum models \cite{Randall:1999ee,Randall:1999vf} the presence of two 4d branes in a 5 dimensional
space induces a metric with an overall normalization or warp factor that is exponentially decreasing with the fifth dimension
and that offers an explanation of the huge hierarchy among the Planck and EW scale in terms of $\mathcal{O}(1)$ fundamental parameters.
When the fermions are allowed to propagate in the fifth dimension, rather than being confined in a brane, their profile in
the fifth dimension determined by the warp factor and a bulk mass term provides exponential factors for the Yukawa couplings
as well, offering an explanation of the flavour pattern in terms of $\mathcal{O}(1)$ fundamental or 5th dimensional parameters \cite{Grossman:1999ra,Gherghetta:2000kr}.
In large extra dimensions theories, submilimiter new spacial directions can provide geometrical factors to explain the hierarchy problem
\cite{ArkaniHamed:1998rs}.
In this scenario, if we live on a ``fat" brane in which the fermion profiles are localized, the mixing among generations is
suppressed by the overlap of this profiles rather than symmetry arguments \cite{ArkaniHamed:1999dc,Mirabelli:1999ks,Branco:2000rb}.
In the extradimensional paradigm in general  therefore the explanation of the hierarchies in flavour is found in
geometry rather than symmetry.

\item \emph{Anarchy} The possibility of the flavour parameters being just random numbers without
any utter reason has been also explored \cite{Hall:1999sn,deGouvea:2003xe}, and even if the recent measurement
of a ``large" $\theta_{13}$ lepton mixing angle favors this hypothesis for the neutrino mass matrix \cite{deGouvea:2012ac},
the strongly hierarchical pattern of masses and mixing of charged fermions is not natural in this framework.
\end{itemize}

These models introduce in general new physics coupled to the flavour sector of the Standard Model,
which means modifying the phenomenological pattern too. This observation applies to other models as well: any new
physics that couples to the SM flavour sector will change the predictions for observables
and shall be contrasted with data. This is examined next.

\section{Flavour Phenomenology}
Once again the effective field theory is put to use,
\begin{equation}
\mathscr{L}=\mathscr{L}_{SM}+ \frac{1}{M}\mathcal{O}^W+\sum \frac{c_i}{\Lambda_f^2}\mathcal{O}^i+\mathcal{O}(1/\Lambda_f^3)
\label{genLag}
\end{equation}
This Lagrangian describes the Standard Model theory, represented by
the first term, plus new 
physics corrections in a very general manner encoded in the two next terms. 
The first correction in Eq.~\ref{genLag} is the Weinberg Operator of Eq.~\ref{WeinOp} which has already been examined and taken into account.
The next corrections have a different scale $\Lambda_f$ motivated by naturalness criteria.
In this category we include the operators that do not break lepton number
nor baryon number, listed in 
\cite{Buchmuller:1985jz} and only recently reduced to the minimum
set via equations of motion \cite{Grzadkowski:2010es}. Therefore they need
not be suppressed by the lepton number violation scale $M$. There are nonetheless contributions
of $1/M^2$ in Eq.~\ref{genLag}, but these are too small for phenomenological
purposes after applying the upper bound from neutrino masses. Let us note that in certain seesaw models
the lepton number and flavour scales are separated \cite{Wyler:1982dd,Mohapatra:1986bd,Branco:1988ex,Raidal:2004vt,Gavela:2009cd}, such that
their low energy phenomenology falls in the description above \cite{Broncano:2002rw,Antusch:2006vwa}.
As a concrete example of a modification to the SM a possible operator at order $1/\Lambda_f^2$ is:
\begin{equation}
c_6\,\mathcal{O}^6= c_{\alpha\beta\sigma\rho}\bar Q_L^{\alpha} \gamma_\mu Q_L^\beta \bar Q_L^\sigma \gamma^\mu Q_L^\rho \,,
\end{equation}
where greek indices run over different flavours and the constants $c_{\alpha\beta\sigma\rho}$
are the coefficients different in general for each flavour combination.
\figuremacroW{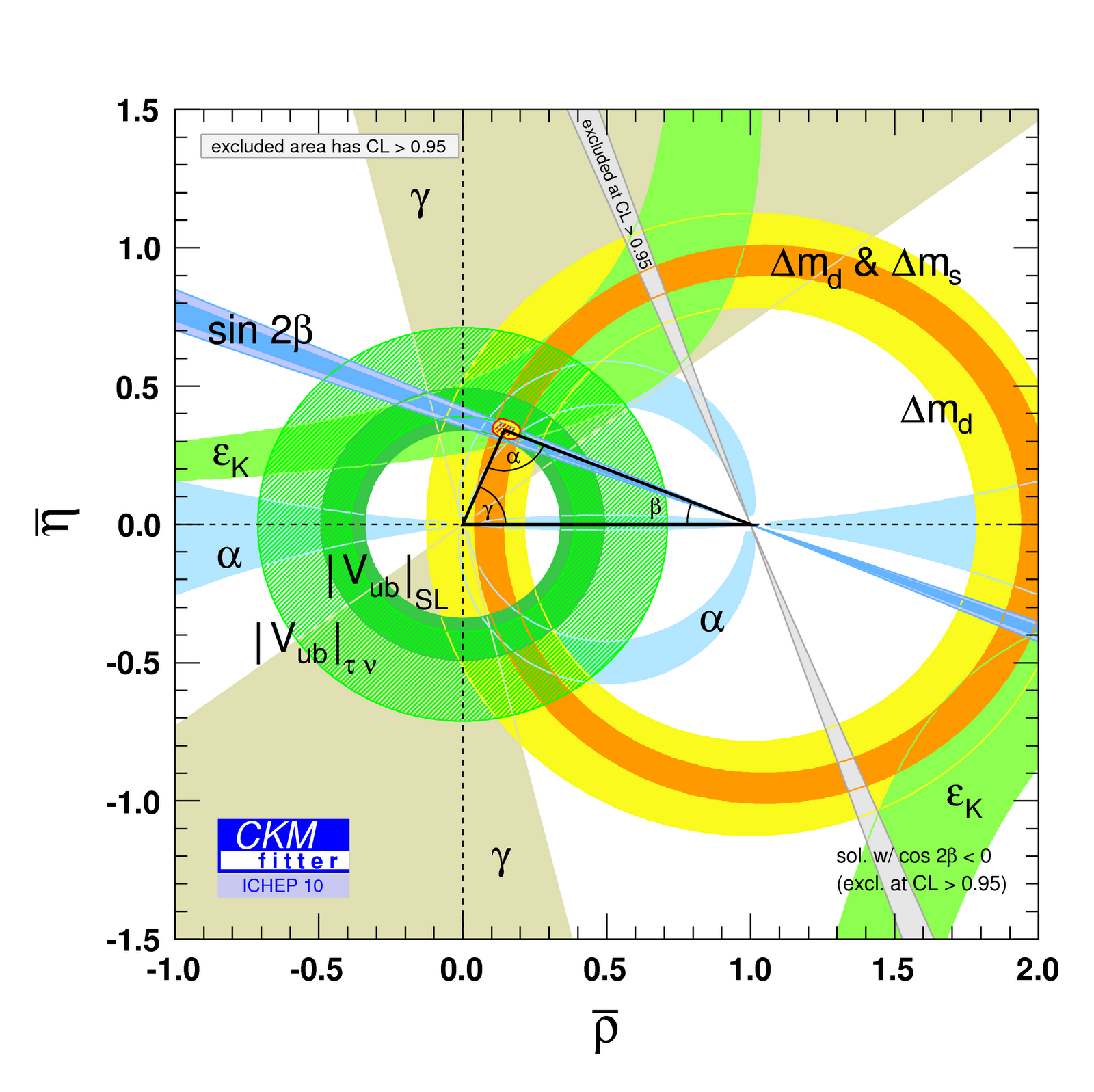}{ Experimentally allowed regions for the CKM mixing parameters $\bar \rho$ and $\bar \eta$}{The overlap of the experimentally allowed regions extracted from kaon and B-meson observables in the $\bar \rho-\bar \eta$ plane, mixing parameters defined in Eq.~\ref{CKMDef}, shows the good agreement of the SM with the flavour data.}{.5}
The modification induced by this term in observable quantities can be computed and compared with data.
A wide and ambitious set of experiments has provided the rich
present amount of flavour data; from the precise branching ratios of B mesons in B factories
to the search for flavour violation in the charged lepton sector, all the D and K meson
observables, and if we include CP violation, the stringent electric dipole moments.

Contrast of the experimental data with expectations has led, in most occasions,
 to a corroboration of the Standard Model in spite of new physics, and at times certain
hints of deviations from the standard theory raised hopes  
(a partial list is \cite{Abazov:2010hv,Aaij:2011in,Aguilar:2001ty})
that either were washed away afterwards, or stand as of today inconclusive.
It is the case then that no clear proof of
physics other than the SM and neutrino masses driving flavour data has been found. 

Indeed the data has been not only enough to determine the
flavour parameters of the SM but also to impose stress tests on the theory, all faintlessly passed.
Fig. \ref{ckmfitter} shows how all experimentally allowed 
regions in the mixing parameter plane of $\bar\rho-\bar \eta$, variables defined in Eq. \ref{CKMDef},
meet around the allowed value.
\begin{table}[h]
\centering
\begin{tabular}{cccc}
Operator & Bounds on $\Lambda_f$ (TeV) & Bounds on $c$ ($\Lambda_f=1$TeV) &  Observables\\
&$c=1$\,\,\,\,\,\,\,\,\,\,\,\,\,\,\,\,$c=i$&$\mathfrak{Re}(c)$\,\,\,\,\,\,\,\,\,\,\,\,\,\,\,\,\,$\mathfrak{Im}(c)$&\\
\hline
$\left(s_L\gamma_\mu d_L\right)^2 $& $ 9.8\times10^2 \,\,\,\,\,\,1.6\times 10^4$ & $9.0\times 10^{-7} \,\,\,\,\,\,3.4\times 10^{-9}$&$\Delta m_K,\epsilon_K$\\
$(s_R d_L)(s_Ld_R) $&$1.8\times10^4 \,\,\,\,\,\,3.2\times10^5$&$6.9\times10^{-9}\,\,\,\,\,\,2.6\times10^{-11} $&$\Delta m_K,\epsilon_K$\\
$(c_L\gamma_\mu u_L)^2 $&$1.2 \times 10^3\,\,\,\,\,\, 2.9 \times 10^3 $&$ 5.6 \times 10^{-7}\,\,\,\,\,\,1.0 \times 10^{-7} $&$\Delta m_D; |q/p|;\phi_D$\\
$(c_R u_L)(c_Lu_R) $&$ 6.2 \times 10^3 \,\,\,\,\,\,1.5 \times 10^4 $&$5.7\times  10^{-8} \,\,\,\,\,\,1.1 \times 10^{-8}$&$ \Delta m_D; |q/p|; \phi_D$\\
$(b_L\gamma_\mu d_L)^2 $&$ 6.6 \times 10^2 \,\,\,\,\,\,9.3 \times 10^2 $&$ 2.3 \times  10^{-6}\,\,\,\,\,\, 1.1 \times 10^{-6}$&$ \Delta m_{B_d} ; S_{\Psi K_S}$\\
$(b_R d_L)(b_Ld_R)$&$  2.5 \times 10^3 \,\,\,\,\,\,3.6 \times 10^3$&$  3.9\times  10^{-7}\,\,\,\,\,\, 1.9 \times 10^{-7} $&$\Delta m_{B_d} ; S_{\Psi K_S}$\\
$(b_L\gamma_\mu s_L)^2 $&$1.4\times 10^2\,\,\,\,\,\, 2.5 \times 10^2 $&$5.0 \times 10^{-5} \,\,\,\,\,\,1.7 \times 10^{-5} $&$\Delta m_{B_s} ; S_{\Psi \Phi} $\\
$(b_R s_L)(b_Ls_R)$&$ 4.8 \times 10^2\,\,\,\,\,\,8.3 \times 10^2 $&$8.8 \times 10^{-6}\,\,\,\,\,\, 2.9 \times 10^{-6}$&$ \Delta m_{B_s} ; S_{\Psi \Phi}$\\
$F^{\mu\nu} \bar \mu_R \sigma_{\mu\nu} e_L$&$6.1 \times 10^4\,\,\,\,\,\,6.1 \times 10^4$&$2.7 \times 10^{-10}\,\,\,\,\,\,2.7 \times 10^{-10}$&$\mu\rightarrow e\gamma$\\
$(\mu_L\gamma_\mu e_L)(u_L\gamma_\mu u_L)$ &$4.9\times10^2\,\,\,\,\,\,4.9\times10^2$&$4.1\times10^{-6}\,\,\,\,\,\,4.1\times10^{-6}$&$\mu\rightarrow e(Ti)$\\
$(\mu_L\gamma_\mu e_L)(d_L\gamma_\mu d_L)$ &$5.4\times10^2\,\,\,\,\,\,5.4\times10^2$&$3.5\times10^{-6}\,\,\,\,\,\,3.5\times10^{-6}$&$\mu\rightarrow e(Ti)$\\
\end{tabular}
\caption{Bounds on the different operators, see text for details.}
\label{BoundsEFT}
\end{table}
The absence of new physics evidence translates in bounds on the new physics scale, reported in table \ref{BoundsEFT}.
When placing the bounds, the magnitude that is constrained is the combination $c/\Lambda_f^2$ as is the one appearing
in the Lagrangian of Eq.~\ref{genLag}.
Naturalness criteria points at constants $c$ of $\mathcal{O}(1)$, a case reported in table \ref{BoundsEFT} both for
CP conservation $c=1$ (second column) and CP violation $c=i$ (third column). On the other hand if the scale is fixed at
the TeV then the constants have severe upper bounds as the fourth and fifth columns in
table \ref{BoundsEFT} show.  The quark bounds are taken from \cite{Isidori:2013ez}
whereas the lepton data is taken from \cite{Adam:2013mnn,Dohmen:1993mp} and computed
with the formulae of \cite{Alonso:2012ji}.


\section{Minimal Flavour Violation}

The bounds on new physics place a dilemma: either giving up new physics
till the thousands of TeVs scale and with it the possibility of any direct test in laboratories,
or assume that the flavour structure of new physics is highly non-generic or fined-tuned.

A solution to this dichotomy is the celebrated Minimal Flavour Violation scheme 
\cite{D'Ambrosio:2002ex,Cirigliano:2005ck,Davidson:2006bd,Alonso:2011jd}
which is predictive, realistic, model independent and symmetry driven.
The previous section showed that flavour phenomenology at present is
explained by the SM plus neutrino masses solely, this is to say that
 the mass terms contain all the known flavour structure and ergo determine the
flavour violation. The conclusion is that the mass terms are the only source for \emph{all} flavour and CP violation
data at our disposal. The minimality assumption of MFV is to upgrade this source
to be the only one in physics Beyond the Standard Model at low energies.

In the absence of the mass terms the theory presents a symmetry which is
formally conserved if the sources of flavour violation are assigned
transformation properties, given in table \ref{YukReps} for the present realization.
\begin{table}[h]
\centering
\begin{tabular}{ccccccc}
& $SU(3)_{Q_L}$ & $SU(3)_{U_R}$ & $SU(3)_{D_R}$ & $SU(3)_{\ell_L}$ & $SU(3)_{E_R}$&$O(3)_{N_R}$\\
\hline
$Y_U$ & 3 &$\bar 3$&1&1&1&1\\
$Y_D$ & 3&1 &$\bar 3$&1&1&1\\
$Y_E$ & 1 &1&1&$3$&$\bar 3$&1\\
$Y_\nu$ & 1 &1&1&$3$&$1$&3\\
\end{tabular}
\caption{Spurious transformations of the Yukawa couplings under $\mathcal{G}_\mathcal{F}$}
\label{YukReps}
\end{table}
The formal restoration of the flavour symmetry applied in the
effective field theory set-up determines the 
flavour constants which shall be such as to form flavour invariant combinations
with the matter fields and built out of the sole sources of flavour violation at low
energies, the Yukawas. The previous operator will serve as example now:
\begin{equation}\label{ExMFV}
c_6\,\mathcal{O}^6= \bar Q_L^{\alpha}\left(Y_UY_U^\dagger\right)_{\alpha\beta} \gamma_\mu Q_L^\beta \bar Q_L^\sigma\left(Y_UY_U^\dagger\right)_{\sigma\rho} \gamma^\mu Q_L^\rho \,.
\end{equation}
where the transformations listed in Tables \ref{FLRepsNAB}\,,\,\ref{YukReps} leave the above construction invariant.
The Yukawa couplings, can be written as in Eqs.~\ref{YukParQ}\,,\,\ref{MixMat}\,,\,\ref{EigenYukQ} and therefore all parameters
entering the example of Eq. \ref{ExMFV} are known; they are just masses and mixings.

It should be underlined that MFV is not a model of flavour and the value of the new dynamical flavour scale $\Lambda_f$ is not fixed, 
however the suppression introduced via the flavour parameters makes this scale compatible with the TeV, see
\cite{Isidori:2012ts} for a recent analysis. What it does predict is precise and constrained relations between different flavour transitions.

\include{XYZ}

\chapter{Spontaneous Flavour Symmetry Breaking} 


\ifpdf
    \graphicspath{{X/figures/PNG/}{X/figures/PDF/}{X/figures/}}
\else
    \graphicspath{{X/figures/EPS/}{X/figures/}}
\fi


The previous chapter illustrated how the entire body of flavour data can be explained through a
single entity, the mass terms. This has been shown to be the only culprit of flavour violation.
If we pause and look at the previous sentence, 
it is interesting to see how the jargon itself already assumes that there is something to
be violated and implicitly a breaking idea. 
It has been shown that the symmetry of the matter content of the free theory here considered is the product of the gauge and flavour symmetries;
$\mathcal{G}\times\mathcal{G_F}$, and that Yukawa terms do not respect $\mathcal{G_F}$. Subgroups of this
group could also be considered, here the full $\mathcal{G_F}$ is adopted in the general case,
although in certain cases the axial abelian factors $U(1)_{A}$ will be dropped\footnote{Or alternatively broken by a different mechanism, like a Froggat-Nielsen model.}. The case of conservation of the full $\mathcal{G_F}$ group is also denoted
\emph{axial conserving case}, whereas assuming that the $U(1)_A$ symmetries are not exact will constitute the \emph{explicitly axial breaking}
case $\mathcal{G}_\mathcal{F}^{\slashed{A}}\sim SU(3)^5\times SO(3)$. In all cases the full non-abelian group is considered.

The ansatz of MFV showed the usefulness of assigning spurious
transformation properties to the Yukawa couplings and having a formal flavour conservation
at the phenomenological level.
It is only natural to take the next step and assume the flavour symmetry is exact at some
high energy scale $\Lambda_f$ and the Yukawa couplings are the remains
of fields that had real transformations properties under this symmetry.
The underlying idea of dynamical Yukawa couplings is depicted in Fig.~\ref{DibujoFlavour} 
which resembles similar diagrams in Froggat-Nielsen theories.
The basic assumption is indeed already present in the literature; for example in the first formulation of MFV by 
Chivukula and Georgi~\cite{Chivukula:1987py}, the Yukawa couplings corresponded to a fermion condensate.
It should be mentioned that a flavour breaking mechanism with different 
 continuos non-abelian groups than the here considered has been explored in the quark \cite{Maiani:2011zz,Michel:1970mua,Michel:1971th,Berezhiani:2005tp,Berezhiani:2001mh} and lepton \cite{Jenkins:2009dy,Hanany:2010vu,Blankenburg:2012nx} sectors,
whereas the invariant pieces needed to construct a potential were made explicit and analyzed for quarks and the group $\mathcal G_\mathcal F^q$
 in Refs.~\cite{Feldmann:2009dc,Jenkins:2009dy,Hanany:2010vu}.
The quantum corrections to the work of Ref.~\cite{Alonso:2011yg} were studied in Refs.~\cite{Nardi:2011st,Espinosa:2012uu}.

The analysis of a two generation case will serve as illustration and guide in the next chapter,
for this reason it is useful and compact to introduce $n_g$ for the number of generations.
The straight-forward generalization of the flavour group is then:
\begin{align}\nonumber
&\qquad\qquad\qquad\qquad\qquad\qquad\mathcal{G_F}=\mathcal{G}_\mathcal F^q\times \mathcal{G}_\mathcal F^l\, , \\
\mathcal{G}_\mathcal F^q=&SU(n_g)_{Q_L}\times SU(n_g)_{U_R}\times SU(n_g)_{D_R}\times U(1)_B \times U(1)_{A^U}\times U(1)_{A^D}\, , \\
\mathcal{G}_\mathcal F^l=& SU(n_g)_{\ell_L}\times SU(n_g)_{E_R}\times O(n_g)_{N}\times U(1)_L \times  U(1)_{A^l} \,.
\end{align}
\figuremacroW{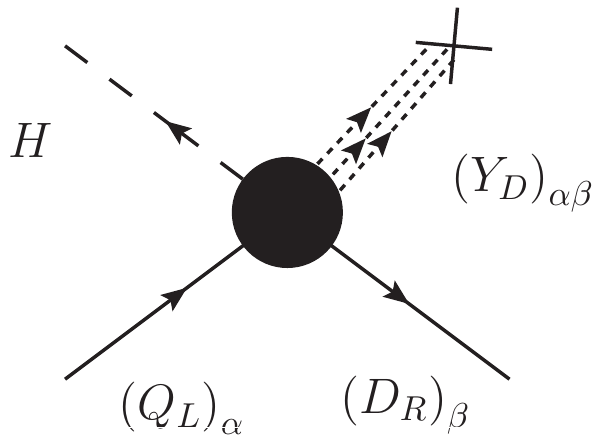}{Yukawa Couplings as vevs of flavour fields}{}{0.4}
\section{Flavour Fields Representation}

The starting point is rendering the Yukawa interaction explicitly invariant under the flavour symmetry.
At the scale $\Lambda_f$ of the new fields responsible for flavour breaking, the
Yukawa couplings will be dynamical themselves, implying the mass dimension of the \emph{Yukawa
Operator} is now $>4$. 
\subsection*{Scalar Flavour Fields in the Bi-Fundamental}
In the effective field theory expansion, the leading term is 
dimension 5\footnote{The expansion now differs from the EFT in the SM context since we have introduced new scalar fields.}:
\begin{equation}
\mathscr{L}_{Yukawa}=\overline{Q}_L\dfrac{\mathcal{Y}_D}{\Lambda_f}D_RH+\overline{Q}_L\dfrac{\mathcal{Y}_U}{\Lambda_f}U_R\tilde{H}+\overline{\ell}_L\dfrac{\mathcal{Y}_E}{\Lambda_f}E_RH+\overline{\ell}_L\dfrac{\mathcal{Y}_\nu}{\Lambda_f}N_R\tilde H+h.c.\,,
\label{YukOps5}
\end{equation}
where there is the need to introduce the cut-off scale $\Lambda_f$,\footnote{The equation above could have, in more generality, coupling constants different for the
up and down sector or equivalently a different scale, here the scale is chosen the same for simplicity.}
the scalar fields $\mathcal{Y}_D$, $\mathcal{Y}_U$, $\mathcal{Y}_E$  and $\mathcal{Y}_\nu$ are dynamical fields in the bi-fundamental representation as detailed in tables \ref{BIF-SFQ},\ref{BIF-SFL},
\begin{table}[h]
\centering
\begin{tabular}{ccccccc}
& $SU(n_g)_{Q_L}$ & $SU(n_g)_{U_R}$ & $SU(n_g)_{D_R}$ & $U(1)_B$ & $U(1)_{A^U}$& $U(1)_{A^D}$\\
\hline
$\mathcal{Y}_U$ & $n_g$ &$ \bar n_g $&1&0& 2& 1 \\
$\mathcal{Y}_D $ & $n_g$ &1&$ \bar n_g $&0& 1& 2 \\
\end{tabular}
\caption{$\mathcal{G}_\mathcal{F}^q$ representation of the quark sector bi-fundamental scalar fields for $n_g$ fermion generations}
\label{BIF-SFQ}
\end{table}%
\begin{table}[h]
\centering
\begin{tabular}{cccccc}
& $SU(n_g)_{\ell_L}$ & $SU(n_g)_{E_R}$ & $O(n_g)_{N_R}$ & $U(1)_L$ & $U(1)_{A^l}$\\
\hline
$\mathcal{Y}_E$ & $n_g$ &$ \bar n_g $&1&0& 2 \\
$\mathcal{Y}_\nu $ & $n_g$ &1&$ n_g $&1& 1 \\
\end{tabular}
\caption{$\mathcal{G}_\mathcal{F}^l$ representation of the lepton sector bi-fundamental scalar fields for $n_g$ fermion generations}
\label{BIF-SFL}
\end{table}%
and the relation to ordinary Yukawas is:
\begin{equation}
Y_D \equiv \dfrac{\left\langle\mathcal{Y}_D\right\rangle}{\Lambda_f}\,, \qquad \quad
Y_U \equiv \dfrac{\left\langle\mathcal{Y}_U\right\rangle}{\Lambda_f}\,, \qquad \quad
Y_E \equiv \dfrac{\left\langle\mathcal{Y}_E\right\rangle}{\Lambda_f}\,,  \qquad \quad
Y_\nu \equiv \dfrac{\left\langle\mathcal{Y}_\nu\right\rangle}{\Lambda_f}\,.
\label{BiFConnection}
\end{equation}
This case is hereby labeled \emph{bi-fundamental} scenario, and the fields can be thought
of as matrices whose explicit transformation is:
\begin{align}\label{TransBIFQ}
\mathcal{Y}_U(x)&\xrightarrow{\mathcal{G_F}} \Omega_{Q_L}\mathcal{Y}_U(x)\,\Omega_{U_R}^\dagger\,,&
\mathcal{Y}_D(x)&\xrightarrow{\mathcal{G_F}} \Omega_{Q_L}\mathcal{Y}_D(x)\,\Omega_{D_R}^\dagger\,,\\ \label{TransBIFL}
\mathcal{Y}_E(x)&\xrightarrow{\mathcal{G_F}} \Omega_{\ell_L}\mathcal{Y}_E(x)\,\Omega_{E_R}^\dagger\,,&
\mathcal{Y}_\nu(x)&\xrightarrow{\mathcal{G_F}} \Omega_{\ell_L}\mathcal{Y}_\nu(x)\, O_{N_R}^T\,,
\end{align}
$\Omega_{\psi}$ ($O_{N_R}$) being a unitary (real orthogonal) matrix of the corresponding $\mathcal{G_F}$ subgroup: $\Omega_\psi \Omega_\psi^\dagger=\Omega_\psi^\dagger\Omega_\psi=1$, $\psi=Q_L\,...\,E_R$ ($O_{N_R}O_{N_R}^T=O_{N_R}^TO_{N_R}=1$).
\subsection*{Scalar Flavour Fields in the Fundamental}
The next order in the effective field theory is a $d=6$ Yukawa operator, involving generically 
two scalar fields in the place of the Yukawa couplings,
 \begin{eqnarray}
\mathscr{L}_{Yukawa}=\overline{Q}_L\frac{\chi_D^L\chi_D^{R\dagger}}{\Lambda_f^2}D_RH+\overline{Q}_L\frac{\chi_U^L\chi_U^{R\dagger}}
{\Lambda_f^2}U_R\tilde{H}+\overline{\ell}_L\frac{\chi_{E}^L\chi_E^{R\dagger}}{\Lambda_f^2}E_RH+\overline{\ell}_L\frac{\chi_{\nu}^L\chi_\nu^{R\dagger}}{\Lambda_f^2}N_R\tilde H\, ,
\label{YukLagrangian_Fund}
\end{eqnarray}
which provide the following relations between Yukawa couplings and vevs:
\begin{eqnarray}
Y_D \equiv \dfrac{\left\langle\chi_D^L\chi_D^{R\dagger}\right\rangle}{\Lambda_f^2}\,, \qquad 
Y_U \equiv \dfrac{\left\langle\chi_U^L\chi_U^{R\dagger}\right\rangle}{\Lambda_f^2}\,,\qquad 
Y_E \equiv \dfrac{\left\langle\chi_E^L\chi_E^{R\dagger}\right\rangle}{\Lambda_f^2}\,, \qquad 
Y_\nu \equiv \dfrac{\left\langle\chi_\nu^L\chi_\nu^{R\dagger}\right\rangle}{\Lambda_f^2}\,, \qquad 
\label{FConnection}
\end{eqnarray}
The simplest assignation of charges or transformation properties of these fields is
to consider each of them in the fundamental representation of a given $SU(3)_{\psi}$ subgroup
as specified in tables \ref{F-SFQ} , \ref{F-SFL}.

\begin{table}[h]
\centering
\begin{tabular}{ccccccc}
& $SU(n_g)_{Q_L}$ & $SU(n_g)_{U_R}$ & $SU(n_g)_{D_R}$ & $U(1)_B$ & $U(1)_{A^U}$& $U(1)_{A^D}$\\
\hline
 $\chi^L_U$ & $n_g$  & 1  &1&0& 1& 1 \\
 $\chi^L_D$ &  $n_g$ & 1  &1&0& 1& 1 \\
$\chi^R_U$  & 1         & $n_g $ &1&0& -1& 0 \\
$\chi^R_D$  &1          & 1  &$ n_g $&0& 0& -1 \\
\end{tabular}
\caption{Representation of the quark sector fundamental scalar fields for $n_g$ fermion generations}
\label{F-SFQ}
\end{table}%
\begin{table}[h]
\centering
\begin{tabular}{cccccc}
& $SU(n_g)_{\ell_L}$ & $SU(n_g)_{E_R}$ & $O(n_g)_{N_R}$ & $U(1)_L$ & $U(1)_{A^l}$\\
\hline
$\chi^L_E$   &$n_g$   &1          & 1     &0& 1 \\
$\chi^L_\nu$ &$n_g$  & 1          &1      &0& 1 \\
$\chi^R_E$ &   1         &$n_g$   &1       & 0&-1\\
$\chi^R_N$  &1            &1          &$n_g$& 0  & 0\\
\end{tabular}
\caption{Representation of the lepton sector fundamental scalar fields for $n_g$ fermion generations}
\label{F-SFL}
\end{table}%
These fields are then complex $n_g$-vectors whose transformation under
the flavour group is just a unitary or real rotation;
$\chi_\psi \xrightarrow{\mathcal{G_F}}\Omega_\psi \chi_\psi \,,\,\chi_N^R \xrightarrow{\mathcal{G_F}}O_{N_R} \chi_N^R$.
From the group theory point of view this is the
decomposition in the irreducible pieces needed to
build up invariant Yukawa operators, and as we
shall see their properties translate into an easy and
clear extraction of the flavour structure.

The third case of a Yukawa operator of mass
dimension 7 could arise from a condensate of fermionic 
fields $Y\sim \left\langle\overline \Psi\Psi\right\rangle/\Lambda_f^3$ \cite{Chivukula:1987py},
or as the product of three scalar fields. In both cases the
simplest decomposition falls trivially into one of the previous or the assignation of representations is an otherwise 
unnecessarily complicated higher dimensional one.

Notice that realizations in which the Yukawa couplings correspond to the vev of an aggregate of fields,  
rather than to a single field, are not the simplest realization of MFV as defined in Ref.~\cite{D'Ambrosio:2002ex}, 
while still corresponding to the essential idea that the Yukawa spurions may have a dynamical origin. 

Finally, another option of dependence of the Yukawa couplings on the dynamical fields is an inverse one:
\begin{equation}
Y_D \equiv \dfrac{\Lambda_f}{\left\langle\mathcal{Y}_D\right\rangle}\,, \qquad
Y_U \equiv \dfrac{\Lambda_f}{\left\langle\mathcal{Y}_U\right\rangle}\,, \qquad
Y_E \equiv \dfrac{\Lambda_f}{\left\langle\mathcal{Y}_E\right\rangle}\,, \qquad
Y_\nu \equiv \dfrac{\Lambda_f}{\left\langle\mathcal{Y}_\nu\right\rangle}\,.
\label{BiFInverseConnection}
\end{equation}
a case in which the vev of the field rather than the scale $\Lambda_f$ entering the relation is the larger one.
This interesting case arises in models of gauged flavour symmetry \cite{Grinstein:2010ve,Feldmann:2010yp,Guadagnoli:2011id},
in which the anomaly cancellation requirements call for the introduction of fermion fields, whose
interaction in a renormalizable Lagrangian with the scalar fields and ordinary fermions suffice to 
constitute a self consistent theory that after the integration of the heavy states yields the relation above. The
transformation properties of the fields are the same as in the bi-fundamental case.

For simplicity in the group decomposition
and since they appear as the two leading terms in the effective field theory approach, we will focus the
analysis here in the fundamental and bi-fundamental cases, or the dimension 5 and 6 Yukawa operators, the
former nonetheless also applies to relation~\ref{BiFInverseConnection}.

\section{The Scalar Potential}
The way in which the scalar fields $\mathcal{Y}\,,\,\chi$ acquire a vev is through a scalar potential. This potential
must be invariant under the gauge group of the SM $\mathcal{G}$ and the flavour group $\mathcal{G_F}$.
The study is focused on the potential constituted by the flavour fields only, 
even if there might be some mixing with the singlet combination
$H^\dagger H$ of the Higgs field, an exploration of this last case can be
found in \cite{Lopez-Honorez:2013wla} in which the flavour scalar fields are postulated as Dark Matter. Resuming, the coupling with the Higgs doublet
 would add to the hierarchy problem but make no difference in the
determination of the flavour fields minimum since the mass scale of the latter is taken
larger than the Higgs vev: $\Lambda_f^2\gg v^2$.

The goal of this work is therefore to address the problem of the determination and analysis of the general $\mathcal{G_F}$-invariant scalar potential and its 
minima for the flavour scalar fields denoted above by $\mathcal{Y}$ and $\chi$. 
The central question is whether it is possible to obtain the SM Yukawa pattern - i.e. the observed values of quark 
masses and mixings- with a ``natural" potential. 

It is worth noticing that the structure of the scalar potentials constructed here is more general than the particular 
effective realization  in Eqs.~\ref{BiFConnection} and \ref{FConnection} and it would apply also
for Eq.~\ref{BiFInverseConnection} as it relies exclusively 
on invariance under the symmetry $\mathcal{G_F}$ and on the flavour field representation, bi-fundamental or fundamental. 

This observation is relevant, because the case of gauged flavour symmetry leading
to Eq.~\ref{BiFInverseConnection} addresses two problems that this approach has.
Namely the presence of Goldstone bosons as a result of the spontaneous breaking
of a continuous symmetry and the constraints placed on the presence of new particles
carrying flavour and inducing potentially dangerous FCNC effects. 

The Goldstone bosons
in a spontaneously broken flavour gauge symmetry are eaten by the flavour group vector bosons which become massive.
These particles even if massive would induce dangerous flavour changing processes which we expect
to be suppressed by their scale. The case of gauged flavour symmetries is however such that
 the inverse relation of Yukawas of Eq. \ref{BiFInverseConnection} translates also to the particle masses, 
 so that the new particles inducing flavour changing in the lightest
generations are the heaviest in the new physics spectrum \cite{Buras:2011wi}. These two facts conform a possible acceptable
and realistic scenario where to embed the present study.

\subsection{Generalities on Minimization}

The variables in which we will be minimizing are the parameters of the scalar fields modulo a $\mathcal{G_F}$
transformation. 
The discussion of which are those variables in the bi-fundamental case is familiar to the particle physicist: 
they are the equivalent of masses and mixing angles. Indeed we can substitute in Eq. \ref{BiFConnection}
the explicit formula for the Yukawas, Eqs. \ref{YukParQ}\,-\ref{EigenYukL}, and express the variables
of the scalar field at the minimum in terms of flavour parameters. 

The equation obtained in this way is the condition of the vev of the scalar fields fixing the masses and mixings
\emph{that are measured}. It is not clear at all though that a spontaneous breaking mechanism can yield the very
values that Yukawas actually have. To find this out the minimization of the potential has to be completed, 
such that for the next two chapters masses and mixing will be treated as variables roaming
all their possible range. The question is whether at the minimum of the potential these variables can
take the values corresponding to the known spectrum and if so to which cost.


The $\mathcal{G_F}$ invariants, out of which the potential is built, will be denoted generically  by $I_{j}$, while
$y_i$ stand for the physical variables of the scalar fields connected explicitly to masses and mixing.
Let us call $n$ the number of physical parameters that suffice to describe the general vev of the flavour fields,
that is to say there are $n$ variables $y_i\,,i=1,2,...,n$. 

A simple result is that there are $n$ independent invariants $I_j$, since the inversion of the relation of the latter
in terms of the variables\footnote{Inverse relation which is unique up to discrete choices \cite{Jenkins:2009dy}.} allows to 
express any new invariant $I'$ in terms of the independent set $\{I_j\}$; $I'=I'(y_i)=I'(y_i(I_j))$.

In terms of the set of invariants $\{I_j\}$ the stationary or extremal points of the potential, among them the true vacuum, are the solutions
to the equation,

\begin{equation}
\sum_j \frac{\partial I_j}{\partial y_i} \frac{\partial V}{\partial I_j}=0\,,
\label{EqsMin}
\end{equation}
where $V$ stands for the general potential.
These $n$ equations will fix the $n$ parameters. One can regard this array of equations as a matrix $J_{ij}=\partial I_j/\partial y_i$,
which is just the Jacobian of the change of ``coordinates" $I_j=I_j(y_i)$,
times a vector $\partial V/\partial I_j$.

This system, if the Jacobian has rank $n$, has only the solution of a null vector $\partial V/\partial I_j=0$, which 
is the case for example for the Higgs potential of the SM. 

When
the Jacobian has rank smaller than $n$, the system of Eqs. \ref{EqsMin} simplifies to a 
number of equations equal to the rank of the Jacobian. The extreme case would be a rank 0 Jacobian, 
which is the trivial, but always present, symmetry preserving case. This link of
the smallest rank with the largest symmetry can be extended; indeed in general terms
the reduction of the rank implies the appearance of symmetries left unbroken. A conjectured theorem by Michel and Radicati \cite{Michel:1971th,Michel:1970mua}, translated to
the notation used here, states that the maximal unbroken subgroup cases, given the fields that break the symmetry,
are insured an stationary point when the values of the fields are confined to a compact region.
N. Cabibbo and L. Maiani completed the study of an explicit example of the above theorem \cite{Maiani:2011zz} while introducing the
tool of the Jacobian analysis as it is used in this thesis together with a geometrical interpretation
outlined next.

For a geometric comprehension of the reduction of the Jacobian's rank the
manifold of possible values for the invariants can be considered, 
hereby denoted \emph{$I$-manifold}. 
The \emph{$I$-manifold} can be embedded in a $n$-$th$ dimensional real space $\mathcal{R}^n$. 
Whenever the Jacobian has reduced rank there exist one or more directions
in which a variation in the parameters $y$ has 0 variation in the \emph{I-manifold}, let us denote this displacement
$\delta y_i$, then this statement reads,
\begin{equation}
\delta I_j =\sum_i\frac{\partial I_j}{\partial y_i}\, \delta y_i=0\,.
\end{equation}
 This direction is the 
normal to a boundary of the \emph{$I$-manifold}, as displacements in this direction are not allowed.
 The further
the rank is reduced the more reduced is the dimension of this boundary. Those points for which 
the rank was reduced the most while still triggering symmetry breaking, will be denoted singular here; they are the
maximal unbroken symmetry cases. 

In the general case one can expect to have a combination of both, reduced rank  of the Jacobian
and potential-dependent solutions. In this sense the present study adds to the work of Refs.~\cite{Maiani:2011zz,Michel:1971th,Michel:1970mua}
two points through the study of an explicit general potential: i) we will be able to determine under what conditions the singular points (or
maximal unbroken configurations) correspond to absolute minima; ii) the exploration of the general case will reveal whether
other than singular minima are allowed or not.
It is in any case worth examining first the Jacobian, as it is
done in the next chapters.

Another relevant issue is the number of invariants that enter the potential. If one is to stop the analysis at 
a given operator's dimensionality as it is customary in effective field theory some of the invariants are left out.
Does this mean that there are parameters left undetermined by the potential, i. e. flat directions?
We shall see that these flat directions are related to the presence of unbroken symmetries 
and therefore are unphysical, so rather than the potential in such cases
being unpredictive is quite the opposite, it imposes symmetries in the low energy
spectrum.

\include{XYZ}

\chapter{Quark Sector} 


\ifpdf
    \graphicspath{{X/figures/PNG/}{X/figures/PDF/}{X/figures/}}
\else
    \graphicspath{{X/figures/EPS/}{X/figures/}}
\fi


This chapter will concern the analysis of flavour symmetry breaking in  the quark sector
through the study of the general potential in both the bi-fundamental and fundamental representation
cases.

\section{Bi-fundamental Flavour Scalar Fields}

At a scale above the electroweak scale and around $\Lambda_f$ we assume that the Yukawa interactions
are originated by a Yukawa operator with dimension $=5$ as made explicit in Eq.~\ref{YukOps5}, the
connection to masses and mixing of the new scalar fields given in Eqs.~\ref{YukParQ},\ref{EigenYukQ},\ref{BiFConnection}.
The analysis of the potential for the bi-fundamental scalar fields
is split in the two and three generation cases.

\subsection{Two Family Case}
\label{sec:2F}

The discussion of the general scalar potential starts by illustrating the two-family case,  
postponing the discussion of three families to the next section. Even if restricted to a simplified case, with a smaller 
number of Yukawa couplings and mixing angles, it is a very reasonable starting-up scenario, that corresponds to the 
limit in which the third family is decoupled, as suggested by the hierarchy between quark masses and the smallness 
of the CKM mixing angles\footnote{We follow here the PDG~\cite{Beringer:1900zz} conventions for the CKM matrix 
parametrization.} $\theta_{23}$ and $\theta_{13}$. In this section, moreover, most of the conventions 
and ideas to be used later on for the three-family analysis will be introduced.

The number of variables that suffice for the description of the physical degrees of
freedom of the scalar fields $\mathcal{Y}$ is the starting point of the analysis.
Extending the bi-unitary parametrization for the Yukawas given in the first terms of Eq.~\ref{BIU1}
to the scalar fields and performing a $\mathcal{G_F}$ rotation as in Eq.~\ref{TransBIFQ}, the algebraic objects left
are a unitary matrix, and two diagonal matrices of eigenvalues. Out of the 4 parameters of a general
unitary $2\times2$ matrix, three are complex phases which can be rotated away via diagonal phase rotations
of $\mathcal{G_F}$. The remaining
variables are therefore an angle in the mixing matrix and 4 eigenvalues arranged in two diagonal matrices:
a total of $n=5$ following the notation introduced.
This is nothing else than the usual discussion of physical parameters in the Yukawa couplings, applicable
to the flavour fields since the underlying symmetry is the same.

The explicit connection of scalar fields variables and flavour parameters is,
\begin{eqnarray}
\left\langle\mathcal{Y}_D\right\rangle=\Lambda_f\mathbf{y}_D=\Lambda_f \left(
        \begin{array}{cc}
           y_d  & 0  \\
            0 & y_s  \\
        \end{array}
\right),\quad & 
\left\langle\mathcal{Y}_U\right\rangle=\Lambda_f V_C^\dagger \mathbf{y}_U=\Lambda_fV_C^\dagger\left(
        \begin{array}{cc}
           y_u  & 0  \\
            0 & y_c  \\
        \end{array}\right),\, \label{SpurionsVEVs2F} 
\end{eqnarray}
where
\begin{eqnarray}
 V_C =\left(
        \begin{array}{cc}
           \cos\theta_c & \sin\theta_c  \\
            -\sin\theta_c & \cos\theta_c  \\
        \end{array}
\right),
\label{Cabibbo}
\end{eqnarray}
is the usual Cabibbo rotation among the first two families.

From the transformation properties in Eq.~\ref{TransBIFQ}, it is straightforward to write 
the list of independent invariants that enter in the scalar potential. For
the case of two generations that occupies us now, five independent invariants can be constructed
respecting the whole $\mathcal{G}_\mathcal{F}^q$ group \cite{Feldmann:2009dc,Jenkins:2009dy}: 
\begin{align}\label{Invariants2F_BiFundAC1}
I_U=&\mbox{Tr}\left(\mathcal{Y}_U\mathcal{Y}_U^\dagger\right)\,,& I_D=&\mbox{Tr}\left(\mathnormal{\mathcal{Y}_D}\mathcal{Y}_D^\dagger\right)\,,\\ \label{Invariants2F_BiFundAC2}
I_{U^2}=&\mbox{Tr}\left(\mathcal{Y}_U\mathcal{Y}_U^\dagger
\mathcal{Y}_U\mathcal{Y}_U^\dagger\right)\,,& I_{D^2}=&\mbox{Tr}\left(\mathcal{Y}_D\mathcal{Y}_D^\dagger
\mathcal{Y}_D\mathcal{Y}_D^\dagger\right)\,,\\
I_{UD}=&\mbox{Tr}\left(\mathcal{Y}_U\mathcal{Y}_U^\dagger\mathcal{Y}_D\mathcal{Y}_D^\dagger\right)\,.&&
\label{Invariants2F_BiFundAC3}
\end{align}
The vevs of these invariants expressed in terms of masses and mixing angles are\footnote{Let us drop the vev symbols in $\left\langle I \right\rangle$ for simplicity in notation.}: 
\begin{gather}\label{Inv2FQ1}
{I_U}=\Lambda_f^2\, (y_u^2+y_c^2)\,,\qquad \qquad \qquad {I_D}=\Lambda_f^2\, (y_d^2+y_s^2)\,,\\ \label{Inv2FQ2}
I_{U^2}=\Lambda_f^4\, (y_u^4+y_c^4)\,,\qquad \qquad \qquad I_{D^2}=\Lambda_f^4\, (y_d^4+y_s^4)\,,\\ \label{Inv2FQ3}
{I_{UD}}=\Lambda_f^4 \left[\left(y_c^2-y_u^2\right)\left(y_s^2-y_d^2\right)\cos2\theta_c+
        \left(y_c^2+y_u^2\right)\left(y_s^2+y_d^2\right)\right]/2\,.
\end{gather}

The previous counting of parameters made use of the full $\mathcal{G}_\mathcal{F}^q$ group;
the absence of $U(1)_A$ factors does not allow for overall phase redefinitions
and therefore in the explicitly axial breaking case ($\mathcal{G}_\mathcal{F}^{\slashed A, q}\sim SU(n_g)^3$) two more parameters appear: the overall phases of the scalar fields. 
In the axial breaking case therefore the number of variables is $n=7$.

This case allows for two new invariants of dimension 2,
\begin{equation}
\begin{array}{ll}
I_{\tilde U}=\det\left(\mathcal{Y}_U\right)\,,& \qquad I_{\tilde D}=\det\left(\mathnormal{\mathcal{Y}_D}\right)\,,\\[2mm]
\end{array}
\label{dets2F}
\end{equation}
the two extra parameters appearing in this case are the complex phase of the determinant for each $\mathcal{Y}$ field.

The two complex determinants together with the previous 5 operators of Eq. \ref{Invariants2F_BiFundAC1}-\ref{Invariants2F_BiFundAC3} add up to 9 real quantities
which points to two invariants being dependent on the rest.
Indeed the Cayley-Hamilton relation in 2 dimensions reads:
\begin{align}
\mbox{Tr}\left(\mathcal{Y}_U\mathcal{Y}_U^\dagger
\mathcal{Y}_U\mathcal{Y}_U^\dagger\right)=&\mbox{Tr}\left(\mathcal{Y}_U\mathcal{Y}_U^\dagger\right)^2-2\det\left(\mathcal{Y}_U\right)\det\left(\mathcal{Y}_U^\dagger\right),
\label{det-tr-2GU}
\\
\mbox{Tr}\left(\mathcal{Y}_D\mathcal{Y}_D^\dagger
\mathcal{Y}_D\mathcal{Y}_D^\dagger\right)=&\mbox{Tr}\left(\mathcal{Y}_D\mathcal{Y}_D^\dagger\right)^2-2\det\left(\mathcal{Y}_D\right)\det\left(\mathcal{Y}_D^\dagger\right).
\label{det-tr-2GD}
\end{align}
The two determinants in terms of the variables read:
\begin{equation}
\begin{array}{l}
 {I_{\tilde U}}=\Lambda_f^2\, y_u\,y_c\,e^{i\phi_U}\,,\qquad \qquad \qquad I_{\tilde D}=\Lambda_f^2\, y_d\,y_s e^{i\phi_D}\,.
\end{array}
\label{InvariantsExplicit2F_BiFund}
\end{equation}
The symmetry matters for the outcome of the analysis, so we shall make clear the differences in
the choices of preserving the axial $U(1)$'s or not.

Notice that the mixing angle appears in both cases exclusively in $I_{UD}$, which is the only operator that mixes 
the up and down flavour field sectors. This is as intuitively expected: {\it the mixing angle describes the relative
misalignment between the up and down sectors basis}. Eq.~\ref{Inv2FQ3}
shows that the degeneracy in any of the two sectors makes the angle unphysical, or, in terms of the scalar fields and
flavour symmetry, reabsorvable via a  $\mathcal{G}_\mathcal{F}^q$ rotation.

Since there is one mixing parameter only in this case this invariant is related to
all possible invariants describing mixing, in particular the Jarlskog invariant for two families,  
\begin{equation}
4\mathcal{J}=4\det\left(\left[Y_UY_U^\dagger,Y_DY_D^\dagger\right]\right)=\left(\sin{2\theta_c}\right)^2
       \left(y_c^2-y_u^2\right)^2\left(y_s^2-y_d^2\right)^2 \,,\nonumber
\end{equation}
is related to $I_{UD}$ via
\begin{equation}
\dfrac{1}{\Lambda_f^4}\dfrac{\partial}{\partial \theta_c}\mbox{Tr}\left(\mathcal{Y}_U\mathcal{Y}_U^\dagger\mathnormal{\mathcal{Y}_D}\mathnormal{\mathcal{Y}_D}^\dagger\right) = -2  \sqrt{\mathcal{J}}\,.
\end{equation}


The lowest dimension invariants that characterize symmetry breaking unmistakably are $I_U$ and $I_D$.
Indeed for $\left\langle I_U \right\rangle\neq 0$ or $\left\langle I_D \right\rangle\neq0$, $\mathcal{G}_\mathcal{F}^q$ is
broken, whereas if $\left\langle I_U \right\rangle=\left\langle I_D \right\rangle=0$, $\mathcal{G}_\mathcal{F}^q$ remains unbroken.
These invariants though only contain information on the overall scale of the breaking and 
make no distinction on hierarchies among eigenvalues. $I_{U,D}$ can be
thought of as radii whose value gives no information on the ``angular" variables. These
variables can be chosen as the differences in eigenvalues, and their value at the minimum
will fix the hierarchies among the different generations.
The invariants that will determine these hierarchies will therefore be those of Eqs. \ref{Invariants2F_BiFundAC2},
\ref{Invariants2F_BiFundAC3}.
\subsubsection{The Jacobian}\label{J2F}
The Jacobian of the change of coordinates from the variables to the invariants of Eqs.\,
\ref{Invariants2F_BiFundAC1}\,,\ref{Invariants2F_BiFundAC3} is a $n\times n$
matrix. We are interested in the determinant for the location of the regions of reduced rank, 
or boundaries of the \emph{I-manifold} \cite{onpreparation}. 
 For this purpose we observe that the Jacobian has
the shape:
\begin{equation}
J=\left(\begin{array}{ccc}
\partial_{\mathbf{y}_U} I_{U^n} & 0 & \partial_{\mathbf{y}_U} I_{UD}\\
0   & \partial_{\mathbf{y}_D} I_{D^n} & \partial_{\mathbf{y}_D} I_{UD}\\
0 & 0 & \partial_{\theta_c } I_{UD}
\end{array}\right)\equiv\left(\begin{array}{ccc}
J_U & 0 & \partial_{\mathbf{y}_U} I_{UD}\\
0   & J_D & \partial_{\mathbf{y}_D} I_{UD}\\
0 & 0 &J_{UD}
\end{array}\right)\,.
\label{FullJac}
\end{equation}
where $I_{U^n}$ ($I_{U^n}$) stands for the set of invariants composed of $\mathcal{Y}_U$ ($\mathcal{Y}_D$) only
and $\mathbf{y}_{U,D}$ are defined in Eq. \ref{SpurionsVEVs2F}.
This structure of the Jacobian implies that the determinant simplifies to:
\begin{equation}
\det{J}=\det J_U \det J_D  \det J_{UD}\,,
\end{equation}
which is a result extensible to the 3 generation case.
The third factor of this product reads\footnote{Hereby the Jacobians will be written dimensionless since the factors of $\Lambda_f$ are irrelevant for the analysis; they could be nonetheless restored by adding a power of $\Lambda_f$ for each power of $y_i$.}:
\begin{equation}
 \det J_{UD}=\sin{2\theta_c}\left(y_c^2-y_u^2\right)\left(y_d^2-y_s^2\right)\, ,
\end{equation}
which signals $\theta_c=0,\pi/2$ as boundaries, both of them corresponding
to no mixing, we will examine this further in the next section.
For the following analysis we select  the $\theta_c=0$ solution for illustration.

\begin{itemize}
\item {\bf Axial Conserving Case: $\mathcal{G}_\mathcal{F}^q\sim U(n_g)^3$ -} The set of invariants in Eq.~\ref{Inv2FQ1} , \ref{Inv2FQ2} yields:
\begin{equation}
J_U=\partial_{\mathbf{y}_U} \left(\mbox{Tr}\left(\mathcal{Y}_U\mathcal{Y}_U^\dagger\right)\,,\,
\mbox{Tr}\left(\mathcal{Y}_U\mathcal{Y}_U^\dagger
\mathcal{Y}_U\mathcal{Y}_U^\dagger\right)\right)=\left(\begin{array}{cc}
2y_u&4y_u^3\\
2y_c & 4 y_c^3\\
\end{array}\right)\, ,
\end{equation}
and
\begin{equation}
J_D=\partial_{\mathbf y_D} \left(\mbox{Tr}\left(\mathcal{Y}_D\mathcal{Y}_D^\dagger\right)\,,\,
\mbox{Tr}\left(\mathcal{Y}_D\mathcal{Y}_D^\dagger
\mathcal{Y}_D\mathcal{Y}_D^\dagger\right)\right)=\left(\begin{array}{cc}
2y_d & 4y_d^3\\
2y_s & 4 y_s^3\\
\end{array}\right)\, ,
\end{equation}
so that:
\begin{equation}
 \det J_U=8\,y_cy_u (y_c^2-y_u^2)\, ,\qquad\qquad \det J_D=8\,y_sy_d (y_s^2-y_d^2)\, .\label{Jac2F}
\end{equation}
The present case allows for explicit illustration of the connection of  boundaries of the \emph{I-manifold}
and vanishing of the Jacobian. The invariants satisfy in general:
\begin{equation}
\frac{1}{2}I_{U}^2\leq I_{U^2}\leq I_{U}^2\,,\qquad\qquad \frac{1}{2}I_{D}^2\leq I_{D^2}\leq I_{D}^2\,. \label{Ineq2F}
\end{equation}
The saturation of the inequalities above occurs at the boundaries. It is now easy to check
via substitution of Eqs.~\ref{Inv2FQ2},\ref{Inv2FQ3} in Eq.~\ref{Ineq2F} that the upper bound
is satisfied for $y_{u,d}=0$ and the lower bound for $y_{c,s}=y_{u,d}$; the two possibilities
of canceling Eqs.~\ref{Jac2F}.

The solutions encoded in this case can be classified according to the symmetry left unbroken,
\begin{enumerate}
\item $\mathcal{G}_\mathcal{F}^q\rightarrow U(1)_V^2\times U(1)_A^2$\,\, \emph{Hierarchical spectrum for
both up and down sectors}
\begin{equation}
\mathcal{Y}_D=\Lambda_f 
\left(\begin{array}{cc}
0&0\\
0&y'\\
\end{array}
\right)\, ,\qquad\qquad
\mathcal{Y}_U=\Lambda_f\left(\begin{array}{cc}
0&0\\
0&y\\
\end{array}
\right)\, .
\label{1-2F}
\end{equation}
\item $\mathcal{G}_\mathcal{F}^q\rightarrow U(1)_V^2\times U(1)_A$\,\, 
\begin{itemize}
\item[a)]\emph{Degenerate down quarks, hierarchical up quarks,}
\begin{equation}
\mathcal{Y}_D=\Lambda_f 
\left(\begin{array}{cc}
y'&0\\
0&y'\\
\end{array}
\right)\,  ,\qquad\qquad
\mathcal{Y}_U=\Lambda_f\left(\begin{array}{cc}
0&0\\
0&y\\
\end{array}
\right)\,.
\label{2a-2F}
\end{equation}
\item[b)] \emph{Degenerate up quarks, hierarchical down quarks,}
\begin{equation}
\mathcal{Y}_D=\Lambda_f 
\left(\begin{array}{cc}
0&0\\
0&y'\\
\end{array}
\right)\, ,\qquad\qquad
\mathcal{Y}_U=\Lambda_f\left(\begin{array}{cc}
y&0\\
0&y\\
\end{array}
\right)\, .
\label{2b-2F}
\end{equation}
\end{itemize}
\item $\mathcal{G}_\mathcal{F}^q\rightarrow SU(2)_V\times U(1)_B$\,\, \emph{Down and Up quarks degenerate}
\begin{equation}
\mathcal{Y}_D=\Lambda_f 
\left(\begin{array}{cc}
y'&0\\
0&y'\\
\end{array}
\right)\,  ,\qquad\qquad
\mathcal{Y}_U=\Lambda_f\left(\begin{array}{cc}
y&0\\
0&y\\
\end{array}
\right)\, .
\label{3-2F}
\end{equation}
\end{enumerate}
The notation is such that $U(1)_V$ denote generation number 
and $U(1)_{A}$ chiral rotations within a generation, explicitly:
\begin{align}\label{U1V}
U(1)_V&\,:\,\Bigg\{\begin{array}{cccc}
U(1)_{c+s}:&  \left(\begin{array}{c}
c_L\\
s_L\\
\end{array}\right)\rightarrow e^{ia}\left(\begin{array}{c}
c_L\\
s_L\\
\end{array}\right),&
c_R\rightarrow e^{ia}c_R\,,&
s_R\rightarrow e^{ia}s_R\,,\\[3mm]
U(1)_{u+d}:& \left(\begin{array}{c}
u_L\\
d_L\\
\end{array}\right)\rightarrow e^{ia}\left(\begin{array}{c}
u_L\\
d_L\\
\end{array}\right),&
u_R\rightarrow e^{ia}u_R\,,&
d_R\rightarrow e^{ia}d_R\,,\\
\end{array}\\
U(1)_A:&\Bigg\{\begin{array}{ccc}
U(1)_{u_A}:& \left(\begin{array}{c}
u_L\\
d_L\\
\end{array}\right)\rightarrow e^{ia}\left(\begin{array}{c}
u_L\\
d_L\\
\end{array}\right),&
u_R\rightarrow e^{-ia}u_R\,,\\[3mm]
U(1)_{d_A}:& \left(\begin{array}{c}
u_L\\
d_L\\
\end{array}\right)\rightarrow e^{ia}\left(\begin{array}{c}
u_L\\
d_L\\
\end{array}\right),&
d_R\rightarrow e^{-ia}d_R\,.\\
\end{array}
\label{U1A}
\end{align}

\figuremacroW{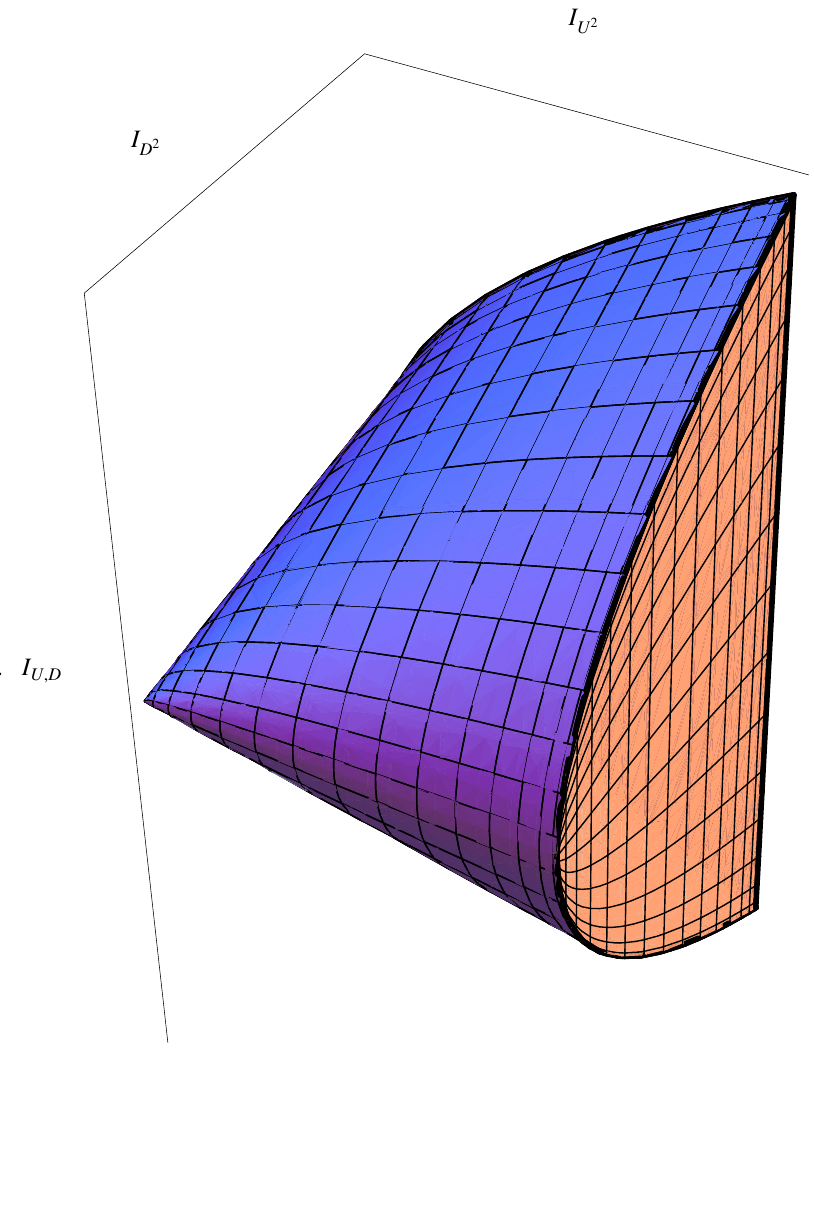}{{\it I-manifold} spanned by $\mathcal{G}_\mathcal{F}^q$ invariants built with $\mathcal Y_{U,D}$ for fixed $I_U$ , $I_D$ and 2 quark generations}{The boundaries of this manifold correspond to configurations of flavour fields that leave and unbroken symmetry. The vertex to the left
is associated to degenerate up and down sectors and a $U(2)$ symmetry. The upper and lower vertexes on the right correspond to a $U(1)^4$ symmetry, hierarchical up
and down sectors and mixing angle  vanishing or $\pi/2$ respectively. The parabola joining these two last points seen (unseen) on the figure corresponds to hierarchical up (down) sector and $0$ or $\pi/2$ mixing angle leaving an $U(1)^3$ unbroken symmetry. These vertexes and  parabolae are the only configurations that the
renormalizable potential allows for.}{.6}
Summarizing, the total Jacobian determinant is:
\begin{equation}
\det J=-64\,y_u y_d y_s y_c\,\sin{2\theta_c}\left(y_c^2-y_u^2\right)^2\left(y_s^2-y_d^2\right)^2\,
\end{equation}
and the two largest subgroups of $\mathcal{G}_\mathcal{F}^q$ are $U(2)$ and $U(1)^4$ associated to two singular points:
the vertex point of the Fig.~\ref{Boundaries2D} and the upper corner of the same figure respectively.
\item {\bf Explicitly axial breaking case: $\mathcal{G}_{\mathcal{F}}^{\slashed A, q}\sim SU(n_g)^3$ -} The invariants differ in this case and so do the Jacobians:
\begin{equation}
J_U=\partial_y \left(\mbox{Tr}\left(\mathcal{Y}_U\mathcal{Y}_U^\dagger\right)\,,\,
\det \mathcal{Y}_U\right)=\left(\begin{array}{cc}
2y_u&y_c\\
2y_c &  y_u\\
\end{array}\right)\,,
\end{equation}
and
\begin{equation}
J_D=\partial_y \left(\mbox{Tr}\left(\mathcal{Y}_D\mathcal{Y}_D^\dagger\right)\,,\,
\det \mathcal{Y}_D\right)=\left(\begin{array}{cc}
2y_d & y_s\\
2y_s &  y_d\\
\end{array}\right)\,,
\end{equation}
so that
\begin{equation}
 \det J_U= 2 (y_u^2-y_c^2)\,,\qquad\qquad \det J_D=2 (y_d^2-y_s^2)\,,
\end{equation}
and the single solution associated to the pattern  $\mathcal{G}_\mathcal{F}^q\rightarrow SU(2)_V\times U(1)_B$
survives since now no axial symmetry is present from the beginning. The
single boundary in this case as opposed to the axial preserving case can
be identified in the general inequalities:
\begin{equation}
|I_{\tilde U}|\leq \frac{1}{2} I_U\,,\qquad\qquad |I_{\tilde D}|\leq \frac{1}{2} I_D\,,\label{ineqD2F}
\end{equation}
which are saturated for degenerate masses only $y_{c,s}=y_{u,d}$.

The third invariant related to the phase $\phi_{U,D}$ can be taken to be
 $\mbox{Arg}\left(\det \mathcal{Y}_{U,D}\right)$, which is no other
than the variable itself. Then this part of the Jacobian is block diagonal and constant,
such that the determinant of the Jacobian stays the same. 

Altogether the Jacobian determinant is:
\begin{equation}
\det J=-4\sin{2\theta_c}\,\left(y_c^2-y_u^2\right)^2\left(y_s^2-y_d^2\right)^2\,,
\end{equation}
and the only maximal subgroup is $U(2)_V$.
\end{itemize}

\subsubsection{The Potential at the Renormalizable Level}\label{potQ2f}
The study of the Jacobian helped identify simple solutions 
in which some subgroup of $\mathcal{G}_\mathcal{F}^q$ was left unbroken corresponding
to boundaries of the \emph{I-manifold}. This analysis will serve as guide in
the evaluation of the general scalar potential at the renormalizable level
and the set of minima it allows for. The following study will reveal features obscured in the Jacobian method and 
will give further insight about the possible configurations and the role of unbroken symmetries.
In particular it will reveal which of the above extrema (boundaries) correspond to minima
and whether the potential allows for solutions outside of the boundaries and of what kind.

 \subsubsection*{Axial preserving case: $\mathcal{G}_\mathcal{F}^q\sim U(n_g)^3$ }
The most general renormalizable potential invariant under the whole flavour symmetry group $\mathcal{G}_F^q$ can be
written in a compact manner by means of
 the introduction of the array:
\begin{align}
X\equiv&\left(I_U, I_D\right)^T=\left(\mbox{Tr}\left(\mathcal{Y}_U\mathcal{Y}_U^\dagger\right),\mbox{Tr}\left(\mathcal{Y}_D\mathcal{Y}_D^\dagger\right)\right)^T\,,
\label{Xvector}
\end{align}
in terms of which:
\begin{align}\nonumber
V^{(4)}=&-\mu^2 \cdot X+ X^T \cdot \lambda \cdot X +g\,\mbox{Tr}\left(\mathcal{Y}_U\mathcal{Y}_U^\dagger
\mathcal{Y}_D\mathcal{Y}_D^\dagger\right)\\& + h_U \mbox{Tr}\left(\mathcal{Y}_U\mathcal{Y}_U^\dagger
\mathcal{Y}_U\mathcal{Y}_U^\dagger\right)+h_D \mbox{Tr}\left(\mathcal{Y}_D\mathcal{Y}_D^\dagger
\mathcal{Y}_D\mathcal{Y}_D^\dagger\right)\,,
\label{Pot2FBi}
\end{align}
where $\lambda$ is a $2\times2$ real symmetric matrix, $\mu^2$ a real 2-vector and $h_{U,D}, g$ three
real parameters: a total of 8 parameters enter this potential.
Strict naturalness criteria would require all dimensionless couplings $\lambda$, $h_{U,D}$ and $g$ to be of 
order $1$,  and the dimensionful $\mu$-terms to be 
of the same order of magnitude of $\Lambda_f$ but below to ensure the EFT convergence. The evaluation of
the possible minima will reveal next nonetheless that even relaxing this condition
the set of possible vacua is severely restricted.

Although it is not the full solution to the minimization procedure, let us consider in a first step and for illustration
the first two terms in \ref{Pot2FBi}, taking the limit $g,h_{U,D}\rightarrow 0$.
We can rewrite this part, if the matrix $\lambda$ is invertible as:
\begin{equation}
-\mu^2 \cdot X+X^T \cdot \lambda \cdot X = \left(X- \frac{1}{2}\lambda^{-1} \cdot \mu^2\right)^T
\lambda\left(X- \frac{1}{2}\lambda^{-1} \cdot \mu^2\right) - \mu^2\cdot \frac{\lambda^{-1}}{4}\cdot\mu^2
\end{equation}
which is the generalization of a mexican-hat potential for two invariants.
It is clear that if the ``vector" $\frac{1}{2}\lambda^{-1} \cdot \mu^2$ takes positive values the minimum would set:

\begin{equation}
\left(\begin{array}{c}
I_U\\
I_D\\
\end{array}\right)=\Lambda_f^2 \left(\begin{array}{c}
y_c^2+y_u^2\\
y_s^2+y_d^2\\
\end{array}\right)=
\frac{1}{2}\lambda^{-1} \cdot \mu^2\label{solIUID}
\end{equation}
This equation sets the order of magnitude of the Yukawa couplings as
$y\sim \mu/(\Lambda_f\sqrt{\lambda})$, which signals the ratio of
the mass scale of the scalar fields  and the high scale $\Lambda_f$.
For generic values of $\mu^2$ and $\lambda$ nonetheless the Yukawa
magnitude of up and down quarks would be the same, so that the two entries
of  $\lambda^{-1} \cdot\mu^2$ should 
accommodate certain tuning, in the two family case under consideration it would imply
a $\mathcal{O}(10\%)$ ratio $y_s/y_c\simeq10^{-1}=\sqrt{(\lambda^{-1} \mu^2)_U}/\sqrt{(\lambda^{-1} \mu^2)_D}.$\footnote{The values $U,D$ label the to entries of $\mu^2$: $(\mu^2_U,\mu^2_D)$}.
However let us recall here that for simplicity the coupling of the up and
down scalar fields in the Yukawa operators were assumed the same, but
if we were to extend this case to a two Higgs doublet scenario for example, the value
of $\tan\beta$ could make this tuning disappear. As shown next, it
is the hierarchies within each up and down sector that the potential
is unavoidably responsible for in this scheme.

For the complete minimization the extension of the above is simple,
the effect of the invariants left out $I_{U,D,UD}$ adds up to
a modified $\lambda$ as shown in the appendix, Sec.~\ref{App1}.

The stepwise strategy for minimization starts with the minimization in
those variables that appear less often in the potential, so that after solving
in their minima equations the left-over potential no longer depends on them.
Then the next variable which appears less often is selected and the process iterated again in this
matrioska like fashion.

The starting point is then the angle variable, appearing in one invariant only,
then follows the minimization of a variable independent from Tr$(\mathcal{Y}_{U,D}^\dagger\mathcal{Y}_{U,D})$, which appears most often
in the potential. The variables used in particular can be taken to be the differences of eigenvalues 
Tr$(\mathcal{Y}_{U,D}(-\sigma_3)\mathcal{Y}_{U,D}^\dagger)=\Lambda_f^2\left(y_{c,s}^2-y_{u,d}^2\right)$.
The values of these variables will determine the hierarchy among the different generations,
whereas Tr$(\mathcal{Y}_{U,D}\mathcal{Y}_{U,D}^\dagger)$ will have an impact on the overall magnitude of
the Yukawas.

This method dictates therefore that we start with the mixing angle, that
appears in the single invariant $I_{UD}$. The equation for the angle is,
\begin{equation}
\dfrac{\partial V^{(4)}}{\partial \theta_c}= g\dfrac{\partial{{I_{UD}}}}{\partial \theta_c}=
- g\, \Lambda_f^4\sin{2\theta_c}\left(y_c^2-y_u^2\right)\left(y_s^2-y_d^2\right)=0\,.
\label{CabibboEq2F_BiFund}
\end{equation}
The minimum of the scalar potential thus occurs 
for $\sin\theta_c=0$ or $\cos\theta_c=0$, for non-degenerate quark masses, which
is the only case in which the angle makes physical sense. For determining which of these options
is selected and to provide a very useful and general understanding of the minimization in
unitary matrices parameters, the {\bf Von Neumann trace inequality} for positive definite hermitian matrices
is here reproduced:

\emph{Let two hermitian positive definite $j\times j$ matrices A and B have eigenvalues of moduli $\alpha_1\leq\alpha_2\leq ...\leq \alpha_j$
and $\beta_1\leq\beta_2\leq...\leq\beta_j$ respectively, then the following inequality holds:}
\begin{equation}
\sum_{i=1}^j\alpha_{j+1-i}\,\beta_i\leq \mbox{Tr}\left(A\,B\right)\leq \sum_{i=1}^j\alpha_i\beta_i\,.
\label{VNTr}
\end{equation}

The usefulness of this inequality is that it tells us that, considering the eigenvalues at a fixed value
 and  varying the rest of parameters in the matrix, that is, the mixing parameters in the unitary matrices, 
the extrema are found for trivial unitary matrices. The inequality applied 
in the case of the invariant $I_{UD}$:
\begin{equation}
y_u^2y_s^2+y_d^2y_c^2\leq \mbox{Tr}\left(V_{C}^\dagger\mathbf{y}_U^2\,V_{C}\,\mathbf{y}_D^2\right)\leq y_u^2y_d^2+y_s^2y_c^2\,.
\label{VNQ2G}
\end{equation}
The two extrema are indeed given by the two solutions for the angle in Eq.~\ref{CabibboEq2F_BiFund}.
Which of these two is selected depends nonetheless on the sign of the coefficient in front of the invariant in the potential:
\begin{itemize}
\item $g>0$ The potential is minimized when $I_{UD}$ is minimized, which through Eq.~\ref{VNQ2G} corresponds to:
\begin{equation}\label{MIX1}
V_C=\left(\begin{array}{cc}
0&1\\
1&0\\
\end{array}\right)\,,\qquad \theta_c=\pi/2\,,
\end{equation}
and the situation is such that the charm quark would couple only to the down type quark and
the up to the strange, in an  `inverted hierarchy" scenario.
\item $g<0$ The potential is minimized when $I_{UD}$ is maximized, so Eq. \ref{VNQ2G} determines:
\begin{equation}\label{MIX2}
V_C=\left(\begin{array}{cc}
1&0\\
0&1\\
\end{array}\right)\,,\qquad \theta_c=0\, ,
\end{equation}
This case is closer to reality, now the Cabibbo angle is set to 0 and the charm only couples 
to the strange quark, and the up to the down.
\end{itemize}
One can check that both these configurations leave an invariant $U(1)_V^2$ as defined in Eq.~\ref{U1V}.

All in all, the straightforward lesson that follows from Eq.~\ref{CabibboEq2F_BiFund}
is that, given the mass splittings observed in nature, {\it the scalar potential for bi-fundamental flavour fields does 
not allow  mixing at the renormalizable level.}

The next step is the minimization in eigenvalues differences.
The first relevant point is that only the invariants $I_{U^2},I_{D^2},I_{U,D}$ of Eqs.~\ref{Inv2FQ2}-\ref{Inv2FQ3}
depend on the differences of eigenvalues squared; this is explicit in Eq.~\ref{Inv2FQ3} for $I_{UD}$ whereas for $I_{U^2,D^2}$,
\begin{align} I_{U^2}=&\Lambda_f^4\left(y_{u}^4+y_{c}^4\right)=\frac{\Lambda_f^4}{2}\left((y_{u}^2+y_{c}^2)^2+(y_{u}^2-y_{c}^2)^2\right)\,,\\
I_{D^2}=&\Lambda_f^4\left(y_{d}^4+y_{s}^4\right)=\frac{\Lambda_f^4}{2}\left((y_{d}^2+y_{s}^2)^2+(y_{d}^2-y_{s}^2)^2\right)\,.
\end{align}
All these invariants appear linearly in the potential, Eq.~\ref{Pot2FBi}. 

Before entering the different possible solutions for the hierarchy of eigenvalues, a intuitive view of the potential behavior
is given to identify the solution which is relevant phenomenologically.

When the operators in Eq.~\ref{Inv2FQ2}
have negative coefficients in Eq.~\ref{Pot2FBi}.  ($h_{U,D}<0$) the potential diminishes towards the hierarchical
configuration, which maximizes $I_{U^2,D^2}$ and minimizes $-|h_{U,D}|\,I_{U^2,D^2}$. 
In the case of $I_{UD}$ after we substitute in Eq.~\ref{Inv2FQ3} and subsequently in Eq.~\ref{Pot2FBi} the two possible solutions
for the mixing at the minimum for each sign of $g$, Eqs.~\ref{MIX1},\ref{MIX2}. The term left does no longer depend on the
angle but it does depend on the product of mass differences:\begin{equation} V \supset gI_{U,D}\Big|_{\left\langle \theta_c \right\rangle}= \frac{\Lambda_f^4}{2}\left[g\left(y_c^2+y_u^2\right)\left(y_s^2+y_d^2\right)-|g|\left(y_c^2-y_u^2\right)\left(y_s^2-y_d^2\right)
        \right]\,,\end{equation} such that it always pushes
towards the hierarchical configuration for both up and down type quarks. Therefore for negative $h_{U,D}$ and $g$ the minimum will
correspond to a hierarchical mass configuration without mixing.
For the resemblance of nature \emph{this configuration}
(associated to case 1 of Eq. \ref{1-2F} in the Jacobian
analysis) {\it is a good first approximation:
only the heaviest family is massive so that $y_u=y_d=0$ and the mixing is vanishing}. 

For completeness all the possible minima and
their connection to the potential parameters are listed below (again for $g<0$):

\begin{itemize}
\item[\bf I] In this configuration a strong hierarchy arises,
\begin{equation}
\mathcal{Y}_D=\Lambda_f \left(\begin{array}{cc}
						0&0\\
						0&y_s\\
						\end{array}\right)\, ,\qquad\qquad
\mathcal{Y}_U=\Lambda_f \left(\begin{array}{cc}
						0&0\\
						0&y_c\\
						\end{array}\right)\, ,
\end{equation}
which presents an unbroken symmetry $\mathcal{G}_\mathcal{F}^q\rightarrow U(1)_V^2\times U(1)_A^2$ and
is just case 1 in the Jacobian analysis, see \ref{1-2F}.
\item[\bf II] This case forbids mass for the up quark 
\begin{equation}
\mathcal{Y}_D=\Lambda_f \left(\begin{array}{cc}
						y_d&0\\
						0&y_s\\
						\end{array}\right)\, ,\qquad\qquad
\mathcal{Y}_U=\Lambda_f \left(\begin{array}{cc}
						0&0\\
						0&y_c\\
						\end{array}\right)\, ,\label{II-2F}
\end{equation}
whereas the mass difference in the down sector is set by the relation 
\begin{equation}\frac{y_s^2-y_d^2}{ y_s^2+y_d^2}=\frac{-g}{2 h_D}\frac{I_U}{I_D}\,,\label{II2Fg}
\end{equation}
and the breaking pattern is
$\mathcal{G}_\mathcal{F}^q\rightarrow U(1)_V^2\times U(1)_A$.
\item[\bf III] The analogous of case {\bf II} for massless down quark reads:
\begin{equation}
\mathcal{Y}_D=\Lambda_f \left(\begin{array}{cc}
						0&0\\
						0&y_s\\
						\end{array}\right)\, ,\qquad\qquad
\mathcal{Y}_U=\Lambda_f \left(\begin{array}{cc}
						y_u&0\\
						0&y_c\\
						\end{array}\right)\, ,\label{III-2F}
\end{equation}
\begin{equation}\frac{y_c^2-y_u^2}{ y_c^2+y_u^2}=\frac{-g}{2 h_D}\frac{I_D}{I_U}\,,\label{III2Fg}
\end{equation}
and again $\mathcal{G}_\mathcal{F}^q\rightarrow U(1)_V^2\times U(1)_A$.

\item[\bf IV] Finally a completely degenerate scenario is possible in region {\bf IV}
\begin{equation}
\mathcal{Y}_D=\Lambda_f \left(\begin{array}{cc}
						y'&0\\
						0&y'\\
						\end{array}\right)\, ,\qquad\qquad
\mathcal{Y}_U=\Lambda_f \left(\begin{array}{cc}
						y&0\\
						0&y\\
						\end{array}\right)\, ,\label{IV-2F}
\end{equation}
having now that the potential triggers $\mathcal{G}_\mathcal{F}^q\rightarrow SU(2)_V\times U(1)_B$
This scenario is very far from reality, but listed for completeness, and the analogous
of case 3 and Eq.~\ref{3-2F} in the Jacobian analysis.
\end{itemize}
These regions are shown in the $h_U-h_D$ plane in fig. \ref{VacuaMapQuarks}.

\figuremacroW{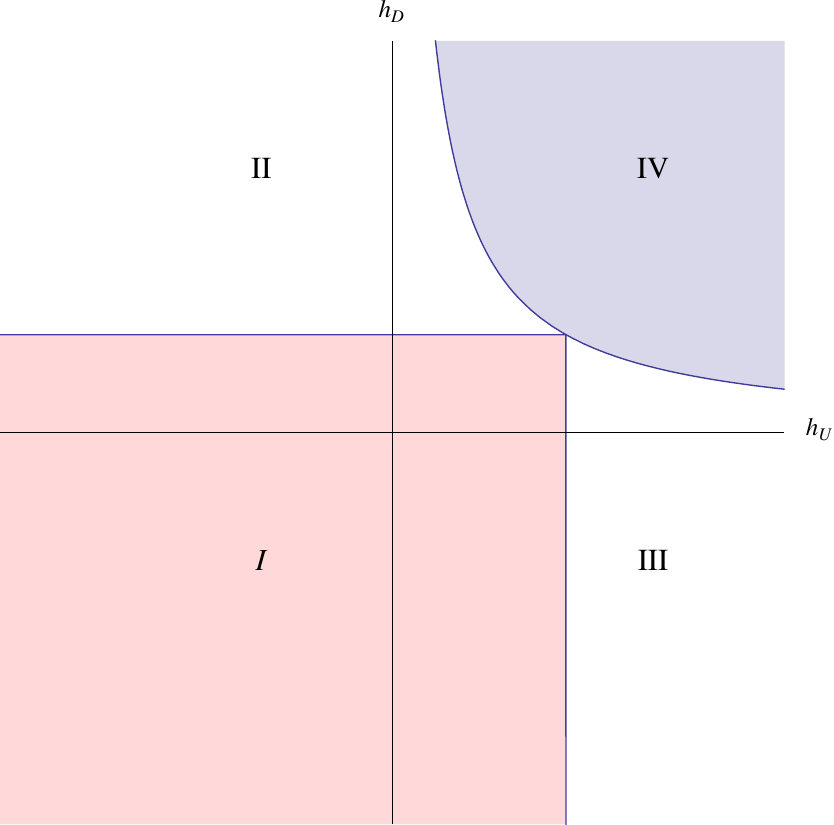}{Regions for the different quark mass configurations allowed at the absolute minimum in the $h_U-h_D$ plane  for $g<0$}{ {\bf I} is the region that yields a hierarchical spectrum for both up and down sectors.  {\bf II} ({\bf III}) presents a hierarchical up (down) spectrum and region {\bf IV} results in degenerate up and down sectors. See appendix, Sec.~\ref{App1} for details.}{.7}

Note that the cases found here are not quite the same as the ones found in
the Jacobian analysis. Case 2.a (Eq.~\ref{2a-2F}) and 2.b (Eq.~\ref{2b-2F}) are only present in the limiting
case $g\rightarrow 0$ of {\bf II} (Eqs.~\ref{II-2F},\ref{II2Fg}) and {\bf III} (Eqs.~\ref{III-2F},\ref{III2Fg}), so those are fine tuned cases.
The reason for this is found in the symmetries,
indeed cases 2.a and {\bf II} and 2.b and {\bf III} have the same symmetry, so from this point of
view there is nothing special on having two eigenvalues degenerate in one sector when in the other
sector one entry is 0.
The connection of the up and down sector is due to the common group transformation properties
under $SU(3)_{Q_L}$ of $\mathcal{Y}_{U,D}$ and indeed this correlation disappears if the mixing invariant is neglected $g\rightarrow 0$,
as can be checked on Eqs.~\ref{II2Fg},\ref{III2Fg}.

The singular point solutions, {\bf I} and {\bf IV}, are present as the absolute 
minima under certain conditions detailed in the appendix (\ref{App1}) but are not the only possibilities.

\subsubsection*{Explicitly axial breaking case: $\mathcal{G}_\mathcal{F}^{\slashed A,q}\sim SU(n_g)^3$}

The set-up will change with the introduction of the determinants in Eq. \ref{dets2F}
when choosing to \emph{violate} $U(1)_{A^U}\times U(1)_{A^D}$ \emph{explicitly}.
By making use of the analogous of $X$ in this case,
\begin{align}
\tilde X=&\left(I_U, I_D, I_{\tilde U}, I_{\tilde D} \right)^T\\
=&\left(\mbox{Tr}\left(\mathcal{Y}_U\mathcal{Y}_U^\dagger\right),\mbox{Tr}\left(\mathcal{Y}_D\mathcal{Y}_D^\dagger\right), \det\left(\mathcal{Y}_U\right)\, ,\,\det\left(\mathcal{Y}_D\right)\ \right)^T\,,
\end{align}
the potential reads:
\begin{equation}
V^{(4)}=-\tilde \mu^2 \cdot \tilde X+ \tilde X^T \cdot \tilde \lambda \cdot \tilde X +h.c.+g\, I_{UD}
\label{PotF}
\end{equation}
where $\lambda$ is a matrix and $\mu^2$ a 4-vector, the entries of these two structures are complex
when they involve the determinants. The number of parameters has increased now to 15 (out of which 9 are complex), since
the flavour symmetry is less restrictive.
Nonetheless the phases of the determinants are variables not observable at low energies and their
minimization is of no interest here; suffice then to assume that they are set to their minimum
values. One can then effectively consider all parameters in Eq. \ref{PotF} real.

Parallel to the axial conserving case we have that, in the limit $g\rightarrow0$, the minimum sets
\begin{equation}
\left\langle\tilde{X}\right\rangle=\frac{1}{2}\tilde \lambda^{-1} \cdot\tilde \mu^2\,,
\end{equation}
if the entries of such vector are in the inside of the \emph{I-manifold}.
This now requires two conditions in the entries of $\tilde \lambda^{-1}\cdot \tilde \mu^2/2$.
First, the two first entries have to be positive, since $I_{U,D}$ contained in $\tilde X$ are always positive;
second, the condition of Eq.~\ref{ineqD2F} must be
satisfied by the associated entries of $\tilde \lambda^{-1} \cdot \tilde \mu^2$.
If this second condition is not realized the minimum
is  at the boundary, that is,  $I_{U}= 2 |I_{\tilde U}|$ ($I_{D}= 2 |I_{\tilde D}|$) or equivalently $y_u=y_c$ ($y_d=y_s$).

Note also that in this case the solutions {\bf I, II} and {\bf III} are not
present just like cases 2.a and 2.b were not either in the Jacobian 
analysis.

These considerations together with the distinct symmetries from which
they arise lead to propose an \emph{ansatz for the explanation of the hierarchy 
among the two generations of quarks}.

First, we consider the whole $\mathcal{G}_\mathcal{F}^q$ group, so that determinants
are forbidden and the minimum is in region {\bf I} where the up and
down are massless at this order.
Then, introduction of a small source of breaking of the $U(1)_A$'s would
allow for the appearance of determinant terms in the potential with a naturally small
 coefficient since it is constrained by a symmetry, and whose impact is to produce 
 small masses for the light family.

This set-up is qualitatively explainable from symmetry considerations. In the axial preserving case
the solution of hierarchical masses was present but the explicit breaking of the axial symmetry 
 does not allow for such solutions. This means that a small perturbation
on the axial symmetry breaking direction produces a small shift in the light quark masses.

\subsubsection{The Potential at the Non-Renormalizable Level}
The scalar potential at the renormalizable level in the axial preserving case
allows for solutions with a strong hierarchy for both sectors of quarks,
that can be perturbed via a small breaking of the axial $U(1)_{A^U,\,A^D}$
to displace the minimum and lift the zero masses of the lightest quarks.  
The Cabibbo angle was unavoidably set to 0. In this 
section we explore whether non-renormalizable terms in the potential may
complete the picture and produce a small Cabibbo angle.

Consider the addition of non-renormalizable operators to the scalar potential, $V^{(i>4)}$. It is interesting to 
notice that this does {\it not} require the introduction of  new invariants beyond those in Eqs.~\ref{Invariants2F_BiFundAC1}-\ref{Invariants2F_BiFundAC3}:
all higher order traces and determinants can in fact be expressed in terms of that basis of five ``renormalizable'' 
invariants. 

The lowest higher dimensional contributions to the scalar potential have dimension six. 
At this order, the only terms involving the mixing angle are
\begin{equation}
V^{(6)}\supset \dfrac{1}{\Lambda_f^2}\left(\alpha_UI_{UD} I_{U}+\alpha_DI_{UD} I_{D}+\tilde \alpha_UI_{UD} I_{\tilde U}+\tilde \alpha_DI_{UD} I_{\tilde D}\right)\,.
\end{equation}
These terms, however, show the same dependence on the Cabibbo angle previously found  in Eq.~(\ref{CabibboEq2F_BiFund}) 
and, consequently, they can simply be absorbed in the redefinition of the lowest order parameter $g$. 
 To find a 
non-trivial angular structure it turns out that terms in the potential of dimension eight (or higher) have to be 
considered, that is
\begin{equation}
V^{(8)}\supset \frac{\alpha}{\Lambda_f^4} I_{UD}^2\,,
\end{equation}
with which the possibility of a mexican hat-like potential for $I_{UD}$ becomes possible
\begin{equation}
V^{(8)}\supset \frac{\alpha}{\Lambda_f^4}\left( I_{UD}-\frac{g}{2\alpha}\Lambda_f^4\right)^2\,,
\end{equation}
which would set
\begin{equation}
\sin^2\theta\simeq \dfrac{g}{2\, y^2_c y^2_s \alpha}\,.
\end{equation}
Using the experimental values of the Yukawa couplings $y_s$ and $y_c$, a realistic value for $\sin\theta$ can be 
obtained although at the price of assuming a highly fine-tuned hierarchy between the dimensionless coefficients of $d=4$ and 
$d=8$ terms, $g/\alpha\sim 10^{-10}$, that cannot be naturally justified in an effective Lagrangian approach. 

The conclusion is therefore that mixing is absent in a natural 2 generation quark case.
\subsection{Three Family Case}
\label{sec:3F}
In this section we extend the approach discussed in the previous section to the three-family case. 
The two bi-triplets scalars $\mathcal{Y}_{U,D}$ transform explicitly under the flavour symmetry $\mathcal{G}_F^q$, as in Eq.~\ref{TransBIFQ}
and the Yukawa Lagrangian is the same as that in Eq.~\ref{YukOps5}. Once the flavons develop a vev 
the flavour symmetry is broken and one should recover the observed fermion masses and CKM matrix 
given by Eq.~\ref{BiFConnection}.
 
While most of the procedure follows the steps of the 2 generation case, a few differences shall
be underlined. First, the number of variables and therefore independent invariants differs. As in
the two family case, we can absorb three unitary matrices with $\mathcal{G_F}$ rotations to
leave two diagonal matrices with 3 eigenvalues each and a unitary matrix. The latter contains
three angles and 6 phases; diagonal complex phase transformations allow to eliminate 5 of these,
so that the unitary matrix contains 4 physical parameters. In total 10 parameters describe the
axial preserving case. This resembles closely the
usual discussion of physical flavour parameters as expected. 

The higher number of variables implies that the set of invariants extends
beyond mass dimension 4 and therefore not all of them will be present at the renormalizable
level.

%
%

The list of invariants now reads  \cite{Feldmann:2009dc,Jenkins:2009dy}:
\begin{align}\label{InvBIF3F1}
I_{U}&=\mbox{Tr}\left[\mathcal{Y}_U \mathcal{Y}_U^\dagger \right]\,, & I_{D}&= \mbox{Tr}\left[\mathcal{Y}_D\mathcal{Y}_D^\dagger \right]\,,\\ \label{InvBIF3F2}
I_{U^2}&=\mbox{Tr}\left[\left(\mathcal{Y}_U\mathcal{Y}_U^\dagger\right)^2\right]\,,& I_{D^2}&= \mbox{Tr}\left[\left(\mathcal{Y}_D\mathcal{Y}_D^\dagger \right)^2\right]\,,\\
I_{U^3}&=\mbox{Tr}\left[\left(\mathcal{Y}_U\mathcal{Y}_U^\dagger\right)^3\right]\,,& I_{D^3}&= \mbox{Tr}\left[\left(\mathcal{Y}_D\mathcal{Y}_D^\dagger \right)^3\right]\,,\label{InvBIF3F3}
\end{align}
these first 6 invariants depend only on eigenvalues while the following 4 contain mixing too,
\begin{align}
I_{U,D}&=\mbox{Tr}\left[\mathcal{Y}_U \mathcal{Y}_U^\dagger \mathcal{Y}_D \mathcal{Y}_D^\dagger \right]\, , & I_{U,D^2}&= \mbox{Tr}\left[\mathcal{Y}_U \mathcal{Y}_U^\dagger \left(\mathcal{Y}_D \mathcal{Y}_D^\dagger\right)^2\right]\, , \\
I_{U^2,D}&=\mbox{Tr}\left[\mathcal{Y}_U \mathcal{Y}_U^\dagger \left(\mathcal{Y}_D \mathcal{Y}_D^\dagger\right)^2 \right]\, ,& I_{(U,D)^2}&= \mbox{Tr}\left[\left(\mathcal{Y}_U \mathcal{Y}_U^\dagger \mathcal{Y}_D \mathcal{Y}_D^\dagger\right)^2 \right]\,.
\end{align}
Explicitly these invariants read\footnote{Here again latin indexes run through down-type quark mass states and greek indexes through up-type quark mass states.}:
\begin{align}
I_{U}&=\Lambda_f^2\sum y_{\alpha}^2\,, & I_{D}&=\Lambda_f^2\sum y_{i}^2\,,\\
I_{U^2}&=\Lambda_f^4\sum y_{\alpha}^4\,,& I_{D^2}&= \Lambda_f^4\sum y_{i}^4\,,\\
I_{U^3}&=\Lambda_f^6\sum y_{\alpha}^6\,,& I_{D^6}&= \Lambda_f^6\sum y_{i}^6\, ,
\end{align}
\begin{align}
I_{U,D}&=\Lambda_f^4\sum y_{\alpha}^2V_{\alpha i }y_i^2 V^*_{\alpha i}\, , & I_{U,D^2}&= \Lambda_f^6\sum y_{\alpha}^2V_{\alpha i }y_i^4 V^*_{\alpha i}\, , \\
I_{U^2,D}&=\Lambda_f^6\sum y_{\alpha}^4V_{\alpha i }y_i^2 V^*_{\alpha i}\, ,& I_{(U,D)^2}&= \Lambda_f^8\sum \,y_\alpha^2 \,V_{\alpha i} \, y_i^2  \,V_{\beta i}^* \, y_\beta^2  \,V_{\beta j}\, y_j^2 \,V_{\alpha j}^*\, ,
\end{align}
In the explicitly axial breaking case two complex phases add to the previous number of
parameters so that 12 altogether conform the total.
In this case the determinants 
\begin{align}
I_{\tilde U}&=\det\left(\mathcal{Y}_U\right)\,, & I_{\tilde D}&= \det\left(\mathcal{Y}_D\right)\,,
\end{align}
substitute the invariants in Eq. \ref{InvBIF3F3} since they are connected through the relations:
\begin{align}\nonumber
\mbox{Tr}\left(\left(\mathcal{Y}_U^\dagger \mathcal{Y}_U\right)^3\right)=&\frac{3}{2}\mbox{Tr}\left(\left(\mathcal{Y}_U^\dagger \mathcal{Y}_U\right)^2\right)\mbox{Tr}\left(\mathcal{Y}_U^\dagger \mathcal{Y}_U\right)\\&-\frac{1}{2}\left(\mbox{Tr}\left(\mathcal{Y}_U^\dagger \mathcal{Y}_U\right)\right)^3+3\det \mathcal{Y}_U\det \mathcal{Y}_U^\dagger\\\nonumber
\mbox{Tr}\left(\left(\mathcal{Y}_D^\dagger \mathcal{Y}_D\right)^3\right)=&\frac{3}{2}\mbox{Tr}\left(\left(\mathcal{Y}_D^\dagger \mathcal{Y}_D\right)^2\right)\mbox{Tr}\left(\mathcal{Y}_D^\dagger \mathcal{Y}_D\right)\\
&-\frac{1}{2}\left(\mbox{Tr}\left(\mathcal{Y}_D^\dagger \mathcal{Y}_D\right)\right)^3+3\det \mathcal{Y}_D\det \mathcal{Y}_D^\dagger
\label{CHthrm3}
\end{align} 
and they read in terms of the variables,
\begin{align}
I_{\tilde U}&=\Lambda_f^3 e^{i\phi_U}\prod y_{\alpha}\,, & I_{\tilde D}&= \Lambda_f^3 e^{i\phi_U} \prod y_{i}\,,
\end{align}
which makes clear that the determinants of the fields change
to mass dimension 3 in the present 3 family case.
\subsubsection{The Jacobian}\label{Jac3GQ}

The study of the Jacobian is developed next. The Jacobian has an structure as in Eq. \ref{FullJac}. For the mass terms the analysis was first carried out in \cite{Maiani:2011zz,Maiani:2013lma},
while for the mixing we refer to \cite{onpreparation}.

Let's turn first to the mixing Jacobian $J_{UD}$. We know that 4 parameters suffice to
describe the mixing. Rather than choosing a parametrization for $V_{CKM}$, let us use the properties of a unitary matrix, substituting Eq.
\ref{SpurionsVEVs2F} in $I_{U,D}$:
\begin{eqnarray}
\Lambda_f^{-4}I_{U,D}&=&\sum_{\alpha,i}^3 y_{\alpha}^2\,V_{\alpha i}\,y_{i}^2\,V_{\alpha i}^*\,,\\
&=&\sum_{\alpha,i}^{3,2} y_{\alpha}^2V_{\alpha i}\left(y_{i}^2-y_b^2\right)V_{\alpha i}^*+y_b^2\sum_{\alpha}y_\alpha^2\,,\\
&=&\sum_{\alpha,i}^{2} \left(y_{\alpha}^2-y_t^2\right)V_{\alpha i}\left(y_{i}^2-y_b^2\right)V_{\alpha i}^*\,,+y_b^2\sum_{\alpha}y_\alpha^2+y_t^2\sum_{i}y_i^2\,,
\end{eqnarray}
where the terms independent of mixing elements are irrelevant for the analysis and will not be kept in the following. 
Note that what is achieved in using the unitarity relations is to rewrite the invariant in terms of 4 mixing elements, namely 
$|V_{ud}|,|V_{us}|,|V_{cd}|$ and $|V_{cs}|$. The choice of these 4 is of course to one's discretion; we can choose
 other 4 by removing the $\alpha'th$ row and the $i'th$ column of $V_{CKM}$.

The same procedure for $I_{U,D^2}$ and $I_{U^2,D}$ yields:
\begin{align}
\Lambda_f^{-6}I_{U,D^2}&=\sum_{\alpha,i}^2 \left(y_{\alpha}^2-y_{t}^2\right)V_{\alpha i}\left(y_{i}^2+y_{b}^2\right)\left(y_{i}^2-y_b^2\right)V_{\alpha i}^*+\cdots\,,\\
\Lambda_f^{-6}I_{U^2,D}&=\sum_{\alpha,i}^2 \left(y_{\alpha}^2+y_{t}^2\right)\left(y_{\alpha}^2-y_{t}^2\right)V_{\alpha i}\left(y_{i}^2-y_b^2\right)V_{\alpha i}^*+\cdots\,,
\end{align}
whereas $I_{(U,D)^2}$ is more involved:
\begin{equation}
\Lambda_f^{-8}I_{(U,D)^2}=\sum_{\alpha,\beta,i,j}^3 \,\left( y_\alpha^2 -y_t^2\right)\,V_{\alpha i} \, \left( y_i^2 -y_b^2\right) \,V_{\beta i}^* \, \left( y_\beta^2 -y_t^2\right) \,V_{\beta j}\,\left( y_j^2 -y_b^2\right)\,V_{\alpha j}^*+\cdots\,,
\end{equation}
this equation differs from the square of $I_{U,D}$, 
in terms in which $\beta\neq\alpha$ and $i\neq j$, which implies they are \emph{all} proportional to
4 different mass differences:
\begin{eqnarray}\nonumber
\Lambda_f^{-8}I_{(U,D)^2}&=&\left(\sum_{\alpha,i}^3 y_{\alpha}^2\,V_{\alpha i}\,y_{i}^2\,V_{\alpha i}^*\right)^2
-2\left(y_u^2-y_t^2\right)
\left(y_c^2-y_t^2\right)\left(y_d^2-y_b^2\right)\left(y_s^2-y_b^2\right)\\ &&\times\left(V_{ud}V_{cs}-V_{us}V_{cd}\right)
\left(V^*_{ud}V^*_{cs}-V^*_{us}V^*_{cd}\right)\, .
\end{eqnarray}
The first part is not relevant as it is a function of a previously categorized invariant. The second though,
has  a peculiar dependence on the mixing parameters. 
To rewrite it in terms of the four independent parameters chosen here the following relation is used:
\begin{equation}
\det\left(V\right)\det\left(V^*\right)=\sum_{\alpha,i}^2 V_{\alpha i}V_{\alpha i}^*-\left(V_{ud}V_{cs}-V_{us}V_{cd}\right)
\left(V^*_{ud}V^*_{cs}-V^*_{us}V^*_{cd}\right)=1\,.
\end{equation}

Resuming, the 4 independent pieces of the invariants:
\begin{align}
 I'_{U,D}&=\sum_{\alpha,i}  \left(y_{\alpha}^2-y_{t}^2\right)\left(y_{i}^2-y_{b}^2\right)V_{\alpha i}V_{\alpha i}^*\,,\\
I'_{U,D^2}&=\sum_{\alpha,i} \left(y_{\alpha}^2-y_{t}^2\right)\left(y_{i}^2+y_{b}^2\right)\left(y_{i}^2-y_b^2\right)V_{\alpha i}V_{\alpha i}^*\,,
 \\
 I'_{U^2,D}&=\sum_{\alpha,i}\left(y_{\alpha}^2+y_{t}^2\right)\left(y_{\alpha}^2-y_{t}^2\right)\left(y_{i}^2-y_b^2\right)V_{\alpha i}V_{\alpha i}^*\,,
\\
 I'_{(U,D)^2}&= \prod_{\beta}\left(y_{\beta}^2-y_{t}^2\right)\prod_j\left(y_{j}^2-y_{b}^2\right)\sum_{\alpha,i}^2 V_{\alpha i}V_{\alpha i}^*\,,
\end{align}
build up the Jacobian
\begin{equation}
J_{UD}=\frac{\partial I'}{\partial |V_{\alpha,i}|}\propto\left(
\begin{array}{cccc}
 |V_{ud}|& \left(y_{d}^2+y_{b}^2\right) |V_{ud}| &  \left(y_{u}^2+y_{t}^2\right) |V_{ud}|&  \left(y_{c}^2-y_{t}^2\right)\left(y_{s}^2-y_{b}^2\right)|V_{ud}|\\[1mm]
  |V_{us}|& \left(y_{s}^2+y_{b}^2\right) |V_{us}| &  \left(y_{u}^2+y_{t}^2\right) |V_{us}|&  \left(y_{c}^2-y_{t}^2\right)\left(y_{d}^2-y_{b}^2\right)|V_{us}|\\[1mm]
   |V_{cd}|& \left(y_{d}^2+y_{b}^2\right) |V_{cd}| &  \left(y_{c}^2+y_{t}^2\right) |V_{cd}|&  \left(y_{u}^2-y_{t}^2\right)\left(y_{s}^2-y_{b}^2\right)|V_{cd}|\\[1mm]
    |V_{cs}|& \left(y_{s}^2+y_{b}^2\right) |V_{cs}| &  \left(y_{c}^2+y_{t}^2\right) |V_{cs}|&  \left(y_{u}^2-y_{t}^2\right)\left(y_{d}^2-y_{b}^2\right)|V_{cs}|
\end{array}\right)\label{JUD}
\end{equation}
where the proportionality constant is different for each row, namely the product $\left(y_{\alpha}^2-y_{t}^2\right)\left(y_{i}^2-y_{b}^2\right)$. 
The determinant of $J_{UD}$ is
\begin{align}\nonumber
\det\left(J_{UD}\right)=& \left(y_u^2-y_t^2\right)
\left(y_t^2-y_c^2\right)\left(y_c^2-y_u^2\right)\left(y_d^2-y_b^2\right)\left(y_b^2-y_s^2\right)\left(y_s^2-y_d^2\right)\\&\times|V_{ud}||V_{us}||V_{cd}||V_{cs}|
\end{align}
The analysis has turned out to be as simple as it could be. The determinant vanishes if \emph{any} of the mass differences does,
or if any of the entries of $V_{CKM}$ vanishes. The rank is reduced the most for three  vanishing mixing elements, which
corresponds to (a permutation of) the identity.

Next, the analysis of the invariants containing eigenvalues solely is
presented; the axial breaking case was analyzed in \cite{Maiani:2011zz}
but is reproduced here for completeness.
\begin{itemize}
\item{\bf Axial conserving case: $\mathcal{G}_\mathcal{F}^{q}\sim U(n_g)^3$.}  The Jacobians are in this case,
\begin{equation}
J_U=\partial_{\mathbf{y}_U} \left(\mbox{Tr}\,\mathcal{Y}_U\mathcal{Y}_U^\dagger\,,\,
\mbox{Tr}\,(\mathcal{Y}_U\mathcal{Y}_U^\dagger)^2\,,\,
\mbox{Tr}\,(\mathcal{Y}_U\mathcal{Y}_U^\dagger)^3\right)=\left(\begin{array}{ccc}
2y_u&4y_u^3&6y_u^5\\[2mm]
2y_c & 4 y_c^3& 6 y_c^5\\[2mm]
2y_t &4y_t^3&6y_t^5\\[2mm]
\end{array}\right)\,,
\end{equation}
and 
\begin{equation}
J_D=\partial_{\mathbf{y}_D} \left(\mbox{Tr}\,\mathcal{Y}_D\mathcal{Y}_D^\dagger\,,\,
\mbox{Tr}\,(\mathcal{Y}_D\mathcal{Y}_D^\dagger)^2\,,\,
\mbox{Tr}\,(\mathcal{Y}_D\mathcal{Y}_D^\dagger)^3\right)=\left(\begin{array}{ccc}
2y_d&4y_d^3&6y_d^5\\[2mm]
2y_s & 4 y_s^3&6 y_s^5\\[2mm]
2y_b & 4y_b^3&6y_b^5\\[2mm]
\end{array}\right),
\end{equation}
so that:
\begin{equation}
 \det J_U=48\,y_cy_uy_t (y_u^2-y_c^2)(y_c^2-y_t^2)(y_u^2-y_t^2)\,,
\end{equation}
\begin{equation}
 \det J_D=48\,y_dy_sy_b (y_d^2-y_s^2)(y_s^2-y_b^2)(y_d^2-y_b^2)\,.
\end{equation}
There are now 6 possibilities to cancel each determinant above with ordered
eigenvalues. These can be shorted in those who reduce the rank of the Jacobian to 2,
\begin{equation}
\mathcal{Y}\sim
\left(\begin{array}{ccc}
0&0&0\\
0&y&0\\
0&0&y'\\
\end{array}\right),\quad
\left(\begin{array}{ccc}
y&0&0\\
0&y&0\\
0&0&y'\\
\end{array}\right),
\quad
\left(\begin{array}{ccc}
y&0&0\\
0&y'&0\\
0&0&y'\\
\end{array}\right),
\label{JacSol3f}
\end{equation}
and those that yield a rank 1 Jacobian
\begin{equation}\mathcal{Y}\sim
\left(\begin{array}{ccc}
0&0&0\\
0&0&0\\
0&0&y\\
\end{array}\right),\quad
\mathcal{Y}\sim
\left(\begin{array}{ccc}
0&0&0\\
0&y&0\\
0&0&y\\
\end{array}\right),\quad
\mathcal{Y}\sim\left(\begin{array}{ccc}
y&0&0\\
0&y&0\\
0&0&y\\
\end{array}\right).
\end{equation}
These configurations correspond to boundaries which can also be extracted from the
general inequalities:
\begin{align}
I_{U^3}\geq&\frac{I_U}{2}\left(3I_{U^2}-I_U^2\right)\\
\left(3I_{U^2}-I_U^2\right)^3\geq&2\left(9I_{U^3}-9I_{U^2}I_U+2I_U^3\right)^2
\end{align}
and analogously for the down-type invariants. The different boundaries are depicted
in Fig. \ref{BoundariesGraphU(3)}.

\figuremacroW{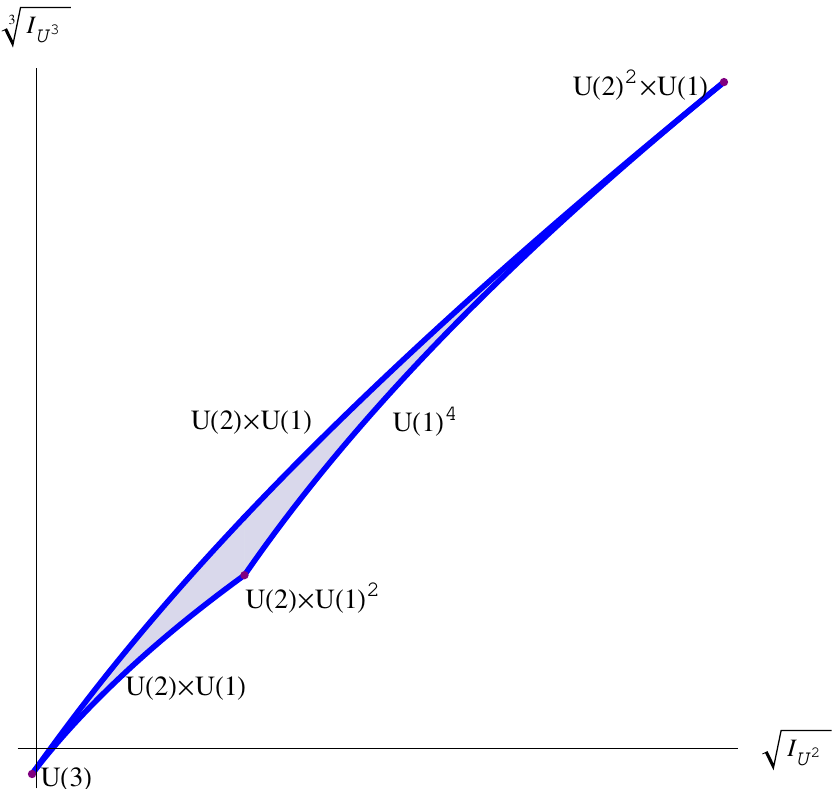}{{\it I-manifold} spanned by $\mathcal {G}_\mathcal {F}^q$ invariants built with $\mathcal Y_{U}$ for fixed $I_U$ and 3 quark generations}{The study of the shape of the allowed region for the invariants $I_{U^2}$, $I_{U^3}$ reveals the different
possible unbroken symmetries at the vacuum.}{.6}

We will not list all the possible combinations of the up and down sector configurations 
but display the three that result in the largest dimension maximal unbroken subgroups or singular points:
\begin{enumerate}
\item $\mathcal{G}_\mathcal{F}^q\rightarrow SU(3)_V\times U(1)_B$\,\, \emph{Down and Up quark sectors degenerate}
\begin{equation}
\mathcal{Y}_U =\Lambda_f \left(\begin{array}{ccc}
y&0&0\\
0&y&0\\
0&0&y\\
\end{array}\right)\,,
\qquad\quad
\mathcal{Y}_D=\Lambda_f \left(\begin{array}{ccc}
y'&0&0\\
0&y'&0\\
0&0&y'\\
\end{array}\right)\,.
\end{equation}
\item $\mathcal{G}_\mathcal{F}^q\rightarrow U(2)^3\times U(1)_{t+b}$\,\, \emph{Down and Up quark sectors hierarchical}
\begin{equation}
\mathcal{Y}_U =\Lambda_f\left(\begin{array}{ccc}
0&0&0\\
0&0&0\\
0&0&y\\
\end{array}\right)\,,\qquad\quad
 \mathcal{Y}_D=\Lambda_f \left(\begin{array}{ccc}
0&0&0\\
0&0&0\\
0&0&y'\\
\end{array}\right)
\end{equation}
\item $\mathcal{G}_\mathcal{F}^q\rightarrow U(2)^2\times U(1)^2$\,\, \emph{Two massive degenerate and one massless fermion for up-type quarks and  hierarchical down sector or vice versa.}
\begin{equation}
\mathcal{Y}_U =\Lambda_f\left(\begin{array}{ccc}
y&0&0\\
0&y&0\\
0&0&0\\
\end{array}\right)\,,\qquad\quad
 \mathcal{Y}_D=\Lambda_f \left(\begin{array}{ccc}
0&0&0\\
0&0&0\\
0&0&y'\\
\end{array}\right)
\end{equation}
\end{enumerate}

\item {\bf  Explicitly axial breaking case: $\mathcal{G}_\mathcal{F}^{\slashed A,q}\sim SU(n_g)^3$} The Jacobians read
\begin{equation}
J_U=\partial_y \left(\det \mathcal{Y}_U\,,\,\mbox{Tr}\,\mathcal{Y}_U\mathcal{Y}_U^\dagger\,,\,
\mbox{Tr}\,(\mathcal{Y}_U\mathcal{Y}_U^\dagger)^2\right)=\left(\begin{array}{ccc}
y_cy_t & 2y_u &4y_u^3\\[2mm]
y_ty_u&2y_c&4 y_c^3\\[2mm]
y_uy_c & 2y_t&4y_t^3\\[2mm]
\end{array}\right)\,,
\end{equation}
\begin{equation}
J_D=\partial_y \left(\det \mathcal{Y}_D\,,\,\mbox{Tr}\,\mathcal{Y}_D\mathcal{Y}_D^\dagger\,,\,
\mbox{Tr}\,(\mathcal{Y}_D\mathcal{Y}_D^\dagger)^2\right)=\left(\begin{array}{ccc}
y_by_s & 2y_d &4y_d^3\\[2mm]
y_dy_b&2y_s&4 y_s^3\\[2mm]
y_sy_d & 2y_b&4y_b^3\\[2mm]
\end{array}\right)\,,
\end{equation}
and the determinant of each Jacobian is
\begin{equation}
 \det J_U=8 (y_u^2-y_c^2)(y_c^2-y_t^2)(y_u^2-y_t^2)\,,
\end{equation}
\begin{equation}
 \det J_D=8(y_d^2-y_s^2)(y_s^2-y_b^2)(y_d^2-y_b^2)\,,
\end{equation}
from where we see that the first case in \ref{JacSol3f} is
no longer a solution. For this case the analogous of Fig. \ref{BoundariesGraphU(3)}
was first shown in \cite{Maiani:2011zz}.
\end{itemize}

\subsubsection{The Potential at the Renormalizable Level}\label{RP3QF}

The following study will determine which of the different unbroken
symmetries (boundaries) are respected (possible) at the different minima of the potential.
The renormalizable scalar potential will contain formally the
same independent invariants as in the two generation case:
however these invariants now depend on a higher number of variables.

\subsubsection*{Axial preserving case: $\mathcal{G}_\mathcal{F}^{q}\sim U(n_g)^3$}
The most general scalar potential at the renormalizable level in this case is just the same formally as
for the 2 family case: Eq.~\ref{Pot2FBi}, using the vector X as defined in \ref{Xvector}.
Next, the results of the minimization process are presented.


First the Von Neumann trace inequality allows for the automatic minimization of the
mixing term, so that two options arise:
\begin{equation} g<0\,,\quad
V_{CKM}=\left(\begin{array}{ccc}
1&0&0\\
0&1&0\\
0&0&1\\
\end{array}\right)\,;\quad\quad g>0\,,\quad
V_{CKM}=\left(\begin{array}{ccc}
0&0&1\\
0&1&0\\
1&0&0\\
\end{array}\right).
\end{equation}
The first is a good approximation to reality, whereas the second one would result
in the top quark coupled only to the down type quark. These solutions 
leave an invariant generation number $U(1)^3_V$ defined as in Eq. \ref{U1V} for generic values of
masses.

The two possibilities above are a reduced number of the various permutation matrices that
the Jacobian analysis singled out. This means that the potential selects some of these boundaries,
concretely those that order in an inverse or direct manner the mass eigenstates of up and down sectors.

With the same procedure as for the two family case we next minimize in
the variables that will determine the hierarchy. These are now the two
possible eigenvalue differences in the up sector and another two in the
down sector.

The potential is formally the same as in the 2 family case and let us note
the ``map" of Fig. \ref{VacuaMapQuarks}
is drawn in terms of invariant magnitudes, as detailed in Sec.~\ref{App1}, such that the dimension
of the matrices involved does not enter the computation. In this sense we expect the same map, as it will turn
out (but only for $g<0$). It is only left to determine what are the hierarchies in these regions.

We can anticipate, focusing on the contrast with the observed flavour pattern, that
a hierarchical solution corresponding to region {\bf I} of Fig.~\ref{VacuaMapQuarks}
where only the heaviest family is massive and the mixing matrix is the identity is a
natural possible solution. The intuitive way to guess the presence of this solution
follows the argument of the two family case in the region of negative $h_{U,D}$ and $g$. 
The resemblance with nature in this case
is good in a first sketch; {\it top and bottom are much heavier than the rest of quarks
and the mix little ($\sim \lambda_c^2$) with them}.

For completeness the set of vacua is listed next for the realistic case of mixing ($g<0$):

\begin{itemize}
\item[\bf I] In this region the equivalent of the hierarchical configuration is now the case of
 vanishing of the lightest 4 eigenvalues,
\begin{equation}
\mathcal{Y}_D =\Lambda_f \left(\begin{array}{ccc}
0&0&0\\
0&0&0\\
0&0&y_b \\
\end{array}\right)\,,\qquad\quad \mathcal{Y}_U =\Lambda_f
\left( \begin{array}{ccc}
0&0&0\\
0&0&0\\
0&0&y_t\\
\end{array}\right)\,,
\end{equation}
and an unbroken $U(2)^3\times U(1)_{t+b}$.
\item[\bf II] In this case we have a hierarchical Yukawa for the up sector
while the two lightest down-type eigenvalues are equal:
\begin{equation}
\mathcal{Y}_D =\Lambda_f \left(\begin{array}{ccc}
y&0&0\\
0&y&0\\
0&0&y_b \\
\end{array}\right)\,,\qquad\quad \mathcal{Y}_U =\Lambda_f
\left( \begin{array}{ccc}
0&0&0\\
0&0&0\\
0&0&y_t\\
\end{array}\right)\,,
\end{equation}
\begin{equation}
y_s=y_d=y\,,\qquad\qquad\frac{y_b^2-y^2}{y_b^2+2y^2}=\frac{-g}{2 h_D}\frac{I_U}{I_D}\,,
\end{equation}
leaving an unbroken an $U(2)_V \times U(2)_{U_R}\times U(1)_{t+b}$.
\item {\bf III} The analogous of the previous case for the up sector is
\begin{equation}
\mathcal{Y}_D =\Lambda_f \left(\begin{array}{ccc}
0&0&0\\
0&0&0\\
0&0&y_b \\
\end{array}\right)\,,\qquad\quad \mathcal{Y}_U =\Lambda_f
\left( \begin{array}{ccc}
y'&0&0\\
0&y'&0\\
0&0&y_t\\
\end{array}\right)\,,
\end{equation}
\begin{equation}
y_c=y_u=y\,,\qquad\qquad\frac{y_t^2-y'^2}{y_t^2+2y'^2}=\frac{-g}{2 h_U}\frac{I_D}{I_U}\,,
\end{equation}
with an unbroken $U(2)_V \times U(2)_{D_R}\times U(1)_{t+b}$.
\item {\bf IV} Finally the degenerate case is simply
\begin{equation}
\mathcal{Y}_D =\Lambda_f \left(\begin{array}{ccc}
y&0&0\\
0&y&0\\
0&0&y \\
\end{array}\right)\,,\qquad\quad \mathcal{Y}_U =\Lambda_f
\left( \begin{array}{ccc}
y'&0&0\\
0&y'&0\\
0&0&y'\\
\end{array}\right)\,,
\end{equation}
respecting a $U(3)_V$ symmetry.
\end{itemize}

Note that none of the solutions have a single vanishing eigenvalue, so
that only the case {\bf I} could be a good approximation to reality. It is the
case that the potential being the same as for two families, the picture
of possible vacua in Fig. \ref{VacuaMapQuarks} is the same, only
the unbroken symmetry is different, but the maximal that we could choose \cite{Michel:1970mua,Michel:1971th}.

This is not the situation for $g>0$, in such case we expect new solutions like $\mathcal Y_U \sim$Diag$(0,y,y')\,,\,\mathcal Y_D \sim$Diag$(y'',0,0)$,
which even if having an inverted hierarchy could be worth exploring.
\subsubsection*{Explicitly axial breaking case: $\mathcal{G}_\mathcal{F}^{\slashed A,q}\sim SU(n_g)^3$}
The potential is now:
\begin{align}\nonumber
V^{(4)}=&-\mu^2 \cdot X+ X^T \cdot \lambda \cdot X +g\,\mbox{Tr}\left(\mathcal{Y}_U\mathcal{Y}_U^\dagger
\mathcal{Y}_D\mathcal{Y}_D^\dagger\right) + h_U \mbox{Tr}\left(\mathcal{Y}_U\mathcal{Y}_U^\dagger
\mathcal{Y}_U\mathcal{Y}_U^\dagger\right)\\&+h_D \mbox{Tr}\left(\mathcal{Y}_D\mathcal{Y}_D^\dagger
\mathcal{Y}_D\mathcal{Y}_D^\dagger\right)+\left(\tilde\mu_U \det \mathcal{Y}_U+\tilde\mu_D \det \mathcal{Y}_D+h.c.\right)\,.
\label{Pot3FBiAB}
\end{align}
The determinants appear only linearly in this case such that the minimization in the complex phase simplifies, for example in 
the case of the up-type flavour field:
\begin{equation}
\tilde \mu_U \det \mathcal{Y}_U+h.c.=|\tilde \mu_U| |\det\mathcal Y_U| 2 \cos(\phi_U+\phi_{\tilde \mu})\rightarrow  \cos(\left\langle \phi+\phi_{\tilde\mu}\right\rangle)=-1
\end{equation}
so that we can consider effectively a positive determinant, $\det \mathcal{Y}_U>0$, and a negative coefficient, $\tilde\mu_U<0$, see \cite{Alonso:2011yg}.

The inclusion of determinants will not change the possibilities listed 
as {\bf I , II , III , IV}, since all of these configurations are also boundaries in this case.
In other words,
part of the symmetries in the solutions above are still left
after removing the $U(1)_{A^U,A^D}$ factors,
 namely $SU(2)_{D_R,E_R}$.
This did not happen in the two family case as the
unbroken symmetry was $``U(1)"$ rather than $``U(2)"$.
Note however that a solution like $\mathcal{Y}\sim$Diag$(0,y',y)$
can be perturbed by $U(1)_{A^U,A^D}$ breaking terms to 
produce a small third eigenvalue.

\subsubsection{The Potential at the Non-Renormalizable Level}

The first issue to deal with in this case is the fact that the
order of magnitude of the Yukawa eigenvalues is set by the ratio 
$y\sim \mu/(\Lambda_f\sqrt{\lambda})$ which implies for the top Yukawa
that the vev of the field $\mu/\sqrt{\lambda}$ is around the scale $\Lambda_f$
 signaling a bad convergence of the EFT.
To cope with this first  it is noted that the top Yukawa runs down
with energy whereas the relation $y\sim \mu/(\Lambda_f\sqrt{\lambda})$
does not determine the overall scale $\Lambda_f$. For energies of the order of $10^{8}$ GeV \cite{Degrassi:2012ry}
the top Yukawa is already smaller than the weak coupling constant allowing
the usual expansion in EFT.

The case in which the two scales are of the same order can nonetheless formally be treated in
the same sense as the non-linear $\sigma$-model, see also \cite{Kagan:2009bn} in this regard.
First the isolation in a single invariant of the problematic terms is accomplished by
the set of invariants; $\{\,I_U, I_{U^2}-\left(I_U\right)^2,I_{U^3}-\left(I_U\right)^3 \,\}$
instead of Eqs. \ref{InvBIF3F1}-\ref{InvBIF3F3}, such that the latter two are suppressed by one power of the second
highest eigenvalue: $y_c^2$. Terms in $I_U$ can be summed in
a generic function in the potential $F\left(I_U/\Lambda_f^2\right)\equiv F(y_t'^2)$ where $y_t'$ stands for
the highest eigenvalue of $\mathcal{Y}_U$, different from the top Yukawa
since the connection with Yukawas has also to be revisited
\begin{equation}
Y_U=\frac{\mathcal{Y}_U}{\Lambda_f}+\sum_ic_i\frac{\mathcal{Y}_U\left(\mathcal{Y}_U^\dagger\mathcal{Y}_U\right)^i}{\Lambda_f^{2i+1}}\simeq V_{CKM}^\dagger \left( \begin{array}{ccc}
y_u&0&0\\
0&y_c&0\\
0&0&f(y_t')\\
\end{array}\right).
\end{equation}
The relation of $y_t'$ with the top Yukawa coupling is then $y_t=f(y_t')$. Then substitution in the function $F$ yields the potential
as a function of the top Yukawa coupling $F(f^{-1}(y_t))$. This means certainly a loss in predictivity since the introduced 
functions $F\,,\,f$ are general, however for the present discussion it suffices that $F(f^{-1}(x))$ has a minimum
at $x\simeq1$.

In either case and to conclude this discussion, the symmetry arguments used to identify the possible vacua
hold the same in this ``strong interacting" scenario.

One interesting point is the possibility of non-renormalizable operators correcting the pattern of the renormalizable potential.
It is a priori either a fine-tuned option like in the two family case or unsuccessful since the configurations are protected by a large
 unbroken symmetry. The intuitive reason for this is that for perturbations to displace the minimum they must create a small
 tilt in the potential via linear dependence on the deviations from the 0-order solution; however non-renormalizable terms contain high powers of eigenvalues and therefore the corrections they introduce are not linear in the perturbations.
%
%

\mathversion{bold}
\section{Flavour Scalar Fields in the Fundamental}
\label{sec:Fund2F}
\mathversion{normal}
In the simplest case from the group theory
point of view, each Yukawa corresponds to two scalar fields $\chi$ transforming in the 
fundamental representation and the Yukawa Operator has dimension 6.
This approach would {\it a priori} allow to  introduce  one 
new field for each $SU(n_g)$ component of the flavour symmetry: three fields. However, such a minimal setup leads to an 
unsatisfactory realization of the flavour sector as no physical mixing angle is allowed. The situation changes qualitatively, though, if two $SU(n_g)_{Q_L}$ fundamental representations are introduced, 
one for the up and one for the down quark sectors, the field content is detailed in table \ref{F-SFQ}.

Before discussing the potential, inspection of  Eq. \ref{FConnection}
will illuminate the road ahead. The hypothesis now is that Yukawas are built out 
of two fundamental representations. In linear algebra terms, the Yukawa matrix is made out of two vectors.
This is of course a very strong assumption on the structure of the matrix. First and 
foremost such a matrix has rank 1, so that \emph{by construction, there is one
single eigenvalue per up and down sector different from 0}. Note that this statement
is independent of the number of generations and the scalar potential. The situation is then
a good starting approximation for a hierarchical spectrum.

Second, the number of variables in the flavour fields will now not be the
same as low energy flavour observables.  The scalar fields are fundamental and can be thought of as complex vectors that are ``rotated" under a  flavour symmetry
 transformation. The only physical 
invariants that can be associated to such vectors are the moduli and, if they live in the same space, their relative angles.
Altogether the list of independent invariants and therefore physical variables describing the fields is,
\begin{equation}Z=\left\{
\chi_U^{L\dagger}\chi_U^L\,,\qquad
\chi_U^{R\dagger}\chi_U^R\,,\qquad
\chi_D^{L\dagger}\chi_D^L\,,\qquad
\chi_D^{R\dagger}\chi_D^R\,,\qquad
\chi_U^{L\dagger}\chi_D^L\,\right\}
\end{equation}
where the array $Z$ will be useful for notation purposes; its index runs over the five values: $(U,L)\,,\,(U,R)\,,\,(D,L)\,,\,(D,R)\,,\,(U,D)$.

 There is now also a clear geometrical 
interpretation of the mixing angle: {\it the mixing angle between two generations of quarks is the misalignment 
of the $\chi^L$ flavons in $SU(n_g)_{Q_L}$ space}.

The \emph{I-manifold} in this case possesses a number of boundaries identifiable studying $Z$. Indeed
$Z$ does not cover all 5 dimensional space, but its components satisfy in general,
\begin{align}\label{FBT1}
Z_{(U,L),(D,L),(U,R),(D,R)}\geq0\, ,\\ \label{FBT2}
|Z_{(U,D)}|^2\leq Z_{(U,L)}Z_{(D,L)}\, ,
\end{align}
the boundaries are reached when the above inequalities are saturated.

A word on the phenomenology of this scenario is due as well.
Let us compare the phenomenology expected from bi-fundamental flavons (i.e. $d=5$ Yukawa operator) with that from fundamental 
flavons (i.e. $d=6$ Yukawa operators). For bi-fundamentals,  the  list of effective FCNC operators is exactly the same that 
in the original MFV proposal~\cite{D'Ambrosio:2002ex}. The case of fundamentals presents some differences: higher-dimension invariants 
can be constructed in this case, exhibiting lower dimension than in the bi-fundamental case. 
 For instance, one can compare these two operators: 
\begin{equation}
\overline{D}_R\,\mathnormal{\mathcal{Y}_D}^\dagger\,\mathcal{Y}_U\,\mathcal{Y}_U^\dagger\,Q_L\sim[\text{mass}]^6\qquad \longleftrightarrow \qquad\overline{D}_R\,
\chi_D^{R}\,\chi_U^{L\dagger}\,Q_L\sim[\text{mass}]^5\,,
\label{OperatorsFCNCdimensions}
\end{equation}
where the mass dimension of the invariant is shown in brackets; with  these two types of basic bilinear FCNC 
structures it is possible to build effective operators describing FCNC processes, but differing on the degree of 
suppression that they exhibit. 
This underlines the fact that the identification of Yukawa couplings with aggregates of two or more flavons is a 
setup which goes technically beyond the realization of MFV, resulting possibly in a distinct phenomenology which 
could provide a way to distinguish between fundamental and bi-fundamental origin.

Let us turn now to the construction of the potential.

%
%
\subsection{The Potential at the Renormalizable Level}
\label{sec:Fund2F_ScalarPot}

Previous considerations regarding the scale separation between EW and flavour breaking scale hold also in this case, 
and in consequence the Higgs sector contributions will not be explicitly described. 
The Potential for the $\chi$ fields can be written in the compact manner,
\begin{equation}
V^{(4)}=-\mu^2_f \cdot Z + Z^T\cdot \lambda_f\cdot Z + h.c.\, ,
\end{equation}
The total number of operators that can be introduced at 
the renormalizable level is 20, out of which 6 are complex\footnote{Minimization
in the (unobservable at low energy) phase of $\chi_U^{L\dagger}\chi_D^L$ nonetheless
eliminates all complex phases of the potential.}. However, only 5 different combinations of these will enter the 
minimization equations. The solution
\begin{equation}
\left\langle Z\right\rangle=\frac{1}{2}\lambda^{-1}_f\mu_f^2\, ,
\label{solF}
\end{equation}
exists if the vector $\lambda^{-1}_f\mu_f^2/2$ takes values inside the possible range of $Z$, that is the \emph{I-manifold}.
The case in which this does not happen leads to a boundary of the invariant space. This occurs both when at least one of the first 4 entries
turns negative in $\lambda^{-1}_f\mu_f^2$ and the boundary is of the type of Eq.~\ref{FBT1} \emph{and} when $(\lambda^{-1}_f\mu_f^2)_{(U,L)}(\lambda^{-1}_f\mu_f^2)_{(D,L)}\leq |(\lambda^{-1}_f\mu_f^2)_{(U,D)}|^2$ which corresponds to the boundary that saturates~\ref{FBT2}. This last case corresponds to the two vectors $\chi_{U,D}^L$ aligned, that precludes any  mixing.
This means that the no-mixing case is a boundary to which nonetheless the minima of the potential is not restricted in general.

All these considerations make straightforward the extraction 
of the Yukawa structure. 

\begin{itemize}
\item {\bf Two family case - }
From the expressions for the Yukawa matrices in Eqs.~\ref{FConnection},
and the previous discussion we write that the configuration for the Yukawas is:
\begin{equation}
Y_D=\frac{\left|\chi_D^L\right|\left|\chi_D^R\right|}{\Lambda_f^2} \left(\begin{array}{cc}
 0 & 0\\
 0 & 1\\ 
\end{array}\right)\,,\qquad\quad																	
Y_U=\frac{\left|\chi_U^L\right|\left|\chi_U^R\right|} {\Lambda_f^2} V_{C}
\left(\begin{array}{cc}
 0 & 0\\
 0 & 1\\ 
\end{array}\right)\,,
\label{Yukawas3F_Fund}
\end{equation}
\begin{equation}
V_C=\left(
\begin{array}{cc}
\cos{\theta_c}&\sin{\theta_c}\\
-\sin{\theta_c}&\cos{\theta_c}\\
\end{array}\right)\,,
\end{equation} 
so that quark masses are fixed via Eq. \ref{solF} to:
\begin{equation}
y_c=\sqrt{\frac{\left(\lambda^{-1}_f\mu_f^2\right)_{(U,R)}}{2\Lambda_f^2}\frac{\left(\lambda^{-1}_f\mu_f^2\right)_{(U,L)}}{2\Lambda_f^2}}\, ,\qquad
\quad y_s=\sqrt{\frac{\left(\lambda^{-1}_f\mu_f^2\right)_{(D,R)}}{2\Lambda_f^2}\frac{\left(\lambda^{-1}_f\mu_f^2\right)_{(D,L)}}{2\Lambda_f^2}} \, ,
\end{equation}
\begin{equation}
\cos{\theta_c}=\frac{\left(\lambda^{-1}_f\mu_f^2\right)_{(U,L)}}{\sqrt{\left(\lambda^{-1}_f\mu_f^2\right)_{(U,L)}\left(\lambda^{-1}_f\mu_f^2\right)_{(D,L)}}} \, .
\end{equation}
The vev of the moduli of the $\chi$ fields is of the same order $\mu$ for natural parameters,
so that the cosine of the Cabibbo angle above is typically of $\mathcal{O}(1)$. This means that in
the fundamental case a natural scenario can give rise to both the strong hierarchies in quark
masses and a non-vanishing mixing angle, whereas in the bi-fundamental case the
mixing was unavoidably set to 0.

\item {\bf Three family case - } The extension is simple, the Yukawa matrices
are still of rank one and a single mixing angle arises,
\begin{equation}
Y_D=\frac{\left|\chi_D^L\right|\left|\chi_D^R\right|}{\Lambda_f^2} \left(\begin{array}{ccc}
0 & 0 & 0\\
0 & 0 & 0\\
0 & 0 & 1\\ 
\end{array}\right)\,,\qquad\quad																	
Y_U=\frac{\left|\chi_U^L\right|\left|\chi_U^R\right|} {\Lambda_f^2} V^\dagger_{CKM}
\left(\begin{array}{ccc}
0 & 0 & 0\\
0 & 0 & 0\\
0 & 0 & 1\\ 
\end{array}\right)\,.
\label{Yukawas3F_Fund}
\end{equation}
\begin{equation}
V_{CKM}=\left(
\begin{array}{ccc}
1&0&0\\
0&\cos{\theta_{23}}&\sin{\theta_{23}}\\
0&-\sin{\theta_{23}}&\cos{\theta_{23}}\\
\end{array}\right)\,.
\end{equation} 
with
\begin{equation}
y_t=\sqrt{\frac{\left(\lambda^{-1}_f\mu_f^2\right)_{(U,R)}}{2\Lambda_f^2}\frac{\left(\lambda^{-1}_f\mu_f^2\right)_{(U,L)}}{2\Lambda_f^2}}\, ,\qquad
\quad y_b=\sqrt{\frac{\left(\lambda^{-1}_f\mu_f^2\right)_{(D,R)}}{2\Lambda_f^2}\frac{\left(\lambda^{-1}_f\mu_f^2\right)_{(D,L)}}{2\Lambda_f^2}} \, ,
\end{equation}
\begin{equation}
\cos{\theta_{23}}=\frac{\left(\lambda^{-1}_f\mu_f^2\right)_{(U,L)}}{\sqrt{\left(\lambda^{-1}_f\mu_f^2\right)_{U,L}\left(\lambda^{-1}_f\mu_f^2\right)_{(D,L)}}} \, .
\end{equation}
The flavour field vevs have not  broken completely the flavour symmetry, leaving a residual $U(1)_{Q_L}\times U(2)_{D_R}\times 
U(2)_{U_R}\times U(1)_B$ symmetry group. This can be seen as follows, in the three dimensional space where $SU(3)_{Q_L}$ acts,
the two vectors $\chi_{U,D}^L$ define a plane; perpendicular to this plane there is the
direction of the family that is completely decoupled form the rest, and in the plane we have
the massive eigenstate and the eigenstate that, even if massless, can be told from the other massless
state as it mixes with the massive.
\end{itemize}

If the hierarchies in mass in each up and down sectors are explained here through the
very construction of the Yukawas via fundamental fields, there is still the
hierarchy of masses between the top and bottom for the potential to accommodate,
that is:
\begin{equation}
y_b^2/y_t^2=\frac{\left(\lambda^{-1}_f\mu_f^2\right)_{D,R}\left(\lambda^{-1}_f\mu_f^2\right)_{D,L}}{\left(\lambda^{-1}_f\mu_f^2\right)_{U,R}\left(\lambda^{-1}_f\mu_f^2\right)_{U,L}}\simeq 5.7\times10^{-4}
\end{equation}
Note that the top-bottom hierarchy is explained in this context by the 4th power ratio of
$\mu$-mass scales so that a typical ratio of $(\mu_{f})_D/(\mu_f)_U\simeq 0.15$ suffices to explain the hierarchy.

One of the consequences of the strong hierarchy in masses imposed in this scenario is that it cannot be
corrected with nonrenormalizable terms in the potential to obtain small masses for the lightest families.
The reason is that the vanishing of all but one eigenvalue in the Yukawa matrices
is obtained as a consequence of the scalar field fundamental content.

Nevertheless, the partial breaking of flavour symmetry provided by Eq.~({\ref{Yukawas3F_Fund}}) can open 
quite interesting possibilities from a model-building point of view. Consider as an example the following 
multi-step approach. In a first step, only the minimal number of fundamental fields are introduced: i.e. $\chi^L$, $\chi^R_U$ 
and $\chi^R_D$. Their vevs break $SU(3)^3$ down to $SU(2)^3$, originating non-vanishing Yukawa couplings 
only for the top and the bottom quarks, without any mixing angle (as we have only one left-handed flavour field). As a second step, 
 four new triplet fields $\chi^{\prime L,R}_{u,d}$ are added,  whose contributions to the Yukawa terms are suppressed 
relatively to the previous flavons. If their vevs 
 point in the direction of the unbroken flavour subgroup 
$SU(2)^3$, then the residual symmetry is further reduced. As a result, non-vanishing charm and strange Yukawa couplings 
are generated together with a mixing among the first two generations:
\begin{equation}
\begin{aligned}
&Y_U\equiv \frac{{\chi^L}\,{\chi_U^{R\dagger}}}{\Lambda_f^2}+ \frac{{\chi_U^{\prime L}}\,
{\chi_U^{\prime R\dagger}}}{\Lambda_f^2}=\left(\begin{array}{ccc}
0&\sin{\theta_c}\,y_c&0\\
0&\cos{\theta_c}\,y_c&0\\
0&0&y_t\\
\end{array}\right)\,,\\
&Y_D\equiv \frac{{\chi^L}\,{\chi_D^{R\dagger}}}{\Lambda_f^2}+ \frac{{\chi_D^{\prime L}}\,
{\chi_D^{\prime R\dagger}}}{\Lambda_f^2}=\left(\begin{array}{ccc}
0&0&0\\
0&y_s&0\\
0&0&y_b\\
\end{array}\right)\,.
\end{aligned}
\end{equation}
The relative suppression of the two sets of flavon vevs correspond to the hierarchy between $y_c$ and $y_t$ ($y_s$ and $y_b$)
. Hopefully, a 
refinement of this argument would allow to explain the rest of the Yukawas and the remaining angles.  The 
construction of the scalar potential for such a setup would be quite model dependent though, and beyond the scope of this discussion.

%
%
\section{Combining fundamentals and bi-fundamentals}
\label{sec:Bi_Fund-Fund}

Until now we have considered separately Yukawa operators of dimension $d=5$ and $d=6$. It is, however, interesting 
to explore if some added value from the simultaneous presence of both kinds of operators can be obtained. 
This is a sensible choice from the point of view of effective Lagrangians in which, working at $\mathcal{O}(1/\Lambda_f^2)$, contributions of three types may be included: i) the leading $d=5$ $\mathcal{O}(1/\Lambda_f)$ operators; ii) renormalizable terms stemming from fundamentals (i.e. from $d=6$ $\mathcal{O}(1/\Lambda_f^2)$ operators; iii) other corrections numerically competitive at the orders considered here. We focus here as illustration on the impact of i) and ii):
\begin{equation}
\mathscr{L}_{Yukawa}=\overline{Q}_L\left[\dfrac{\mathnormal{\mathcal{Y}_D}}{\Lambda_f}+
\dfrac{\chi_D^L\chi_D^{R\dagger} }{\Lambda_f^2} \right]D_RH +
\overline{Q}_L\left[\dfrac{\mathcal{Y}_U}{\Lambda_f}+
\dfrac{\chi_U^L \chi_U^{R\dagger}}{\Lambda_f^2}  \right]U_R\tilde{H}+h.c.\,,
\label{LagrangianMixed}
\end{equation}
As the bi-fundamental flavour fields arise at first order in the $1/\Lambda_f$  expansion, it is suggestive to think of the fundamental 
contributions as a ``higher order'' correction.  
Let us then consider the case in which 
the flavons develop vevs as follows:
\begin{equation}
\dfrac{{\mathcal{Y}_{U,D}}}{\Lambda_f}\sim\left(
        \begin{array}{ccc}
           0  & 0 & 0 \\
           0  & 0 & 0 \\
           0  & 0 & y_{t,b} \\
        \end{array}
\right)\,,\qquad\qquad
\dfrac{|{\chi_{U,D}^L}|}{\Lambda_f}\sim 
           y_{c,s} \,,
\label{lastequation}
\end{equation}
and $\chi_{U,D}^R$ acquire arbitrary vev values of order $\Lambda_f$. Finally, if the
left-handed fundamental flavour fields are aligned perpendicular to the bi-fundamental fields and
misaligned by $\theta_c$ among themselves the Yukawas read,
\begin{equation}
\begin{aligned}
&Y_U=\left(\begin{array}{ccc}
0&\sin{\theta_c}\,y_c&0\\
0&\cos{\theta_c}\,y_c&0\\
0&0&y_t\\
\end{array}\right)\,,\qquad
&Y_D=
\left(\begin{array}{ccc}
0&0&0\\
0&y_s&0\\
0&0&y_b\\
\end{array}\right)\,.
\end{aligned}
\end{equation}
This seems an appealing pattern, with masses for the two heavier generations and one sizable mixing angle, that we chose to identify here with the Cabibbo angle\footnote{Similar constructions have been suggested also in other contexts as in \cite{Berezhiani:2005tp}.}. As for the lighter family,  non-vanishing masses for the up and  down quarks could now result from non-renormalizable operators.

The drawback of this combined analysis is that the direct connection between the minima of the potential and the spectrum is lost and the analysis of the potential would be very involved.


\include{XYZ}

\chapter{Lepton Sector} 


\ifpdf
    \graphicspath{{X/figures/PNG/}{X/figures/PDF/}{X/figures/}}
\else
    \graphicspath{{X/figures/EPS/}{X/figures/}}
\fi

Research in the lepton sector is at the moment in a dynamical and
exciting epoch. With the recent
measure of a sizable $\theta_{13}$ mixing angle \cite{An:2012eh,Ahn:2012nd},
all angles of the mixing matrix are determined and the race
for discovery of CP violation in the lepton sector has started \cite{Pascoli:2013wca}.
At the same time there is an ambitious experimental search
for flavour violation in the charged lepton sector \cite{DeGerone:2012iya,Blondel:2013ia,Hungerford:2009zz,Carey:2008zz} which could pour light
in possible new physics beyond the SM, and provide a new probe of the magnitude of the seesaw scale 
\cite{Alonso:2012ji}. On the cosmology side recent data seem to favor 3 only light species of neutrinos \cite{Ade:2013zuv}. Finally neutrinoless double beta decay 
searches \cite{GomezCadenas:2011it} will explore one very fundamental question: are there fermions
in nature which are their own antiparticle?

For the present theoretical analysis the nature of neutrino masses is crucial.
If neutrinos happen to be Dirac particles, the analysis of the flavour symmetry breaking
mechanism is completely analogous  to that for the quark case: all conclusions drawn are directly translated
to the lepton case and negligible mixing would be favored for the simplest set-up in which each Yukawa coupling is associated to a field in the bi-fundamental of the flavour group. Like for the  quark case, sizable mixing  would be allowed, for setups in which the Yukawas are identified with (combinations of) fields in the fundamental representation of the flavour group,  implying a strong hierarchy for neutrinos.

 We turn here instead to the case in which neutrinos are Majorana particles and more concretely generated by a  type I seesaw model. It has been previously found   \cite{Wyler:1982dd,Mohapatra:1986bd,Branco:1988ex,Raidal:2004vt,Gavela:2009cd} that for type I seesaw scenarios which exhibit approximate lepton number conservation, interesting seesaw models arise in which the 
 effective scale of lepton number is distinct from the flavour scale yielding
 a testable phenomenology~\cite{Gavela:2009cd,Alonso:2010wu,Eboli:2011ia,Dinh:2012bp,Sierra:2012yy}. It was first in this setup that we identified the patterns \cite{Alonso:2012fy} to be established with more generality in the next sections. Let us consider in this chapter the general seesaw I scenario with degenerate heavy right-handed neutrinos as outlined in the introduction.

Within the hypothesis of dynamical Yukawa couplings we introduce two
scalar fields in parallel to the two Yukawa matrices that are bi-fundamentals of
$\mathcal{G}_\mathcal{F}^l$
as detailed in table \ref{BIF-SFL}.

\section{Two Family Case}

The counting of physical parameters goes as follows. It is known \cite{Broncano:2002rw} that for two families with heavy degenerate neutrinos, the number of physical parameters describing the lepton sector is eight: six moduli and two phases. 

Indeed, after using the freedom to choose the lepton charged matrix diagonal, as in Eq.~\ref{YukParL}, 
 $Y_\nu$ is still a priori a general complex matrix with 8 parameters. Two phases can be absorbed through left-handed field  $U(1)$ rotations and an $O(2)$ rotation on the right of the neutrino Yukawa coupling (see Eq. \ref{TransBIFL})
reduces to five the number of physical parameters in $Y_\nu$, so that
altogether $n=7$ parameters suffice to describe the physical degrees of freedom in the lepton Yukawas. The eight physical parameter is the heavy neutrino mass $M$. Below, for the explicit computation, we will use either the Casas-Ibarra parametrization \cite{Casas:2001sr} or the bi-unitary parametrization alike  the
quark case.
The Casas-Ibarra parametrization is useful to maintain explicit the connection with masses and mixing, here it is reproduced for two families,
\begin{equation}
Y_E=
 \left(
 \begin{array}{cc}
  y_e & 0 \\
  0 & y_\mu \\
  \end{array}
  \right)\,,\quad\!
Y_\nu= \frac{\sqrt{2M}}{v}\,U\,
\left( \begin{array}{cc}
  \sqrt{m_{\nu_1}} & 0   \\
  0 &   \sqrt{m_{\nu_2}} \\
  \end{array}\right)
\,R\,,
\label{eq:YukawaVevs}
\end{equation}
\begin{equation}
U=\left( \begin{array}{cc}
\cos{\theta}&\sin{\theta}\\
-\sin{\theta}&\cos{\theta}\\
\end{array}\right)
\left( \begin{array}{cc}
e^{i\alpha}&0\\
0&e^{-i\alpha}\\
\end{array}\right)\,,
\qquad R=\left( \begin{array}{cc}
\cosh{\omega}&-i\sinh{\omega}\\
i\sinh{\omega}&\cosh{\omega}\\
\end{array}\right)\,. 
\end{equation}
 In order to extend the parametrization above to the fields $\mathcal{Y}_E$, $\mathcal{Y}_\nu$, it is convenient to use  the definitions
\begin{equation}
\hat y_{\nu_i}^2\equiv\frac{2M}{v^2}m_{\nu_i}\,,
\end{equation}
leading to 
\begin{equation}
\mathcal{Y}_\nu= \Lambda_f \left( \begin{array}{cc}
\cos{\theta}&\sin{\theta}\\
-\sin{\theta}&\cos{\theta}\\
\end{array}\right)
\left( \begin{array}{cc}
e^{i\alpha}\hat y_{\nu_1}&0\\
0&e^{-i\alpha}\hat y_{\nu_2} \\
\end{array}\right)
\left( \begin{array}{cc}
\cosh{\omega}&-i\sinh{\omega}\\
i\sinh{\omega}&\cosh{\omega}\\
\end{array}\right)\,,
\label{eq:YnuRHbasis}
\end{equation}
\begin{equation}
\mathcal{Y}_E=\Lambda_f\mathbf{y}_E=\Lambda_f\left( \begin{array}{cc}
y_e&0\\
0&y_\mu\\
\end{array}
\right)\,.
\end{equation}
It is the case nonetheless that the minimization procedure is optimized when selecting the second type of parametrization:
the bi-unitary in analogy with quarks (Eq.~\ref{BIU1}):
\begin{align}
\mathcal{Y}_\nu=\Lambda_f\,\mathcal U_L\mathbf{y}_\nu \,\mathcal U_R\,,\qquad \mathcal{Y}_E= \Lambda_f\mathbf{y}_E\,;\qquad\qquad \mathcal U_L\,\mathcal U_L^\dagger=1\, ,\,\,\,\, \mathcal U_R\,\mathcal U_R^\dagger=1\, ,
\label{BI-U}
\end{align}
with $\mathbf{y}_E$ as defined above, $\mathcal U_{L,R}$ being unitary matrices and $\mathbf{y}_\nu$ containing
the eigenvalues of the neutrino Yukawa matrix (for two families: $\mathbf{y}_\nu\equiv\mbox{Diag}(y_{\nu_1},y_{\nu_2})$), distinct from neutrino masses. The connection with the latter
is:
\begin{equation}
U_{PMNS}\,\mathbf{m}_\nu \,U_{PMNS}^T=Y_\nu\frac{v^2}{2M}Y_\nu^T=\frac{v^2}{2M}\mathcal U_L \, \mathbf{y}_\nu \,\mathcal U_R\,\mathcal U_R^T \,\mathbf{y}_\nu \,\mathcal U_L^T\,,
\label{NuMassBI-U}
\end{equation}
where $\mathbf{m}_\nu=$Diag$(m_{\nu_i})$. None of the unitary matrices $\mathcal U_{L,R}$ corresponds to $U_{PMNS}$, rather $U_{PMNS}$ is
the combination of them that diagonalizes the matrix above.

The expression of mixing and masses in terms of the bi-unitary parameters in the general case is involved but
the usefulness of this method is that we will not need it. The potential will select particularly
simple points of this parametrization with an easy connection to low energy parameters.

In the following we will use the Casas-Ibarra parametrization for the Jacobian and mixing analysis and 
move to the bi-unitary to simplify matters in the mass hierarchy analysis of the potential.

The scalar potential for the  $\mathcal{Y}_E$ and $\mathcal{Y}_\nu$ fields must 
 be invariant under the SM gauge symmetry and the flavour symmetry $\mathcal{G}_\mathcal{F}^l$. 
The possible independent invariants reduce to precisely seven terms,
\begin{align}
I_{E}&=\mbox{Tr}\left[\mathcal{Y}_E \mathcal{Y}_E^\dagger \right] \, ,& I_{\nu}&=\mbox{Tr}\left[\mathcal{Y}_\nu \mathcal{Y}_\nu^\dagger \right] \, ,\\ \label{InvBiF2FL2}
I_{E^2}&=\mbox{Tr}\left[(\mathcal{Y}_E \mathcal{Y}_E^\dagger)^2 \right] \, ,& I_{\nu^2}&=\mbox{Tr}\left[(\mathcal{Y}_\nu \mathcal{Y}_\nu^\dagger)^2 \right] \, ,\\
 I_{\nu'}&= \mbox{Tr}\left[\mathcal{Y}_\nu^\dagger \mathcal{Y}_\nu \mathcal{Y}_\nu^T \mathcal{Y}_\nu^*  \right]\, ,&
 I_{\nu,E}&=\mbox{Tr}\left[\mathcal{Y}_\nu \mathcal{Y}_\nu^\dagger \mathcal{Y}_E \mathcal{Y}_E^\dagger \right]\,, \\
 I_{ \nu',E}&=\mbox{Tr}\left[\mathcal{Y}_\nu \mathcal{Y}_\nu^T \mathcal{Y}_\nu^* \mathcal{Y}_\nu^\dagger \mathcal{Y}_E \mathcal{Y}_E^\dagger \right]\,. &
\end{align}
In terms of the variables of the Casas-Ibarra parametrization, the invariants read:
\begin{align}
I_{E}=&\Lambda_f^2\left(y_e^2+ y_\mu^2\right) ,\qquad I_{\nu}=\Lambda_f^2\left(\hat y_{\nu_1}^2+\hat y_{\nu_2}^2\right)\cosh2\omega \,,\\
 I_{E^2}=&\Lambda_f(y_e^4+y_\mu^4)\,, \qquad I_{\nu^2}=\Lambda_f^4((\hat y_{\nu_1}^2-\hat y_{\nu_2}^2)^2+(\hat y_{\nu_1}^2+\hat y_{\nu_2}^2)^2\cosh 4\omega)/2\,,\\
I_{\nu'}=&\Lambda_f^4\left(\hat y_{\nu_1}^4+\hat y_{\nu_2}^4\right) \, ,\\
\nonumber
  I_{\nu,E}=&\Lambda_f^4 [\left(y_\mu^2-y_e^2\right)\left(\hat y_{\nu_1}^2-\hat y_{\nu_2}^2\right)\cos2\theta \cosh 2\omega+ \left(y_e^2+y_\mu^2\right)\left(\hat y_{\nu_1}^2+\hat y_{\nu_2}^2\right)\\ \label{ILMIX}
&+2\left(y_\mu^2-y_e^2\right)\hat y_{\nu_1}\hat y_{\nu_2} \sin2\alpha \sin2\theta \sinh2\omega]/2\,,  \\
 I_{\nu',E}=&\Lambda_f^6 \left[\left(y_\mu^2-y_e^2\right)\left(\hat y_{\nu_1}^4-\hat y_{\nu_2}^4\right)\cos2\theta+
        \left(y_e^2+y_\mu^2\right)\left(\hat y_{\nu_1}^4+\hat y_{\nu_2}^4\right)\right]/2\,. 
\end{align}
These results apply to a seesaw I construction with heavy degenerate neutrinos,
for a general seesaw see \cite{onpreparation2}. Note the different dependence in the mixing angle in Eq.~\ref{ILMIX}.
Crucial to this difference are non trivial values of $\omega$ and $\alpha$ ($\omega\neq0$ , $\sin2 \alpha\neq0$), which 
will be shown in the next section to be natural minima of the system. 

For the explicitly axial breaking case ($\mathcal{G}_\mathcal{F}^{\slashed A,l}\sim SU(n_g)^2\times SO(n_g)$) two new invariants would appear,
\begin{align}
I_{\tilde E}&=\det\left(\mathcal{Y}_E\right) \, ,& I_{\tilde \nu}&= \det\left(\mathcal{Y}_\nu \right)\, ,
\label{DetsL2G}
\end{align}
which would substitute the invariants in Eq. \ref{InvBiF2FL2} as for the quark case, see Eqs. \ref{det-tr-2GU}-\ref{det-tr-2GD}.

Finally, the determinants in Eqs. \ref{DetsL2G} can be expressed as 
\begin{align}
I_{\tilde E}&=\Lambda_f^2y_e y_\mu e^{i\phi_E} ,& I_{\tilde \nu}&=\Lambda_f^2 \hat y_{\nu_1}\hat y_{\nu_2} e^{i\phi_\nu}.
\end{align}

\subsection{The Jacobian}

The Jacobian can be factorized as follows:
\begin{equation}
J=\left(\begin{array}{ccc}
\partial_{\mathbf{y}_E}I_{E^n} & 0 & \partial_{\mathbf{y}_E}  I_{(\nu,E),(\nu',E)}\\
0   & \partial_{\hat{\mathbf{y}}_\nu,\omega}I_{\nu^n} & \partial_{\hat{\mathbf{y}}_\nu,\omega}  I_{(\nu,E),(\nu',E)}\\
0 & 0 & \partial_{\theta,\alpha } I_{(\nu,E),(\nu',E)}
\end{array}\right) \equiv\left(\begin{array}{ccc}
J_E & 0 & \partial_{\mathbf{y}_E}  I_{(\nu,E),(\nu',E)}\\
0   & J_\nu & \partial_{\hat{\mathbf{y}}_\nu,\omega}  I_{(\nu,E),(\nu',E)}\\
0 & 0 & J_{\theta,\alpha}
\end{array}\right). 
\end{equation}
The sub-Jacobian involving the mixing parameters is given by,
\begin{align}
J_{\theta,\alpha}=&\partial_{\theta , \alpha}\left(\mbox{Tr}\left[\mathcal{Y}_\nu \mathcal{Y}_\nu^\dagger \mathcal{Y}_E \mathcal{Y}_E^\dagger \right]\,,\,\mbox{Tr}\left[\mathcal{Y}_\nu \mathcal{Y}_\nu^T \mathcal{Y}_\nu^* \mathcal{Y}_\nu^\dagger \mathcal{Y}_E \mathcal{Y}_E^\dagger \right]\right)\\ \nonumber
\propto&\left(\begin{array}{cc}
2\hat y_{\nu_1}\hat y_{\nu_2}\sinh2\omega\sin2\alpha \cos2\theta-\left(\hat y_{\nu_1}^2-\hat y_{\nu_2}^2\right)\cosh2\omega \sin2\theta &\left(\hat y_{\nu_1}^4-\hat y_{\nu_2}^4\right)\sin 2\theta\\
2\hat y_{\nu_1}\hat y_{\nu_2}\sinh2\omega \sin2\theta\cos2\alpha&0\\
\end{array}\right)\,
\end{align}
with determinant
\begin{equation}
\det{J_{\theta,\alpha}}=2\,\hat y_{\nu_1}\hat y_{\nu_2}\left(y_\mu^2-y_e^2\right)^2\left(\hat y_{\nu_1}^4-\hat y_{\nu_2}^4\right)\sinh 2\omega\, \sin^2 2\theta\,\cos2\alpha\,.
\end{equation}
This last equation shows the fundamental difference with respect to the quark (or more in general Dirac) case: reducing the rank can be accomplished by
choosing $\alpha=\pi/4$. It will be shown later on, through an explicit example, how this solution comes
along with mass degeneracy for light neutrinos.

Let us next consider the analysis of the Jacobian for the mass sector
\begin{itemize}
\item {\bf Axial preserving case: $\mathcal{G}_\mathcal{F}^{l}\sim U(n_g)^2\times O(n_g)$ -} In this case the subJacobian $J_\nu$ is built as follows,
\begin{align}
J_\nu=&\partial_{\hat{\mathbf{y}}_\nu,\omega}\left(\mbox{Tr}\left[\mathcal{Y}_\nu \mathcal{Y}_\nu^\dagger \right]\,,\, \mbox{Tr}\left[(\mathcal{Y}_\nu \mathcal{Y}_\nu^\dagger)^2 \right]\,,\, \mbox{Tr}\left[\mathcal{Y}_\nu \mathcal{Y}_\nu^T\mathcal{Y}_\nu^* \mathcal{Y}_\nu^\dagger \right]\right)\\
=&
\left(\begin{array}{ccc}
2\hat y_{\nu_1}\cosh 2\omega & 4\hat y_{\nu_1}^3\cosh^22\omega+4\hat y_{\nu_1}\hat y_{\nu_2}^2\sinh^22\omega&4\hat y_{\nu_1}^3\\
2\hat y_{\nu_2}\cosh 2\omega &4\hat y_{\nu_2}\hat y_{\nu_1}^2\sinh^22\omega+4\hat y_{\nu_2}^3\cosh^22\omega&4\hat y_{\nu_2}^3\\
2(\hat y_{\nu_1}^2+\hat y_{\nu_2}^2)\sinh2\omega&2(\hat y_{\nu_1}^2+\hat y_{\nu_2}^2)^2\sinh4\omega &0\\
\end{array}\right)\,,\nonumber
\end{align}
and its determinant is,
\begin{equation}
\det{J_\nu}=32\,\hat y_{\nu_1}\hat y_{\nu_2}(\hat y_{\nu_1}^2+\hat y_{\nu_2}^2)^2(\hat y_{\nu_1}^2-\hat y_{\nu_2}^2)\sinh2\omega\,,
\end{equation}
whereas for charged leptons it results, in analogy with the quark case:
\begin{equation}
\det J_E=8\,y_e y_\mu \left(y_e^2-y_\mu^2\right)\,.
\end{equation}
The configurations that reduce the most the range of the Jacobian $J_\nu$ involve $\omega=0$;
given this, the Jacobian $J_{\theta,\alpha}$ is vanishing for either degenerate mass states or $\theta=0$. The
naive singular points are therefore:
\begin{equation}
1.)\qquad \mathcal{Y}_E=\Lambda_f \left(\begin{array}{cc}
						0&0\\
						0&y_\mu\\
						\end{array}\right),\qquad\qquad
\mathcal{Y}_\nu =\Lambda_f \left(\begin{array}{cc}
						0&0\\
						0& y_{\nu_2}\\
						\end{array}\right),\label{1SP2GL}
\end{equation}
with an unbroken $U(1)_e^2$ and 
\begin{equation}
2.)\qquad\mathcal{Y}_E=\Lambda_f \left(\begin{array}{cc}
						y&0\\
						0&y\\
						\end{array}\right),\qquad\qquad
\mathcal{Y}_\nu =\Lambda_f \left(\begin{array}{cc}
						y_\nu&0\\
						0& y_\nu\\
						\end{array}\right),\label{2SP2GL}
\end{equation}
where there is a diagonal $SO(2)$ unbroken. These are the two singular points that
can be identified at this level.
We shall see however that the present parametrization
obscures the presence of different singular points, some of which lead to maximal mixing, a seemingly
absent situation at present. This will be clarified in the study of the renormalizable potential.

\item {\bf Explicitly Axial breaking case: $\mathcal{G}_\mathcal{F}^{\slashed A,l}\sim SU(n_g)^2\times SO(n_g)$ -} The Jacobian reads now,
\begin{align}
J_\nu=&\partial_{\hat{\mathbf{y}}_\nu,\omega}\left( \det\mathcal{Y}_{\nu}\,,\, \mbox{Tr} \left[\mathcal{Y}_\nu \mathcal{Y}_\nu^\dagger \right]\,,\,
 \mbox{Tr}\left[\mathcal{Y}_\nu \mathcal{Y}_\nu^T\mathcal{Y}_\nu^* \mathcal{Y}_\nu^\dagger \right]\right)\, ,\\
=&\left(\begin{array}{ccc}
\hat y_{\nu_2}&2\hat y_{\nu_1}\cosh 2\omega &4\hat y_{\nu_1}^3\\
\hat y_{\nu_1}&2\hat y_{\nu_2}\cosh 2\omega &4\hat y_{\nu_2}^3\\
0&2(\hat y_{\nu_1}+\hat y_{\nu_2})\sinh2\omega &0\\
\end{array}\right),
\end{align}with determinant
\begin{equation}
\det{J_\nu}=8(\hat y_{\nu_1}^2+\hat y_{\nu_2}^2)^2(\hat y_{\nu_1}^2-\hat y_{\nu_2}^2)\sinh2\omega\, ,
\end{equation}
and for charged leptons
\begin{equation}
\det J_E= 2\left(y_e^2-y_\mu^2\right)\,.
\end{equation}
In this case the only singular point at this level is case 2.) of Eq~\ref{2SP2GL}.
\end{itemize}


\subsection{The Potential at the Renormalizable Level}
In this section the study of the renormalizable potential will
reveal that all possible vacua retain some unbroken symmetry and in turn
correspond to some of the boundaries.
The allowed boundaries at the absolute minimum are however not every possible one
and furthermore some of the configurations found in the study of the potential are boundaries veiled
in the previous Jacobian analysis due to the parametrization.
It is the case here, as for quarks, that not only singular points are allowed at the minimum,
such that certain flavour parameters can me adjusted in terms of the parameters of the potential.
This section will treat by default of the \emph{axial preserving case}, unless stated otherwise.

At the renormalizable level the most general potential respecting $\mathcal{G}_\mathcal{F}^l$ is
\begin{align}
V=&-\mu^2\cdot\mathbf{X}+\mathbf{X}^T\cdot\lambda\cdot\mathbf{X}
 +h_E\,\mbox{Tr}\left(\mathcal{Y}_E\mathcal{Y}_E^\dagger\mathcal{Y}_E\mathcal{Y}_E^\dagger\right)+
g\,\mbox{Tr}\left(\mathcal{Y}_E\mathcal{Y}_E^\dagger\mathcal{Y}_\nu\mathcal{Y}_\nu^\dagger\right)
\label{potential}\\
&+h_\nu\, \mbox{Tr}\left(\mathcal{Y}_\nu\mathcal{Y}_\nu^\dagger\mathcal{Y}_\nu\mathcal{Y}_\nu^\dagger\right)+
h_\nu'\, \mbox{Tr}\left(\mathcal{Y}_\nu\mathcal{Y}_\nu^T\mathcal{Y}_\nu^*\mathcal{Y}_\nu^\dagger\right)\,.\nonumber
\end{align}
In this equation $\mathbf{X}$ is a two-component vector defined by
\begin{equation}
 \mathbf{X}\equiv\left(
                     \mbox{Tr}\left(\mathcal{Y}_E\mathcal{Y}_E^\dagger\right)\,,
					\mbox{Tr}\left( \mathcal{Y}_\nu^\dagger \mathcal{Y}_\nu\right)
					\right)^T\,,
\nonumber
\end{equation}
$\mu^2$ is a real two-component vector, $\lambda$ is a $2\times 2$ real and symmetric matrix and all other coefficients are real: a total of 9 parameters, one more than in the quark case since the
new invariant $I_{\nu'}$ is allowed by the symmetry. The full scalar potential includes in addition  Higgs-$\mathcal{Y}_E$ and Higgs-$\mathcal{Y}_\nu$ cross-terms, but they do not affect the flavour pattern and will thus be obviated in what follows.

Consider  first minimization in the mixing parameters. Since  mixing  arises from the misalignment in flavour space of the charged lepton and the neutrino flavour scalar fields, the only relevant invariant at the renormalisable level  is $I_{\nu,E}$. The explicit dependence of $I_{\nu,E}$ on mixing parameters is shown in \ref{ILMIX} and
we reproduce it here,
\begin{align}\nonumber\mbox{Tr}\left(\mathcal{Y}_E\mathcal{Y}_E^\dagger\mathcal{Y}_\nu\mathcal{Y}_\nu^\dagger\right)=&\Lambda_f^4 [\left(y_\mu^2-y_e^2\right)\left(\hat y_{\nu_1}^2-\hat y_{\nu_2}^2\right)\cos2\theta \cosh 2\omega+        \left(y_e^2+y_\mu^2\right)\left(\hat y_{\nu_1}^2+\hat y_{\nu_2}^2\right)\\ 
  &+2\left(y_\mu^2-y_e^2\right)\hat y_{\nu_1}\hat y_{\nu_2} \sin2\alpha \sin2\theta \sinh2\omega]/2\,,
  \label{genericmixing}
\end{align}
for comparison with the quark case analogous,
\begin{align}\mbox{Tr}\left(\mathcal{Y}_D\mathcal{Y}_D^\dagger\mathcal{Y}_U\mathcal{Y}_U^\dagger\right)=\Lambda_f^4 \left[\left(y_c^2-y_u^2\right)\left(y_s^2-y_d^2\right)\cos2\theta+
        \left(y_c^2+y_u^2\right)\left(y_s^2+y_d^2\right)\right]/2\,.
        \label{quarkmixing}
\end{align}
The first term in Eq.~\ref{genericmixing} for leptons corresponds to that for quarks in Eq.~\ref{quarkmixing}. The second line in Eq.~\ref{genericmixing} has a strong impact on the localisation of the minimum of the potential and is responsible for the different results in the quark and lepton sectors. In particular, it contains the Majorana phase $\alpha$ and therefore \emph{connects the Majorana nature of neutrinos to their mixing}.

Eq.~\ref{genericmixing} also shows explicitly the relations expected on physical grounds between the mass spectrum and non-trivial mixing: i) the dependence on the mixing angle disappears in the limit of degenerate charged lepton masses; ii) it also vanishes for degenerate neutrino masses if and only if $\sin2\alpha=0$; iii) on the contrary, for  $\sin2\alpha\ne 0$ the dependence on the mixing angle remains, as it is physical even for degenerate neutrino masses;  iv) the $\alpha$ dependence vanishes when one of the two neutrino masses vanishes or in the absence of mixing, as $\alpha$ becomes then unphysical.

The minimization with respect to the Majorana phase and the mixing angle leads to the constraints:
\begin{gather}
\sinh 2\omega\sqrt{m_{\nu_2}m_{\nu_1}}\sin2\theta\cos2\alpha=0\,,
\label{cos2alpha}\\
\mbox{tg}2\theta=\sin2\alpha \tanh 2\omega
\frac{2\sqrt{m_{\nu_2}m_{\nu_1}}}{m_{\nu_2}-m_{\nu_1}}\,,
\label{tan2theta}
\end{gather}
where we have restored neutrino masses explicitly.
The first condition predicts that the {\it Majorana phase is maximal, $\alpha=\{\pi/4,3\pi/4\}$, for non-trivial mixing angle}. The relative Majorana phase between the two neutrinos is therefore $2\alpha=\pm\pi/2$ which implies no CP violation due to Majorana phases. On the other hand, Eq.~\ref{tan2theta} establishes  a link between the mixing strength and the type of spectrum, which indicates {\it a maximal angle for degenerate neutrino masses, and  a small angle for strong mass hierarchy.} In \cite{onpreparation2} this equation is generalized
to the generic type I seesaw.

Using the Von Neumann trace inequality we have that the previous result corresponds to the configurations
in which the eigenvalues of  $\mathcal{Y}_E\mathcal{Y}_E^\dagger$ and  $\mathcal{Y}_\nu\mathcal{Y}_\nu^\dagger$, 
are coupled in direct or inverse order:
\begin{equation}
\begin{cases}
I_{\nu,E}\Big|_{\left\langle\theta\right\rangle,\left\langle\alpha\right\rangle}\propto m_e^2\, m_+ +m_\mu^2\,m_-\,,&\quad g>0\,,\\[2mm]
I_{\nu,E}\Big|_{\left\langle\theta\right\rangle,\left\langle\alpha\right\rangle}\propto m_e^2\, m_- +m_\mu^2\,m_+\,,&\quad g<0\,,
\end{cases}
\label{trace-minimum}
\end{equation}
where the eigenvalues of  $\mathcal{Y}_\nu\mathcal{Y}_\nu^\dagger$ are,
\begin{align}
&m_\pm \equiv a_\nu\pm\sqrt{a_\nu^2-c_\nu^2}\,,\label{eq:Defmpm}   \\
&a_\nu=(m_{\nu_2}+m_{\nu_1})\cosh 2\omega\,,\,\, c_\nu=2\sqrt{m_{\nu_2}m_{\nu_1}}\,.\nonumber
\end{align}
This two family scenario resulted in a remarkable connection of mass degeneracy
and large angles, for an attempt at a realistic case nonetheless the three family case shall be studied
as is done in Sec. \ref{3FL}.


The minimization of the rest of the potential will fix masses and $\omega$ but, even if being
an involved process, it yields simple results and very constrained patterns. In particular there
are two types of solutions, a class with $\omega=0$ which through Eq.~\ref{tan2theta} results in vanishing
mixing analogously to the quark case and a second type with non-vanishing
$\omega$ and necessarily degenerate neutrino masses $\hat y_{\nu_1}=\hat y_{\nu_2}\equiv \hat y_\nu$. The latter case
 corresponds, through Eq. \ref{tan2theta}, to maximal mixing and Majorana phase ($\theta=\pi/4$, $\alpha=\pi/4$) and
 the neutrino flavour field has the structure\footnote{Up to an overall and unphysical complex phase.}:
\begin{equation}
\mathcal{Y}_{\nu}=\Lambda_f
\left(\begin{array}{cc}
y_{\nu_1}(\omega, \hat y_\nu)&0\\
0&y_{\nu_2}(\omega, \hat y_\nu)\\
\end{array}\right)
\frac{1}{\sqrt{2}}\left(\begin{array}{cc}
1&i\\
-1&i\\
\end{array}\right)
\label{YukMaxAngle}
\end{equation}
which leads to the neutrino mass matrix,
\begin{equation}
U\,\mathbf{m}_\nu \,U^T=\frac{v^2}{2M} \left(\begin{array}{cc}
0&y_{\nu_1} y_{\nu_2}\\
y_{\nu_2} y_{\nu_1}&0\\
\end{array}\right),
\end{equation}
where the degeneracy and maximal angle become evident when diagonalizing.
It is important to note that, even if the neutrino states have the same absolute mass
in this configuration, the maximal Majorana phase still allows for distinction among them
and therefore a meaningful physical angle.

The structure in \ref{YukMaxAngle} reminds of the bi-unitary parametrization and
indeed $y_{\nu_{1,2}}$ are the parameters of Eq.~\ref{BI-U} and the right-hand side
matrix can be associated with a maximal angle and complex $\mathcal U_R$. This is pointing
towards the bi-unitary as a better suited parametrization for minimization; a fact that will be made explicit
and exploited in the three family case.

Before we discuss the possible vacua, let us pause for
examining more closely \ref{YukMaxAngle}. Is there
something special about such a configuration? There is,
it leaves certain symmetry unbroken. For determining it
we perform a transformation of $O(2)_{N_R}$:
\begin{align}
\mathcal{Y}_{\nu} \xrightarrow{O(2)} \mathcal{Y}_{\nu}\, e^{i\sigma_2 \varphi}=&\left(\begin{array}{cc}
 \frac{y_{\nu_1}}{\sqrt{2}}& \frac{iy_{\nu_1}}{\sqrt{2}}\\[2mm]
- \frac{y_{\nu_2}}{\sqrt{2}}& \frac{iy_{\nu_2}}{\sqrt{2}}\\
\end{array}\right)
\left(\begin{array}{cc}
\cos\varphi&\sin\varphi  \\
-\sin\varphi&\cos\varphi \\
\end{array}\right)\\=&
\left(\begin{array}{cc}
e^{-i\varphi}&0\\
0&e^{i\varphi}\\
\end{array}\right)\left(\begin{array}{cc}
 \frac{y_{\nu_1}}{\sqrt{2}}& \frac{iy_{\nu_1}}{\sqrt{2}}\\[2mm]
- \frac{y_{\nu_2}}{\sqrt{2}}& \frac{iy_{\nu_2}}{\sqrt{2}}\\
\end{array}\right).
\end{align}
It is clear now that a simultaneous rotation of the
left handed group $SU(2)_{\ell_L}$ generated by
$\sigma_3$ can compensate the complex phases on the left. Therefore an 
unbroken $U(1)$
is present which in the following is labeled $SO(2)_V$
since it would be the equivalent of $SU(2)_V$ in the 
quark case. 

The minimization process is however still incomplete.
The allowed ratios of eigenvalues both in the charged lepton
and neutrino sectors in the absolute minimum are constrained like in the
quark case. 

Selecting among the different possibilities the one resembling the closest the observed
flavour pattern one finds,
\begin{equation}
\mathcal{Y}_E=\Lambda_f \left(\begin{array}{cc}
						0&0\\
						0&y_\mu\\
						\end{array}\right),\qquad\qquad
\mathcal{Y}_\nu =\Lambda_f \left(\begin{array}{cc}
 \frac{y_{\nu_1}}{\sqrt{2}}& \frac{iy_{\nu_1}}{\sqrt{2}}\\[2mm]
- \frac{y_{\nu_2}}{\sqrt{2}}& \frac{iy_{\nu_2}}{\sqrt{2}}\\
\end{array}\right),
\end{equation}
with a breaking pattern $\mathcal{G}_\mathcal{F}^l\rightarrow U(1)_{e_R}\times SO(2)_V$.
{\it In this scenario the electron is massless and the two neutrinos have the
same absolute value for the mass while the mixing angle is maximal $\theta=\pi/4$ in a tantalizing 
first approximation to the lepton flavour pattern}. Let us stress here that the same framework for the
quark sector led to hierarchies and vanishing mixing.

All possible vacua are listed in what follows for completeness:

\begin{itemize}
\item[\bf I] This hierarchical solution sets the electron massless and forbids Majorana
masses for the light neutrinos,
\begin{equation}
\mathcal{Y}_E=\Lambda_f \left(\begin{array}{cc}
						0&0\\
						0&y_\mu\\
						\end{array}\right),\qquad\qquad
\mathcal{Y}_\nu =\Lambda_f \left(\begin{array}{cc}
						0&0\\
- \frac{y_{\nu_2}}{\sqrt{2}}& \frac{iy_{\nu_2}}{\sqrt{2}}\\
						\end{array}\right),\label{ISP2GL}
\end{equation}

since the breaking pattern is $\mathcal{G}_\mathcal{F}^l \rightarrow U(1)_{LN} \times U(1)_e \times U(1)_{A}$. Even
if there is no Majorana mass for the neutrinos, the muon neutrino mixes with the heavy right handed and produces
flavour effects.
The spectrum has then a massless neutrino, which is mostly active and a heavy sterile Dirac neutrino.

\item[\bf II]  The two leptons have a mass and the neutrino sector has a single massive Dirac fermion,
\begin{equation}
\mathcal{Y}_E=\Lambda_f \left(\begin{array}{cc}
						y_e&0\\
						0&y_\mu\\
						\end{array}\right),\qquad\qquad
\mathcal{Y}_\nu =\Lambda_f \left(\begin{array}{cc}
						0&0\\
- \frac{y_{\nu_2}}{\sqrt{2}}& \frac{iy_{\nu_2}}{\sqrt{2}}\\
						\end{array}\right),
\end{equation}
satisfying
\begin{equation}
\frac{y_{\mu}^2-y_{e}^2}{y_{\mu}^2+y_{e}^2}=\frac{-g}{2 h_E}\frac{I_\nu}{I_E}\, ,
\end{equation}
the unbroken symmetry is $U(1)_e \times U(1)_{LN}$.

\item[\bf III] 
This case yields  a massless electron and two light degenerate Majorana neutrinos,
\begin{equation}
\mathcal{Y}_E=\Lambda_f \left(\begin{array}{cc}
						0&0\\
						0&y_\mu\\
						\end{array}\right),\qquad\qquad
\mathcal{Y}_\nu =\Lambda_f \left(\begin{array}{cc}
 \frac{y_{\nu_1}}{\sqrt{2}}& \frac{iy_{\nu_1}}{\sqrt{2}}\\[2mm]
- \frac{y_{\nu_2}}{\sqrt{2}}& \frac{iy_{\nu_2}}{\sqrt{2}}\\
\end{array}\right),
\end{equation}
with the relation:
\begin{equation}
\frac{y_{\nu_2}^2-y_{\nu_1}^2}{y_{\nu_2}^2+y_{\nu_1}^2}=\frac{-g}{2 (h_\nu -|h'_\nu |)}\frac{I_E}{I_\nu}\,,
\end{equation}
and the symmetry pattern; $\mathcal{G}_\mathcal{F}^l\rightarrow U(1)_{e_R}\times SO(2)_V$.

\item[\bf IV] The degenerate case corresponds to 
a configuration of the Yukawas of the type
\begin{equation}
\mathcal{Y}_E=\Lambda_f \left(\begin{array}{cc}
						y&0\\
						0&y\\
						\end{array}\right),\qquad\qquad
\mathcal{Y}_\nu =\Lambda_f\, y_\nu\left(\begin{array}{cc}
 \frac{1}{\sqrt{2}}& \frac{i}{\sqrt{2}}\\[2mm]
- \frac{1}{\sqrt{2}}& \frac{i}{\sqrt{2}}\\
\end{array}\right),
\end{equation}
which preserves $SO(2)_V$.
\end{itemize}

At this point contrast with the Jacobian analysis reveals that not only it missed
certain singular points but that these have a larger symmetry: configuration {\bf I} (Eq.~\ref{ISP2GL})
has a symmetry that both contains and extends that of case 1 in Eq.~\ref{1SP2GL}. This
problem is not related to the Jacobian method, it has to do with the parametrization; indeed case {\bf I}
corresponds to $\hat y_{\nu_1}=\hat y_{\nu_2}$ and $\omega\rightarrow \infty $ while
keeping $\hat y_{\nu_2}\cosh \omega$ constant. This is certainly an unintuitive limit to take in the Casas-Ibarra
parametrization, but a much evident one in for example the bi-unitary parametrization, to which
we restrict in the following. 

It is worth noticing a difference with quarks, case {\bf III} contains a symmetry
that is larger than that of case {\bf IV}, whereas the quark case is the opposite (Eqs.~\ref{III-2F},\ref{IV-2F}).
From this point of view, when having a maximal angle and degenerate Majorana masses, the
most ``natural" case is a hierarchical charged lepton spectrum.

From this set of possible minima we learn that all the vacua found at the renormalizable 
level have an unbroken symmetry. 
Like in the quark case the introduction of determinants will disrupt those
configurations that have a chiral $U(1)_A$. This fact can be
used to lift the zero eigenvalues through a small determinant coefficient
like in the quark case.

Finally, we remark that all cases with nontrivial mixing result in  
\emph{sharp predictions: a maximal mixing angle and degenerate neutrinos with a $\pi/2$ relative Majorana phase}.

%
%
\section{Three Family case}\label{3FL}
%

The scalar fields are taken to be bi-triplets as detailed in table \ref{BIF-SFL} and are connected proportionally
to Yukawas as seen in Eq.~\ref{BiFConnection}.

The number of parameters that suffice to describe the scalar fields modulo the
flavour symmetry $\mathcal{G}_\mathcal{F}^l$ is discussed next\footnote{A way to determine the number is
to subtract the dimension of the group (dim$(\mathcal{G}_\mathcal{F}^l)$=21) from the number of degrees of freedom of the fields ($2\times 18$).}. Starting as in the 2 family case from diagonal $\mathcal{Y}_E$, then
 $\mathcal{Y}_\nu$ is a complex matrix with a
priori 18 parameters. A $O(3)_{N_R}$ rotation can eliminate 3 of these, and there
is still the residual symmetry of complex phase redefinitions to absorb 3 complex phases,
 leaving 12 parameters in $\mathcal{Y}_\nu$~\cite{Broncano:2002rw}. These parameters in the low energy Lagrangian can be encoded in
3 masses for the light neutrinos, 3 mixing angles and 3 complex phases in $U_{PMNS}$ extractable from
oscillation data, double beta decay and tritium decay
and the 3 remaining parameters control the three charged flavour violating processes ($\mu-e$, $\tau-\mu$, $\tau-e$). These
last three parameters can be taken to be imaginary angles in the Casas-Ibarra $R$-matrix.

The parametrization better suited for minimization nonetheless  is the bi-unitary parametrization of Eq.~\ref{BI-U}, where now
$\mathbf{y}_\nu\equiv\mbox{Diag}(y_{\nu_1},y_{\nu_2},y_{\nu_3})$.
The parameters in \ref{BI-U} for three families are distributed as follows;  4 in the CKM-like matrix $\mathcal U_L$, 3 in $\mathcal U_R$, the three moduli of the eigenvalues in $\mathbf{y}_\nu$ and two relative complex phases of these eigenvalues.

The list of 15 invariants can be split in three groups. The first one comprises the 6 invariants,
\begin{align}
I_{E}&=\mbox{Tr}\left[\mathcal{Y}_E \mathcal{Y}_E^\dagger \right]\, , & I_{\nu}&= \mbox{Tr}\left[\mathcal{Y}_\nu \mathcal{Y}_\nu^\dagger   \right]\, ,\\
I_{E^2}&=\mbox{Tr}\left[\left(\mathcal{Y}_E\mathcal{Y}_E^\dagger\right)^2\right]\, ,& I_{\nu^2}&= \mbox{Tr}\left[\left(\mathcal{Y}_\nu \mathcal{Y}_\nu^\dagger \right)^2  \right]\, ,\\
I_{E^3}&=\mbox{Tr}\left[\left(\mathcal{Y}_E\mathcal{Y}_E^\dagger\right)^3\right]\, ,& I_{\nu^3}&= \mbox{Tr}\left[\left(\mathcal{Y}_\nu \mathcal{Y}_\nu^\dagger \right)^3  \right]\, ,\label{MELAP}
\end{align}
which depend on eigenvalues only. The following 7 correspond to the second group,
\begin{align}
 I_{L}&=\mbox{Tr}\left[\mathcal{Y}_\nu\mathcal{Y}_\nu^\dagger \mathcal{Y}_E\mathcal{Y}_E^\dagger \right]\, ,  & I_{R}&=\mbox{Tr}\left[\mathcal{Y}_\nu^\dagger \mathcal{Y}_\nu\mathcal{Y}_\nu^T \mathcal{Y}_\nu^* \right] \, ,\\
 I_{L^2}&=\mbox{Tr}\left[\mathcal{Y}_\nu \mathcal{Y}_\nu^\dagger \left(\mathcal{Y}_E \mathcal{Y}_E^\dagger\right)^2\right]\, ,& I_{R^2} &=\mbox{Tr}\left[\left(\mathcal{Y}_\nu^\dagger\mathcal{Y}_\nu\right)^2\mathcal{Y}_\nu^T \mathcal{Y}_\nu^* \right]\, ,\\
 I_{L^3}&=\mbox{Tr}\left[\mathcal{Y}_E \mathcal{Y}_E^\dagger \left(\mathcal{Y}_\nu \mathcal{Y}_\nu^\dagger\right)^2\right] \, ,&I_{R^3}&=\mbox{Tr}\left[\left(\mathcal{Y}_\nu^\dagger \mathcal{Y}_\nu\mathcal{Y}_\nu^T \mathcal{Y}_\nu^*\right)^2 \right]\, ,\\
 I_{L^4}&=\mbox{Tr}\left[\left(\mathcal{Y}_\nu \mathcal{Y}_\nu^\dagger \mathcal{Y}_E \mathcal{Y}_E^\dagger\right)^2 \right]\, ,
\end{align}
 and depend on $\mathcal U_L$ and $\mathcal U_R\mathcal U_R^T$ only respectively. The quark analysis for CKM of Sec. \ref{Jac3GQ} goes through the same for these
terms (with the subtlety of considering three elements of $\mathcal U_R\mathcal U_R^T$, as $(\mathcal U_R\mathcal U_R^T)_{ij}=(\mathcal U_R\mathcal U_R^T)_{ji}$) as will be shown next.
Finally the two invariants that will fix the relative complex phases in $\mathbf{y}_\nu$ are
\begin{align}
I_{LR}&=\mbox{Tr}\left[\mathcal{Y}_\nu \mathcal{Y}_\nu^T \mathcal{Y}_\nu^* \mathcal{Y}_\nu^\dagger \mathcal{Y}_E \mathcal{Y}_E^\dagger \right]\, ,& I_{RL}&=\mbox{Tr}\left[\mathcal{Y}_\nu \mathcal{Y}_\nu^T\mathcal{Y}_E^* \mathcal{Y}_E^T  \mathcal{Y}_\nu^* \mathcal{Y}_\nu^\dagger \mathcal{Y}_E \mathcal{Y}_E^\dagger \right]\,,
\end{align}
which completes the list of independent $\mathcal{G}_\mathcal{F}^l$ invariants.
In the axial breaking case two dimension 3 invariants are allowed:
\begin{align}
I_{\tilde E}&=\det\left(\mathcal{Y}_E\right) \, ,& I_{\tilde\nu}&= \det\left(\mathcal{Y}_\nu \right)\, ,
\end{align}
which substitute those of Eq.~\ref{MELAP}, as in the quark case (Eq.~\ref{CHthrm3}).

\subsection{The Jacobian}\label{Jac3FL}

The number of variables and invariants has scaled up to 15, in this sense the Casas-Ibarra parametrization
becomes hard to handle, specially due to the orthogonal matrix. In the context of the bi-unitary parametrization
though we can make use of the previously derived Jacobians in the quark sector. In particular, the unitary relations we employed
for finding the mixing sub-Jacobian $J_{UD}$ in Eq. \ref{JUD} hold for both $\mathcal U_L$ and $\mathcal U_R\mathcal U_R^T$ \cite{onpreparation,onpreparation2}. In this parametrization the structure of the
Jacobian reads:
\begin{gather}
J=\left(\begin{array}{ccccc}
\partial_{\mathbf{y}_E} I_{E^n} & 0&0&\partial_{\mathbf{y}_E} I_{L^n}&\partial_{\mathbf{y}_E} I_{LR}\\
0& \partial_{\mathbf{y}_\nu} I_{\nu^n}&\partial_{\mathbf{y}_\nu} I_{R^n}&\partial_{\mathbf{y}_\nu} I_{L^n}&\partial_{\mathbf{y}_\nu} I_{LR}\\
0&0&\partial_{\mathcal U_R} I_{R^n}& 0&\partial_{\mathcal U_R} I_{LR}\\
0&0&0&\partial_{\mathcal U_L} I_{L^n}& \partial_{\mathcal U_L} I_{LR}  \\
0&0&0&0&\partial_{\mathcal U_L\mathcal U_R} I_{LR}\\
\end{array}\right),\\
\mbox{Diag}(J)\equiv \left(J_E\,,\,J_\nu\,,\, J_{\mathcal U_R}\, ,\, J_{\mathcal U_L}\,,\,J_{LR}\right)
\end{gather}
From the above shape the calculation of the $15\times15$ determinant is reduced
to the product of 5 subdeterminants, those of the diagonal. 

For $J_{\mathcal U_L}$ the calculation of the determinant is just like that of quarks:
\begin{align}
\det\left(J_{\mathcal U_L}\right)= &\left(y_{\nu_1}^2-y_{\nu_2}^2\right)
\left(y_{\nu_2}^2-y_{\nu_3}^2\right)\left(y_{\nu_3}^2-y_{\nu_1}^2\right)\left(y_e^2-y_\mu^2\right)\left(y_\mu^2-y_\tau^2\right)\left(y_\tau^2-y_e^2\right)  \nonumber  \\& |\mathcal U_L^{e1}||\mathcal U_L^{e2}||\mathcal U_L^{\mu 1}||\mathcal U_L^{\mu 2}|\, .
\end{align}
The dependence on $\mathcal U_R$ of the $I_{R^n}$ invariants looks like,
\begin{align}
I_R=&\mbox{Tr}\left(\mathbf{y}_{\nu}^2 \mathcal U_R \mathcal U_R^T \mathbf{y}_{\nu}^2 \mathcal U_R^* \mathcal U_R^\dagger\right)\, , &
I_{R^2}=&\mbox{Tr}\left(\mathbf{y}_{\nu}^4 \mathcal U_R \mathcal U_R^T \mathbf{y}_{\nu}^2 \mathcal U_R^* \mathcal U_R^\dagger\right)\, ,\\
I_{R^3}=&\mbox{Tr}\left(\mathbf{y}_{\nu}^4 \mathcal U_R \mathcal U_R^T \mathbf{y}_{\nu}^4 \mathcal U_R^* \mathcal U_R^\dagger\right)\, ,
\end{align}
and the Jacobian:
\begin{equation}J_{\mathcal U_R}\propto
\left(\begin{array}{ccc}
1&y_{\nu_1}^2+y_{\nu_3}^2 &\left(y_{\nu_1}^2+y_{\nu_3}^2\right)^2 \\
1&y_{\nu_2}^2+y_{\nu_3}^2 &\left(y_{\nu_1}^2-y_{\nu_3}^2\right)^2\\
2&y_{\nu_1}^2+y_{\nu_2}^2+2y_{\nu_3}^3 &2\left(y_{\nu_1}^2+y_{\nu_3}^2\right)\left(y_{\nu_2}^2+y_{\nu_3}^2\right)
\end{array}\right),
\end{equation}
where the proportionality is different for each row and equal to $\left(y_{\nu_1}^2-y_{\nu_3}^2\right)^2$,
$\left(y_{\nu_2}^2-y_{\nu_3}^2\right)^2$ and $\left(y_{\nu_1}^2-y_{\nu_3}^2\right)\left(y_{\nu_2}^2-y_{\nu_3}^2\right)^2$
respectively. The determinant is,
\begin{align}\nonumber
\det J_{\mathcal U_R}=&\left(y_{\nu_1}^2-y_{\nu_2}^2\right)^3\left(y_{\nu_2}^2-y_{\nu_3}^2\right)^3 \left(y_{\nu_3}^2-y_{\nu_1}^2\right)^3\\
&\times|\left(\mathcal U_R \mathcal U^T_R\right)_{11}||\left(\mathcal U_R \mathcal U^T_R\right)_{22}||\left(\mathcal U_R \mathcal U^T_R\right)_{12}|
\end{align}
This means that \emph{the rank is reduced the most for $\mathcal U_{R}\mathcal U_R^T$ being a permutation matrix}, in this
sense $\mathcal U_R$ is the ``square root" of a permutation matrix \emph{such that}, if the permutation matrix
is other than the trivial identity, the matrix $\mathcal U_R$ \emph{contains a maximal angle}. Here is where, technically,
the main difference with the quark case stems. The solution of Eq.~\ref{YukMaxAngle}
in the two family case is indeed that of an $\mathcal U_R$ which is a ``square root" of a permutation, 
such that the Jacobian method allows the identification of all singular points if the bi-unitary parametrization
is employed. 

Next the two invariants $I_{LR,RL}$ read,  in terms of the bi-unitary parametrization,
\begin{align}
I_{LR}=&\mbox{Tr}\left(\mathbf{y}_{\nu} \mathcal U_R \mathcal U_R^T \mathbf{y}_{\nu}^2 \mathcal U_R^* \mathcal U_R^\dagger \mathbf{y}_\nu \mathcal U_L^\dagger
\mathbf{y}_e^2\mathcal U_L\right)\, , \\
I_{RL}=&\mbox{Tr}\left(\mathbf{y}_{\nu} \mathcal U_R \mathcal U_R^T \mathbf{y}_{\nu}\mathcal U_L^T\mathbf{y}_E^2\mathcal U_L^* \mathbf{y_\nu} \mathcal U_R^* \mathcal U_R^\dagger \mathbf{y}_\nu \mathcal U_L^\dagger
\mathbf{y}_e^2\mathcal U_L\right) .
\end{align}
Let's parametrize the two complex phases left as
\begin{equation}
\mathcal U_R\rightarrow   e^{i\alpha_3 \lambda_3} e^{i\alpha_8\lambda_8}\mathcal U_R\,,
\end{equation}
where $\lambda_{3,8}$ are the diagonal Gell-Mann matrices. The Jacobian is built with the four terms:
\begin{align}
\frac{\partial I_{LR}}{\alpha_3}=&i\mbox{Tr}\left(\left[\lambda_3\,\,,\,\, \mathbf{y}_{\nu} \mathcal U_R \mathcal U_R^T \mathbf{y}_{\nu}^2 \mathcal U_R^* \mathcal U_R^\dagger \mathbf{y}_\nu\right] \mathcal U_L^\dagger\mathbf{y}_e^2\mathcal U_L\right)\,, \\
\frac{\partial I_{LR}}{\alpha_8}=&i\mbox{Tr}\left(\left[\lambda_8\,\,,\,\, \mathbf{y}_{\nu} \mathcal U_R \mathcal U_R^T \mathbf{y}_{\nu}^2 \mathcal U_R^* \mathcal U_R^\dagger \mathbf{y}_\nu\right] \mathcal U_L^\dagger\mathbf{y}_e^2\mathcal U_L\right)\,, \\
\frac{\partial I_{RL}}{\alpha_3}=&2i\mbox{Tr}\left(\left[\lambda_3\,\,,\,\, \mathcal U_L^T\mathbf{y}_E^2 \mathcal U_L^*\mathbf{y}_{\nu} \mathcal U_R^* \mathcal U_R^\dagger \mathbf{y}_{\nu}\right] \mathcal U_L^\dagger\mathbf{y}_e^2\mathcal U_L\mathbf{y}_{\nu} \mathcal U_R \mathcal U_R^T \mathbf{y}_{\nu}\right)\,, \\
\frac{\partial I_{RL}}{\alpha_8}=&2i\mbox{Tr}\left(\left[\lambda_8\,\,,\,\, \mathcal U_L^T\mathbf{y}_E^2 \mathcal U_L^*\mathbf{y}_{\nu} \mathcal U_R^* \mathcal U_R^\dagger \mathbf{y}_{\nu}\right] \mathcal U_L^\dagger\mathbf{y}_e^2\mathcal U_L\mathbf{y}_{\nu} \mathcal U_R \mathcal U_R^T \mathbf{y}_{\nu}\right)\,,
\end{align}
and the determinant of this part:
\begin{equation}
J_{LR}=\frac{\partial I_{LR}}{\alpha_8}\frac{\partial I_{RL}}{\alpha_3}-\frac{\partial I_{LR}}{\alpha_3}\frac{\partial I_{RL}}{\alpha_8}
\end{equation}
which vanishes if $ \mathbf{y}_{\nu} \mathcal U_R \mathcal U_R^T \mathbf{y}_{\nu}$, $\mathcal U_L^\dagger\mathbf{y}_e^2\mathcal U_L$ or their
product is diagonal.

The computation of $J_\nu$ and $J_E$ follows the line of
the quark case, 
\begin{itemize}
\item {\bf Axial preserving scenario: $\mathcal{G}_\mathcal{F}^l\sim U(3)^2\times O(3)$ -} The determinants are,
\begin{align}
 \det J_E=&48\,y_{e}y_\mu y_\tau (y_e^2-y_\mu^2)(y_\mu^2-y_\tau^2)(y_\tau^2-y_e^2)\, ,\\
 \det J_\nu=&48\,y_{\nu_1} y_{\nu_2}y_{\nu_3} (y_{\nu_1}^2-y_{\nu_2}^2)(y_{\nu_2}^2-y_{\nu_3}^2)(y_{\nu_3}^2-y_{\nu_1}^2)\, .
\end{align}
\item {\bf Axial breaking scenario: $\mathcal{G}_\mathcal{F}^{\slashed A\, , l}\sim SU(3)^2\times SO(3)$ - } The determinants are, 
\begin{align}
 \det J_E= &8(y_e^2-y_\mu^2)(y_\mu^2-y_\tau^2)(y_\tau^2-y_e^2)\, ,\\
 \det J_\nu= &8(y_{\nu_1}^2-y_{\nu_2}^2)(y_{\nu_2}^2-y_{\nu_3}^2)(y_{\nu_3}^2-y_{\nu_1}^2)\,.
\end{align}
\end{itemize}
The number of boundaries is large in this three family case since there is a variety of
ways to cancel the total determinant $\det J$, let us note simply that the
singular points correspond to either completely degenerate or hierarchical charged lepton and
neutrino spectrum
and $\mathcal U_L$ and $\mathcal U_R\mathcal U_R^T$ corresponding to permutation matrices.
\subsection{The Potential at the Renormalizable Level}\label{potL3g}
The number of boundaries or subgroups of the flavour group has grown sensibly
complicating the Jacobian analysis, the study of the potential will help clarify which
configurations are realized and how at the renormalizable level.

The potential including all possible terms respecting the full flavour group looks just like the
two family case Eq \ref{potential}
and the counting of potential parameters goes
like the same: they add up to 9.
We shall examine next the way in which this potential will fix the
vev of the scalar fields. For the same reason as in the previous
chapter the minimization process will start on those variables that
appear less often in the potential. In this case they are the
parameters of the unitary matrices, which will in turn determine
$U_{PMNS}$.

The left handed matrix $\mathcal U_L$ appears in the term:
\begin{equation}
g\mbox{Tr}\left(\mathcal{Y}_E\mathcal{Y}_E^\dagger\mathcal{Y}_\nu\mathcal{Y}_\nu^\dagger\right)
=g\Lambda_f^4\mbox{Tr}\left(\mathbf{y}_E^2\mathcal U_L\mathbf{y}_\nu^2\mathcal U_L^\dagger\right)\, ,
\end{equation}
through the Von Neumann trace inequality two possible minima are identified,
\begin{align}
g&<0, &\mathcal U_L=&\left(
\begin{array}{ccc}
 1 & 0 & 0 \\
 0 & 1& 0 \\
 0 & 0 & 1
\end{array}
\right); &g\Lambda_f^4\mbox{Tr}\left(\mathbf{y}_E^2\mathcal U_L\mathbf{y}_\nu^2\mathcal U_L^\dagger\right)&\rightarrow g\Lambda_f^4\sum y_{i}^2y_{\nu_i}^2 \,,\\
g&>0,& \mathcal U_L=&\left(
\begin{array}{ccc}
 0 & 0 & 1 \\
 0 & 1& 0 \\
 1 & 0 & 0
\end{array}
\right);&g\Lambda_f^4\mbox{Tr}\left(\mathbf{y}_E^2\mathcal U_L\mathbf{y}_\nu^2\mathcal U_L^\dagger\right)&\rightarrow g\Lambda_f^4\sum y_{i}^2y_{\nu_{4-i}}^2 \,.
\end{align}
Under the same reasoning, $\mathcal U_R$ appears only in:
\begin{equation}
h_\nu'\, \mbox{Tr}\left(\mathcal{Y}_\nu\mathcal{Y}_\nu^T\mathcal{Y}_\nu^*\mathcal{Y}_\nu^\dagger\right)
=h_\nu'\, \mbox{Tr}\left(\mathbf{y}_\nu^2\mathcal U_R\mathcal U_R^T\mathbf{y}_\nu^2 \mathcal U_R^*\mathcal U_R^\dagger\right)
\end{equation}
and takes one of the two discrete possible values at the minimum:
\begin{itemize}
\item[A)] for a negative coefficient,
\begin{align}
h_\nu'&<0\,, &\mathcal U_R\mathcal U_R^T=&\left(
\begin{array}{ccc}
 1 & 0 & 0 \\
 0 & 1& 0 \\
 0 & 0 & 1
\end{array}
\right); &h_\nu'\mbox{Tr}\left(\mathbf{y}_\nu^2\mathcal U_R\mathcal U_R^T\mathbf{y}_\nu^2 \mathcal U_R^*\mathcal U_R^\dagger\right)& \rightarrow h_\nu'\sum_{i=1}^3 y_{\nu_i}^4 \,.\label{A}
\end{align}
\item[B)] for a positive coefficient,
\begin{align}
h_\nu'&>0\,,& \mathcal U_R\mathcal U_R^T=&\left(
\begin{array}{ccc}
 0 & 0 & 1 \\
 0 & 1& 0 \\
 1 & 0 & 0
\end{array}
\right);&h_\nu'\mbox{Tr}\left(\mathbf{y}_\nu^2\mathcal U_R\mathcal U_R^T\mathbf{y}_\nu^2 \mathcal U_R^*\mathcal U_R^\dagger\right)&\rightarrow h_\nu'\sum_{i=1}^3 y_{\nu_i}^2y_{\nu_{4-i}}^2 \,.
\label{B}
\end{align}
\end{itemize}

On the other hand the expression for the neutrino mass matrix in Eq. \ref{NuMassBI-U} contains
precisely the combination $\mathcal U_R\mathcal U_R^T$. A quick look at the four possible combinations
of products of minima for $\mathcal U_{L,R}$, make us realize that they reduce to two, since both configurations of $\mathcal U_L$
leave the neutrino mass matrix unchanged. Nonetheless if the configuration $\mathcal U_R\mathcal U_R^T=\mathbb 1$
corresponds trivially to no mixing, \emph{possibility B for} $\mathcal U_R\mathcal U_R^T$
\emph{implies a maximal angle}. Indeed for the configuration of Eq.~\ref{B} the neutrino flavour field structure is (selecting $\mathcal U_L=\mathbb{1}$),
\begin{equation}
\mathcal{Y}_\nu=\Lambda_f\left(
\begin{array}{ccc}
 \frac{y_{\nu_1}}{\sqrt{2}} & 0 & \frac{iy_{\nu_1}}{\sqrt{2}} \\
 0 & y_{\nu_2} & 0 \\
 -\frac{y_{\nu_3}}{\sqrt{2}} & 0 & \frac{iy_{\nu_3}}{\sqrt{2}}
\end{array}\right)\label{YnuMANG}
\end{equation}
and the neutrino mass matrix reads:
\begin{equation}
\frac{v^2}{M}\left(\begin{array}{ccc}
 0 & 0 & y_{\nu_3}y_{\nu_1} \\
 0 & y_{\nu_2}^2& 0 \\
y_{\nu_3}y_{\nu_1} & 0 & 0
\end{array}\right) =U_{PMNS}\left(
\begin{array}{ccc}
m_{\nu_1}& 0 & 0 \\
 0 & m_{\nu_1}& 0 \\
 0 & 0 & m_{\nu_2}
\end{array}
\right)U_{PMNS}^T\, , \nonumber
\end{equation} 
with a mixing matrix and masses given by,
\begin{equation}\label{mixingabs}
U_{PMNS}=\left(
\begin{array}{ccc}
 \frac{1}{\sqrt{2}}& \frac{i}{\sqrt{2}} & 0  \\
 0  & 0& 1 \\
 -\frac{1}{\sqrt{2}} & \frac{i}{\sqrt{2}}& 0 \\
\end{array}\right),\qquad\quad m_{\nu_1}=\frac{v^2}{M} y_{\nu_1}y_{\nu_3}\, ,\quad m_{\nu_2}=\frac{v^2}{M} y_{\nu_2}^2\,.
\end{equation}
This case may correspond to either normal ($m_{\nu_1}<m_{\nu_2}$) or inverted ($m_{\nu_1}>m_{\nu_2}$) hierarchy in a first rough approximation ($\Delta m_{sol}^2=0$) and the maximal angle lies
always among the two degenerate neutrinos, meaning $\theta_{sol}\simeq \pi/4$; on the other hand, if the 
spectrum is quasi-degenerate, the mixing angle correspondence is unclear and the perturbations for
splitting the masses shall be studied, see Sec. \ref{TRD}.

All these conclusions were drawn from
 the minimization in two terms of the potential
only and they hold quite generally.

Like in the two family case, there is an unbroken symmetry
in configuration B, that is Eq.~\ref{YnuMANG}, since we have,
\begin{equation}
\left(
\begin{array}{ccc}
 \frac{y_{\nu_1}}{\sqrt{2}} & 0 & \frac{iy_{\nu_1}}{\sqrt{2}} \\
 0 & y_{\nu_2} & 0 \\
 -\frac{y_{\nu_3}}{\sqrt{2}} & 0 & \frac{iy_{\nu_3}}{\sqrt{2}}
\end{array}\right)\,e^{i\varphi \lambda_5}=
e^{i\varphi/2 (\lambda_3+\sqrt{3}\lambda_8)}\left(
\begin{array}{ccc}
 \frac{y_{\nu_1}}{\sqrt{2}} & 0 & \frac{iy_{\nu_1}}{\sqrt{2}} \\
 0 & y_{\nu_2} & 0 \\
 -\frac{y_{\nu_3}}{\sqrt{2}} & 0 & \frac{iy_{\nu_3}}{\sqrt{2}}
\end{array}\right)\,,
\label{U1SO}
\end{equation}
a simultaneous rotation in the direction $\lambda_5$ of $O(3)_{N}$
and an opposite sign transformation in the direction $(\lambda_3+\sqrt{3}\lambda_8)/2$
of $SU(3)_{\ell_L}$
constitute a preserved abelian symmetry that we shall denote $U(1)_{\tau-e}$.
It is interesting
to note that on the other hand, the configuration of diagonal $\mathcal{Y}_\nu$ has no symmetry
for generic $y_{\nu_{i}}$ and we shall see how this fits in the general picture of the possible minima.
It is nonetheless evident that for 2 degenerate $y_{\nu_{i}}$ in a diagonal $\mathcal{Y}_\nu$ there is a $SO(2)_V$ symmetry
unbroken and that for a configuration proportional to the identity ($\mathcal{Y}_\nu\propto \mathbb{1}$)
a vectorial $SO(3)_V$ arises. 
One can wonder if this happens for case B, Eq. \ref{YnuMANG} for all eigenvalues degenerate.
The result is that there is an unbroken $SO(3)$ in this case as well.
The two new relations,
\begin{equation}
\left(
\begin{array}{ccc}
 \frac{1}{\sqrt{2}} & 0 & \frac{i}{\sqrt{2}} \\
 0 & 1 & 0 \\
 -\frac{1}{\sqrt{2}} & 0 & \frac{i}{\sqrt{2}}
\end{array}\right)\,e^{i\varphi_2 \lambda_2}=
e^{-i\varphi_2(\lambda_2+\lambda_7)/\sqrt{2}}\left(
\begin{array}{ccc}
 \frac{1}{\sqrt{2}} & 0 & \frac{i}{\sqrt{2}} \\
 0 & 1 & 0 \\
 -\frac{1}{\sqrt{2}} & 0 & \frac{i}{\sqrt{2}}
\end{array}\right),
\end{equation}
\begin{equation}
\left(
\begin{array}{ccc}
 \frac{1}{\sqrt{2}} & 0 & \frac{i}{\sqrt{2}} \\
 0 & 1 & 0 \\
 -\frac{1}{\sqrt{2}} & 0 & \frac{i}{\sqrt{2}}
\end{array}\right)\,e^{i\varphi_3 \lambda_7}=
e^{i\varphi_3(\lambda_1+\lambda_6)/\sqrt{2}}\left(
\begin{array}{ccc}
 \frac{1}{\sqrt{2}} & 0 & \frac{i}{\sqrt{2}} \\
 0 & 1 & 0 \\
 -\frac{1}{\sqrt{2}} & 0 & \frac{i}{\sqrt{2}}
\end{array}\right),
\end{equation}
provide two new directions of conserved symmetry.
This is however not enough to prove that the group is $SO(3)$ and not just $U(1)^3$.
For this purpose the basis,
\begin{gather}
\left\{\frac{1}{2}(\lambda_3+\sqrt{3}\lambda_8)\,, \,-\frac{1}{\sqrt{2}}(\lambda_2+\lambda_7)\,,\,
\frac{1}{\sqrt{2}}(\lambda_1+\lambda_6)\,\right\}\\=
\left\{\,
\left(
\begin{array}{ccc}
 1 & 0 & 0 \\
 0 & 0 & 0 \\
 0 & 0 & -1
\end{array}
\right)\,\,,\,\,
\left(
\begin{array}{ccc}
 0 & \frac{i}{\sqrt{2}} & 0 \\
 -\frac{i}{\sqrt{2}} & 0 & \frac{i}{\sqrt{2}} \\
 0 & -\frac{i}{\sqrt{2}} & 0
\end{array}
\right)\,\, ,\, \,
\left(
\begin{array}{ccc}
 0 & \frac{1}{\sqrt{2}} & 0 \\
 \frac{1}{\sqrt{2}} & 0 & \frac{1}{\sqrt{2}} \\
 0 & \frac{1}{\sqrt{2}} & 0
\end{array}
\right)\,\right\}\label{SOBAS}\,,
\end{gather}
can be shown to have the commutation relations of $SO(3)$, that is structure constants $\epsilon_{ijk}$.

The emphasis will be on case B, Eq.~\ref{B}, since it gives a maximal mixing angle, but first 
a few words on case A are due. If both Yukawas are diagonal, as in case A,
for arbitrary eigenvalues there is no symmetry left unbroken at all. Nonetheless
 $h_\nu'$ is negative for case A and after minimizing in $\mathcal U_R$ the structure of $I_{R}$ (Eq.\ref{A}) is just like
that of $I_{\nu^2}$, so that the effective coupling of $I_{\nu^2}$ can be taken to be
$h_\nu'+h_\nu$. Then the analysis of quarks holds just the same and we find the
type of solutions listed in section \ref{RP3QF}. All of these solutions have at least one pair of eigenvalues
degenerate; this implies that there is indeed always at least one $SO(2)_V$ present at the
minimum.

This same reasoning applied to case B will reveal new freedom in the possible eigenvalues
of the Yukawas, since now the symmetry reported in Eq. \ref{U1SO}, is present for
arbitrary entries in $\mathbf{y}_\nu$.

Before entering the details on the allowed values for the neutrino and charged lepton
eigenvalues at the different vacua, for the reader interested in the closest
solution to the observed flavour pattern we report here a new kind of solution with respect to the quark case:
\begin{equation}
\mathcal{Y}_E =\Lambda_f \left( \begin{array}{ccc}
0&0&0\\
0&y_\mu&0\\
0&0&y_\tau \\
\end{array}\right),\qquad\quad \mathcal{Y}_\nu =\Lambda_f
\left( \begin{array}{ccc}
y_{\nu_1}/\sqrt{2}&0&iy_{\nu_1}/\sqrt{2}\\
0&y_{\nu_2}&0\\
-y_{\nu_3}/\sqrt{2}&0&iy_{\nu_3}/\sqrt{2}\\
\end{array}\right).
\end{equation}
The two different entries for the charged leptons are in agreement with the
larger masses of the muon and tau leptons whereas in the neutrino sector there
is one maximal angle as in Eq.~\ref{mixingabs} and three massive neutrinos, two of them degenerate.
In the limit of three degenerate neutrinos, $y_{\nu_{1}}=y_{\nu_{2}}=y_{\nu_{3}}$,
small corrections to the above pattern give rise to a second large angle, see for instance \cite{Altarelli:2004za} and
references therein.
Unfortunately in the present configuration the two large angles would not correspond to
$\theta_{12}$ and $\theta_{23}$, see Sec. \ref{TRD} for the addressing of this issue.

The list of possible types of vacua for $g<0$ (see appendix, Sec.~\ref{App2} for details) at the renormalizable level is:

\begin{itemize}

\item [\bf I] The hierarchical solution for the eigenvalues translates now into Yukawas of the type,
\begin{equation}
\mathcal{Y}_E =\Lambda_f \left( \begin{array}{ccc}
0&0&0\\
0&0&0\\
0&0&y_\tau \\
\end{array}\right),\qquad\quad \mathcal{Y}_\nu =y_\nu\Lambda_f 
\left( \begin{array}{ccc}
0&0&0\\
0&0&0\\
-1/\sqrt{2}&0&i/\sqrt{2}\\
\end{array}\right),
\end{equation}
and a pattern $\mathcal{G}_\mathcal{F}^l\rightarrow U(2)^2\times U(1)_{LN}$.   There are no light massive neutrinos
in this scenario, but flavour effects are present.

\item [\bf II]The second kind of solution stands the same as in the quark case,
\begin{equation}
\mathcal{Y}_E =\Lambda_f \left( \begin{array}{ccc}
y&0&0\\
0&y&0\\
0&0&y_\tau \\
\end{array}\right),\qquad\quad \mathcal{Y}_\nu =y_\nu\Lambda_f 
\left( \begin{array}{ccc}
0&0&0\\
0&0&0\\
-1/\sqrt{2}&0&i/\sqrt{2}\\
\end{array}\right),
\end{equation}
for now the identity $y_e=y_\mu$ yields the breaking structure
$\mathcal{G}_\mathcal{F}^l\rightarrow U(2)_V\times U(1)_{LN} $, where the unbroken group would be different if the two first eigenvalues
of $\mathcal{Y}_E$ were to differ.

\item [\bf III] The equivalent of case {\bf III} in the 2 family case differs from the extension of this
case in the quark case from 2 to 3 generations. We have now a hierarchical set-up for charged
leptons and arbitrary entries for neutrino Yukawa eigenvalues,
\begin{equation}
\mathcal{Y}_E =\Lambda_f \left( \begin{array}{ccc}
0&0&0\\
0&0&0\\
0&0&y_\tau \\
\end{array}\right),\qquad\quad \mathcal{Y}_\nu =\Lambda_f
\left( \begin{array}{ccc}
y_{\nu_1}/\sqrt{2}&0&iy_{\nu_1}/\sqrt{2}\\
0&y_{\nu_2}&0\\
-y_{\nu_3}/\sqrt{2}&0&iy_{\nu_3}/\sqrt{2}\\
\end{array}\right),
\end{equation}
and the breaking pattern is
$\mathcal{G}_\mathcal{F}^l\rightarrow U(2)_{E_R}\times U(1)_{\tau-e}$. The reason for $y_{\nu_1}\neq y_{\nu_2}$
now is that the degeneracy of these two parameters leads to no extra symmetry, so their equality is not protected.

\item [\bf IV] The completely degenerate configuration is
\begin{equation}
\mathcal{Y}_E =\Lambda_f \left( \begin{array}{ccc}
y&0&0\\
0&y&0\\
0&0&y \\
\end{array}\right),\qquad\quad \mathcal{Y}_\nu =y_\nu\Lambda_f
\left( \begin{array}{ccc}
1/\sqrt{2}&0&i/\sqrt{2}\\
0&1&0\\
-1/\sqrt{2}&0&i/\sqrt{2}\\
\end{array}\right),
\end{equation}
we have now that $\mathcal{G}_\mathcal{F}^l\rightarrow SO(3)_V$ with the vectorial group
as pointed out in Eqs.~\ref{U1SO}-\ref{SOBAS}. In this case nonetheless the mixing loses meaning since
the charged leptons are degenerate.

\item[\bf V] New configurations are now possible as
\begin{equation}
\mathcal{Y}_E =\Lambda_f \left( \begin{array}{ccc}
0&0&0\\
0&y_\mu&0\\
0&0&y_\tau \\
\end{array}\right),\qquad\quad \mathcal{Y}_\nu =\Lambda_f
\left( \begin{array}{ccc}
y_{\nu_1}/\sqrt{2}&0&i y_{\nu_1}/\sqrt{2}\\
0&y_{\nu_2}&0\\
-y_{\nu_3}/\sqrt{2}&0&iy_{\nu_3}/\sqrt{2}\\
\end{array}\right),
\end{equation}
with  $\mathcal{G}_\mathcal{F}^l \rightarrow U(1)_{\tau-e}\times U(1)_{e_R}$. 

\item[\bf VI] The presence of arbitrary charged lepton
masses is possible when two neutrinos are massless,
\begin{equation}
\mathcal{Y}_E =\Lambda_f \left( \begin{array}{ccc}
y_e&0&0\\
0&y_\mu&0\\
0&0&y_\tau \\
\end{array}\right),\qquad\quad \mathcal{Y}_\nu =\Lambda_f
\left( \begin{array}{ccc}
0&0&0\\
0&y_{\nu_2}&0\\
-y_{\nu_3}/\sqrt{2}&0&iy_{\nu_3}/\sqrt{2}\\
\end{array}\right),
\end{equation}
with  $\mathcal{G}_\mathcal{F}^l \rightarrow U(1)_{\tau}\times U(1)_{e} $ since the neutrinos
that the electron and tau couple to are massless.
\item[\bf VII] Case {\bf III} leaves and extended symmetry if two neutrinos are
massless
\begin{equation}
\mathcal{Y}_E =\Lambda_f \left( \begin{array}{ccc}
0&0&0\\
0&0&0\\
0&0&y_\tau \\
\end{array}\right),\qquad\quad \mathcal{Y}_\nu =\Lambda_f
\left( \begin{array}{ccc}
0&0&0\\
0&y_{\nu_2}&0\\
-y_{\nu_3}/\sqrt{2}&0&iy_{\nu_3}/\sqrt{2}\\
\end{array}\right),
\end{equation}
with  $\mathcal{G}_\mathcal{F}^l \rightarrow U(2)_{E_R}\times U(1)_{e}\times{U(1)_\tau}$.

\item[\bf VIII] Finally the case {\bf V} leaves and extended symmetry if one neutrino Yukawa vanishes
and one charged lepton is massless,
\begin{equation}
\mathcal{Y}_E =\Lambda_f \left( \begin{array}{ccc}
0&0&0\\
0&y_{\mu}&0\\
0&0&y_\tau \\
\end{array}\right),\qquad\quad \mathcal{Y}_\nu =\Lambda_f
\left( \begin{array}{ccc}
0&0&0\\
0&y_{\nu_2}&0\\
-y_{\nu_3}/\sqrt{2}&0&iy_{\nu_3}/\sqrt{2}\\
\end{array}\right),
\end{equation}
with  $\mathcal{G}_\mathcal{F}^l \rightarrow  U(1)_{e}\times U(1)_\tau\times U(1)_{e_R}$. 
\end{itemize}

\figuremacroW{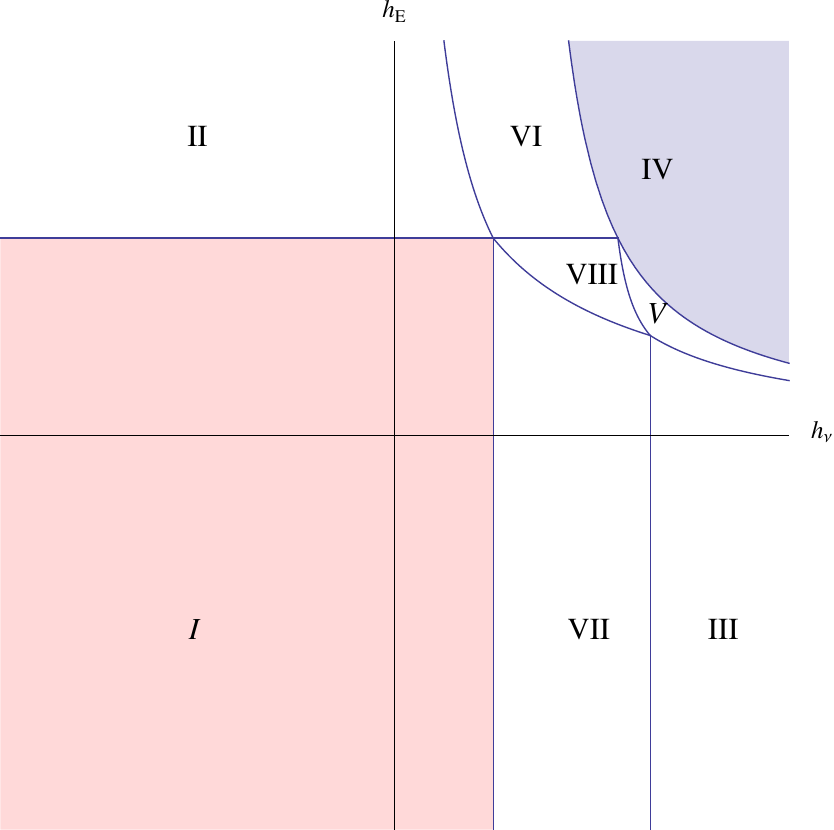}{Regions for the different lepton mass configurations allowed at the
absolute minimum in the $h_\nu-h_E$ plane for $g<0$}{ The different regions correspond to different mass
patterns distinct from the quark case, see text and the appendix \ref{App2} for details.
}{.7}

The possibilities for the vacua have grown sensibly. This is related to
the flavour group $\mathcal{G}_\mathcal{F}^l$ which, in contrast with the quark case, 
allows for the new invariant at the renormalizable level
$I_R$. This invariant gives rise to the maximal angle solution
and produces new configurations for the values
of flavour fields eigenvalues at the minimum. Indeed in the limit
$h_\nu'\rightarrow 0$ all these different cases recombine in the ones for
the quark case, see Sec. \ref{App2} of the appendix.

As for the possible combinations of charged lepton and neutrino eigenvalues
scenarios with (at least) two degenerate neutrino masses can come along with
hierarchical (case {\bf III}) or semi-hierarchical
(case {\bf V}) charged lepton spectrum. Lifting the electron mass from 0 is possible in case {\bf V}
via the introduction of small breaking terms of the axial symmetry, giving the lightest lepton a
naturally smaller mass.

The general conclusion is therefore that 
in first approximation a \emph{maximal mixing angle is obtained
in the lepton sector}
whereas \emph{for the quark case no mixing is allowed} in this same
level of approximation. This stands as a tantalizing framework for
explaining the differences in mixing matrices in the two sectors in
\emph{a symmetry framework comprising both quarks and leptons}. The solution of the \emph{maximal angle} can be
traced back to the \emph{presence of an orthogonal group} in the flavour 
symmetry of the lepton sector, which is in turn \emph{related to the Majorana
nature of neutrino masses}.

\subsection{Realistic mixing and spectrum}\label{TRD}

The study of flavour symmetry breaking for the quark sector in the bi-fundamental case yielded
as a possibility a hierarchical up and down-type fermion spectrum with no mixing,
which is a good approximation to the Yukawa couplings to order $\lambda_c^2$.
The lepton case resulted in a possible flavour pattern with a maximal mixing
angle, a hierarchical charged lepton spectrum and at least two degenerate neutrinos. 
Nonetheless even if the neutrino mass spectrum allows for such a situation at present 
it is the case that the mixing matrix in leptons presents two large angles, so the question
arises of whether the present framework can accommodate a second large angle
and a good first approximation (at least as good as for quarks) to nature . The present section
presents an ansatz to address this question.

The Jacobian analysis pointed to permutation matrices for $\mathcal U_R\mathcal U_R^T$ as candidates
for extremal points, however in the renormalizable potential only the identity and the ``antidiagonal" configuration
were allowed as absolute minima. Let us postulate that a different permutation matrix can be a
local long-lived-enough minimum of the present potential or a minimum of a different, possibly non-renormalizable,
potential. We assume furthermore a degenerate neutrino spectrum\footnote{Unlike case {\bf IV} there is no symmetry reason
to have degenerate neutrinos if the charged leptons are not, note however that if the neutrino sector is consider separately complete degeneracy implies a $SO(3)_V$ symmetry.}, and a semihierarchical charged lepton
spectrum as in case {\bf V}, that is,

\begin{equation}
\mathcal{Y}_E =\Lambda_f \left( \begin{array}{ccc}
0&0&0\\
0&y_\mu&0\\
0&0&y_\tau \\
\end{array}\right),\qquad\mathcal{Y}_\nu =\Lambda_f \,y_{\nu}
\left( \begin{array}{ccc}
1&0&0\\
0&1/\sqrt{2}&-i/\sqrt{2}\\
0&1/\sqrt{2}&i/\sqrt{2}\\
\end{array}\right),
\end{equation}
and a neutrino mass matrix,
\begin{equation}
 U_{PMNS}\mathbf{m}_\nu U_{PMNS}^T=\frac{y^2v^2}{2M}\left(
\begin{array}{ccc}
1&0 & 0 \\
0 & 0 & 1\\
 0 &1 & 0
\end{array}\right)\,.
\end{equation}
In this configuration even if the three neutrinos are degenerate in mass one can
tell one of them from the rest from the relative maximal Majorana phase, as previously.
The angle among the two completely degenerate neutrino states nonetheless is undetermined
at this point such that small perturbations may orient it in an arbitrary direction \cite{Altarelli:2004za}.

As an explicit example let us input the following perturbations to the neutrino matrix, the general case
can be found in Sec. \ref{App3} of the appendix,
\begin{equation}
U_{PMNS}\mathbf{m}_\nu U_{PMNS}^T=\frac{y^2v^2}{2M}\left(
\begin{array}{ccc}
1 + \delta&  (\epsilon+\tilde \epsilon)/2&  (\epsilon-\tilde \epsilon)/2 \\
 (\epsilon+\tilde \epsilon)/2 & \delta & 1\\
(\epsilon-\tilde \epsilon)/2 & 1&\delta\\
\end{array}
\right)\,.
\end{equation}
which lead to the mixing matrix:
\begin{gather}
U_{PMNS}=\left(\begin{array}{ccc}
\frac{1}{\sqrt{2}}&-\frac{1}{\sqrt{2}}&\frac{\tilde\epsilon}{2\sqrt{2}}\\[2mm]
\frac{1}{2}+\frac{\tilde\epsilon}{4\sqrt{2}}&\frac{1}{2}-\frac{\tilde\epsilon}{4\sqrt{2}}&\frac{1}{\sqrt{2}}\\[2mm]
-\frac{1}{2}+\frac{\tilde\epsilon}{4\sqrt{2}}&-\frac{1}{2}-\frac{\tilde\epsilon}{4\sqrt{2}}&\frac{1}{\sqrt{2}}
\end{array}\right)\left(\begin{array}{ccc}
1&0&0\\
0&1&0\\
0&0&i\\
\end{array}
\right)+\mathcal{O}(\epsilon^2,...)\,,\\
\theta_{12}\simeq \pi/4\,,\qquad \theta_{23}\simeq\pi/4\,,\qquad \theta_{13}\simeq \frac{\tilde\epsilon}{2\sqrt{2}}\,.
\end{gather}
and a mass spectrum, to first order:
\begin{gather}
m_{\nu_1}\simeq\frac{y^2v^2}{2M}\left(1+\delta-\frac{\epsilon}{\sqrt{2}}\right)\,,
\quad m_{\nu_2}\simeq\frac{y^2v^2}{2M}\left(1+\delta+\frac{\epsilon}{\sqrt{2}}\right)\,,\\
|m_{\nu_3}|\simeq\frac{y^2v^2}{2M}\left(1-\delta\right)\,,
\end{gather}
which stands as an acceptable first order approximation to the lepton flavour pattern.
\emph{It is then possible to achieve a first sketch of both quark and lepton flavour patterns with
a similar $(\sim\mathcal{O}(10\%))$ approximation}. The origin of the perturbations introduced is
however yet to be specified consistently and constitutes work in progress.

Finally a phenomenological remark on the fate of this scenario; the degenerate case
for neutrino masses is at present very close to the present upper limits of neutrino mass,
such that new data from experiments such as neutrinoless double-beta decay may rule
out or boost the explanation for flavour here proposed, see \cite{onpreparation,onpreparation2}.

\include{XYZ}

\chapter{Resumen y Conclusiones} 


\ifpdf
    \graphicspath{{7/figures/PNG/}{7/figures/PDF/}{7/figures/}}
\else
    \graphicspath{{7/figures/EPS/}{7/figures/}}
\fi
En esta tesis la estructura de sabor de las part\'iculas elementales ha sido examinada
desde el punto de vista de una posible simetr\'ia de sabor impl\'icita. La simetr\'ia de sabor 
considerada es la simetr\'ia global que presenta el ME en ausencia de masa para los fermiones.
La extensi\'on necesaria del ME para acomodar masas de neutrinos introduce no obstante
una dependencia en el modelo elegido. Por simplicidad el escenario del Seesaw con neutrinos
pesados (conocido como tipo I) es considerado cuando se trata de leptones, asumiendo
la existencia de $n_g$ generaciones ligeras y pesadas. La simetr\'ia de sabor es entonces seleccionada como
la mayor simetr\'ia posible de la teor\'ia libre, esquem\'aticamente ~$\mathcal{G_F}\sim U(n_g)^5\times O(n_g)$, en d\'onde $O(n_g)$
est\'a asociado a neutrinos pesados degenerados, cuya masa es la \'unica presente en la teor\'ia libre,
mientras que cada factor $U(n_g)$ corresponde a cada campo con distinta carga en el ME.

Sin espicificar un modelo de sabor es posible explorar la posibilidad de que, a bajas energ\'ias,
los Yukawas sean las fuentes de sabor en el ME y la teor\'ia que lo completa; esta suposici\'on est\'a en acuerdo con
los datos experimentales y se encuetra en el centro del \'exito fenomenol\'ogico de la hip\'otesis de MFV,
implementada a trav\'es de t\'ecnicas de Lagrangianos efectivos. Prosiguiendo este camino, hemos explorado
las consecuencias de un car\'acter din\'amico de los acoplos de Yukawa mediante la determinaci\'on, en
una base general, de los posibles extremos del conjunto de invariantes (gauge y de sabor) que
pueden ser construidos con \'estos. Existen tantos invariantes independientes como par\'ametros f\'isicos,
y un conjunto de invariantes completo e independiente ha sido determinado. Los extremos son identificados mediante
el estudio del Jacobiano de cambio de base de los par\'ametros f\'isicos a los invariantes. Hemos demostrado que, mientras para quarks los extremos de los invariantes apuntan hacia la ausencia de mezcla, para
leptones grandes \'angulos correlacionados con un car\'acter de Majorana no trivial
resultan ser los extremos naturales. En particular, una configuraci\'on posible presenta tres neutrinos
degenerados, un \'angulo atmosf\'erico m\'aximo ($\theta_{23}=\pi/4$) y una fase de Majorana m\'axima ($\pi/2$). Esta estrutura,
al ser perturbada, desarrolla un \'angulo solar ($\theta_{12}$) gen\'ericamente grande, dado que esta variable
parametriza una direcci\'on plana a primer orden, y un \'angulo reactor ($\theta_{13}$) perturbativo. \'Este puede ser un motivador y sugerente primer paso en la empresa
del entendimiento del origen de sabor, dado que este esquema resulta muy similar al obesrvado en la 
naturaleza y puede ser testeado en el futuro cercano \cite{onpreparation}.

Un verdadero origen din\'amico de los acoplos de Yukawa sugiere un paso m\'as: considerar que corresponden
a campos din\'amicos, o agregados de \'estos, que poseen sabor y han adquirido un vev. La simetr\'ia de 
sabor ser\'ia manifiesta en el Lagrangiano total de alta energ\'ia, a una escala $\Lambda_f$. Tras la
rotura espont\'anea de simetr\'ia, los acoplos de Yukawa de bajas energ\'ias resultar\'ian de operadores 
efectivos de dimension $d>4$ invariantes bajo la simetr\'ia de sabor, que involucran uno o mas campos de
sabor junto con los campos usuales del ME.

Solo un escalar (o cunjunto de campos en una configuracion escalar) puede tomar un vev, que deber\'a
corresponder al m\'inimo de un potencial. >Cu\'al es el potencial escalar para estos campos escalares
de sabor? >Puede alguno de sus m\'inimos corresponder naturalmente al espectro observado de masas
y \'angulos? Estas preguntas son respondidas en el presente trabajo. El an\'alisis del potencial est\'a
relacionado con los extremos de los invariantes mencionados antes, pero va mas all\'a dado que el potencial
no tiene necesariamente que compartir los puntos extremos del an‡lisis de los invariantes ni presentar Žstos
como m'nimos absolutos del potencial.

La realizaci\'on mas simple de este tipo se obtene via una correspondencia uno a uno de cada acoplo de
Yukawa (up, down, elecr\'on y neutrino) con un \'unico campo escalar perteneciente a la representaci\'on
bi-fundamental del grupo de sabor $\mathcal{G_F}$. En el lenguaje de Lagrangianos efectivos este caso
corresponde al orden m\'as bajo en la expansi\'on de sabor: operadores de Yukawa de dimension $d=5$
construidos por un campo escalar y los campos del ME usuales. El potencial escalar general para
campos escalares bi-fundamentales ha sido construido para quarks y leptones en el caso de dos y tres
familias. Formalmente, se construye con los invariantes mencionados arriba y no obstante de su combinaci\'on
 surgen nuevos m\'inimos.
 
 Al determinar el potencial escalar, primero se demostr\'o que imponer la simetr\'ia de sabor 
 representa una condici\'on muy restrictiva: al nivel renormalizable s\'olo ciertos t\'erminos
 son permitidos en el potencial, e incluso al nivel renormalizable estructuras constre\~nidas
 deben ser respetadas.
 
 En el caso de quarks, al nivel renormalizable, en el m\'inimo del potencial solo \'angulos nulos
 son permitidos. Respecto a jerarqu\'ias de masa, uno de los posibles m\'inimos presenta
 masas nulas para todos los quarks excepto los pertenecientes a la familia m\'as pesada, esto
 es, un quark tipo down y otro tipo up con masa solamente tanto en dos como en tres familias.
 Existe por lo tanto una soluci\'on incial que se asemeja en primera aproximaci\'on a la naturaleza:
 un espectro jer\'arquico sin mezcla. Dicha soluci\'on puede ser pertubada al nivel renormalizable
 para obtener masas para las familia m\'as ligera mediante t\'erminos de rotura expl\'icita de la
 parte abeliana de $\mathcal{G_F}^q$, es decir $U(1)^3$.
La introducci\'on de t\'erminos no renormalizables en el potencial permite una rotura
 mayor de la simetr\'ia, al precio de enormes ajustes finos, que son inaceptables en nuestra opini\'on
 en el esp\'iritu de la teor\'ia efectiva de campos.
 
 En el sector lept\'onico la misma realizaci\'on de correspondencia  Yukawa-campo,  escalares
 bi-fundamentales, condujo a resultados soprendentemente diferentes. En el caso de dos y tres familias,
 fases de Majorana y \'angulos de mezcla no triviales pueden ser seleccionados por el m\'inimo del potencial,
indicando una nueva conexi\'on en la estructura de masas de neutrinos: i) grandes \'angulos de mezcla
son posibles; ii) hay una fuerte correlaci\'on entre \'angulos de mezcla grandes y espectro degenerado de
masas; iii) la fase de Majorana relativa es predicha como m\'axima, $2\alpha= \pi/2$, aunque no
implica violaci\'on de conjugaci\'on de carga y paridad observable.

Las soluciones exactas del potencial renomalizable condujentes a mezcla no trivial muestran un \'unico \'angulo
m\'aximo entre dos neutrinos degenerados pero distinguibles tanto para el caso de dos como el de tres familias.
Esto conduce, para el caso de jerarqu\'ia normal e invertida, a el \'angulo m\'aximo siendo el solar en lugar
del atmosf\'erico, num\'ericamente compatible con un valor m\'aximo. El caso de tres neutrino ligeros
degenerados y \'angulos de mezcla grandes para $\theta_{12}$ y $\theta_{23}$ identificado en el an\'alisis
de extremos de los invariantes no aparece no obstante
como m\'inimo absoluto del potential renormalizable; podr\'ia ser un m\'inimo local de dicho potencial o
el m\'inimo absoluto de un potencial no-renormalizable.

Otra avenida explorada en este trabajo asocia dos campos a cada acoplo de Yukawa, esto es
$Y\sim \chi^L\chi^{R\dagger}/\Lambda_f^2$. Esta situaci\'on es atrayente dado que mientras que los Yukawas
son objetos compuestos, los nuevos campos est\'an en la representaci\'on fundamental. Dichos campos
podr\'ian ser escalares o fermi\'onicos: aqui nos centramos exclusivamente en escalares. Desde el punto
de vista de Lagrangianos efectivos, este caso podr\'ia corresponder al siguiente al primer orden en la
expansi\'on: operadores de Yukawa efectivos de dimension 6, como fuentes totales o parciales de los Yukawas de
baja energ\'ia. Hemos constru\'ido el potencial escalar general para campos escalares en la representaci\'on fundamental
para los casos de dos y tres familias de quarks, aunque las conclusiones se transladan de manera directa
a leptones. Por construcci\'on este escenario resulta inevitablemente en una fuerte jerarqu\'ia de masas:
solamente un quark en cada sector up y down obtiene masa: los quarks top y bottom.
Una mezcla no trivial requiere dos campos escalares de sector up y down (neutrino y electr\'on)
transformando bajo el grupo $SU(3)_{Q_L}$. En consequencia el contenido m\'inimo es de cuatro campos
$\chi^L_{U\,(\nu)}$, $\chi^L_{D\,(E)}$, $\chi^R_{U\,(\nu)}$ and $\chi^R_{D\,(E)}$ y la mezcla surge
de la interacci\'on entre los dos primeros. En resumen, para escalares en la fundamental en un 
modo natural se obtiene: i) una fuerte jerarqu\'ia entre quarks de la misma carga, se\~nalando un
quark distinguible por su mayor masa en cada sector; ii) un \'angulo de mezcla no trivial, que puede ser identificado tanto para quarks como para leptones con el del sector $23$ en el caso de tres familias.

Finalmente, como una posible correcci\'on a los patrones discutidos previamente, se ha
discutido brevemente la posibilidad de introducir simult\'aneamente escalares bi-fundamentales y fundamentales.
Es una posibilidad muy sensata, desde el punto de vista de Lagrangianos efectivos, considerar 
operadores de Yukawa de orden $d=5$ y $d=6$ trabajando a orden $\mathcal{O}(1/\Lambda_f^2)$.
Sugiere que el t\'ermino de $d=5$, que acarrea bi-fundamentales, podr\'ia proporcionar la contribuci\'on
dominante, mientras que el operador de $d=6$, que trae consigo los campos en la fundamental,
proporciona correcciones para inducir masas no nulas para las dos familias ligeras junto con \'angulos
no triviales.

En general, es destacable que el requisito de invarianza bajo la simetr\'ia de sabor
constri\~na fuertemente el potencial escalar y consequentemente los m\'inimos y 
patrones de ruptura de simetr\'ia. De entre los resultados obtenidos uno sobresale de
entre los dem\'as. En el m\'inimo del potencial, al nivel renormalizable, los \'angulos
de mezcla para quarks son nulos a primer orden, mientras que la mezcla en los leptones
resulta ser m\'axima. La presencia de mezcla m\'axima es debida al factor $O(n_g)$
del grupo de sabor, que est\'a a su vez relacionado con la naturaleza Majorana de los neutrinos.
La explicaci\'on de la diferente estructura de mixing entre quarks y leptones en este escenario
es, en \'ultima instancia, la distinta naturaleza de los dos tipos de fermiones: Dirac y Majorana.

\chapter{Summary and Conclusions} 


\ifpdf
    \graphicspath{{7/figures/PNG/}{7/figures/PDF/}{7/figures/}}
\else
    \graphicspath{{7/figures/EPS/}{7/figures/}}
\fi

In this dissertation the flavour pattern of the elementary particles
was examined from the point of view of its possible underlying flavour symmetry.   The flavour symmetry considered is the global flavour symmetry which the SM possesses in the limit of massless fermions. The necessary extension of the SM to accommodate Majorana neutrino masses introduces nevertheless a model dependence in the neutrino sector; for simplicity the seesaw scenario  with heavy right-handed neutrinos (known as type I) is considered here when dealing with leptons, assuming  $n_g$ generations in both the light and heavy sectors. \emph{ The largest possible flavour symmetry of the free theory}  for both quark and lepton sectors is then,   schematically, ~$\mathcal{G_F}\sim U(n_g)^5\times O(n_g)$, with $O(n_g)$
associated to heavy degenerate neutrinos, whose mass is the only one present in the free theory, and each $U(n_g)$ factor for each SM fermion field\footnote{The flavour
group can alternatively be defined as the largest flavour group in the absence of Yukawa interactions.}.

Without particularizing to any concrete flavour model, it is possible to explore the possibility that, at low energies, the  Yukawas may be the sources of flavour in the SM and beyond;  this assumption is well in agreement with data and lies at the heart of the phenomenological success of the MFV ansatz, implemented through effective Lagrangian techniques. Walking further on this path, we have explored the consequences of an hypothetical dynamical character for the Yukawa couplings themselves by determining, on general grounds, the possible extrema of the (gauge and flavour) invariants that can be constructed out of them. There are as many independent invariants as physical parameters, and a complete set of independent invariants has been determined. The extrema are identified via the study of the Jacobian of the change of basis from the physical parameters to invariants. We have shown that, while for quarks the extrema of the invariants point to no mixing, for leptons maximal mixing angles and Majorana phases correlated with neutrino mass degeneracy turn out to be natural extrema. In particular, a possible configuration presents three degenerate neutrinos, 
a maximal ($\pi/4$) atmospheric angle ($\theta_{23}$) and a maximal relative Majorana phase ($\pi/2$). This last setup when perturbed presents a generically large solar angle ($\theta_{12}$), since this
variable parametrizes a flat direction at first order, and a
perturbative reactor angle ($\theta_{13}$) together with small neutrino mass splittings.
This may be a very encouraging and suggestive first step in the quest for the understanding of the origin of flavour, as these patterns resemble closely the mixings observed in nature
and the degeneracy of neutrino masses will be tested in the near future \cite{onpreparation}.


A true dynamical origin for the Yukawa couplings suggests a further step: to consider them as corresponding to dynamical fields, or aggregate of fields, that carry flavour and have taken a vev. Flavour would be a manifest symmetry of the total, high energy Lagrangian, at a flavour scale $\Lambda_f$. After spontaneous symmetry breaking, 
the low-energy Yukawa interactions would result from effective operators of 
dimension $d>4$ invariant under the flavour symmetry, which involve one or more flavour fields together with the usual SM fermionic and Higgs fields.




Only a scalar field (or an aggregate of fields in a scalar configuration) can get a vev, which should correspond 
to the minimum of a potential. What is the scalar potential for those scalar flavour fields? 
May some of its minima naturally correspond to the observed spectra of masses and mixing angles? 
These questions  have been addressed in this work.  
The analysis of the potential is related to the extrema of the invariants mentioned above,
but it goes beyond since the potential need neither share the extremal points of the invariant analysis
nor present these extremal points as absolute minima.



The simplest realization of this kind is obtained by a one-to-one correspondence of  each Yukawa coupling with a single scalar 
field transforming 
in the bi-fundamental of the flavour group $\mathcal{G_F}$. In the language of effective Lagrangians, this may correspond to the 
lowest order terms in the flavour expansion: $d=5$ effective Yukawa operators made out of one flavour field plus  
the usual SM fields. The general scalar potential for bi-fundamental flavor scalar fields was constructed 
for quark and leptons in the two and three family case. Formally, it can be simply built out of the same Yukawa invariants mentioned above:  from their combination new minima may a priori follow.

When determining the scalar potential, it was first shown that the underlying flavour symmetry 
is a very restrictive constraint: at the renormalizable level only a few terms are allowed in the potential, 
and even at the non-renormalizable level quite constrained patterns have to be respected.

For the quark case at the renormalizable level, at the minimum of the potential only vanishing
mixing angles are allowed. Regarding mass hierarchies, one of the possible minima allows
 vanishing Yukawa couplings for all quarks but those in the heaviest family, both for the two and
 three generation cases. There is therefore an starting solution in the quark case which
 resembles in first approximation nature: a hierarchical spectrum with no mixing.
 This type of solution can be perturbed  at the
 renormalizable level to provide masses for the
 lightest family, by means of small explicit breaking terms of the abelian part of $\mathcal{G}_\mathcal{F}^q$, that is $U(1)^3$. 
The introduction of non-renormalizable terms in the potential allowed for 
 further breaking of the symmetry, at the price of large fine-tunings, which are in our opinion unacceptable in the spirit of  and effective field theory
 approach.

For the lepton sector, the same realization one-Yukawa-one-field, that is, of scalar bi-fundamental fields led to strikingly different results. In the two and three family cases non-trivial Majorana phases and mixing angles may be selected by the potential minima, configurations
contained in the invariants extrema analysis. The differences with the quark case are: i) large mixing angles are possible; ii) there is a strong correlation between mixing strength and mass spectrum; iii) the relative Majorana phase among the two massive neutrinos is predicted to be maximal, $2\alpha= \pi/2$, for non-trivial mixing angle; moreover, although the Majorana phase is maximal, it does not lead to experimental signatures of CP violation, as it exists a basis in which all  terms in the Lagrangian are real. 

The exact solutions of the renormalizable potential leading to non-trivial mixing showed one maximal mixing
angle only among two degenerate in mass but distinct (since their relative Majorana phase is maximal) neutrinos for both two and three generations. This scenario leads in the
case of normal or inverted hierarchies to the maximal angle being the
solar instead of the atmospheric angle. The case of all three neutrinos degenerate, large $\theta_{12}$ and maximal $\theta_{23}$ identified
in the invariants extrema analysis turns out not to be present as an absolute minimum of the renormalizable potential; 
it could be a local minimum of the renormalizable potential or and absolute minimum of a nonrenormalizable potential.

Another avenue explored in this work  associates two vector flavour fields to each Yukawa spurion, i.e. a Yukawa 
$Y\sim \chi^L\chi^{R\dagger}/\Lambda_f^2$. This is an attractive scenario in that while 
Yukawas are composite objects, the new fields are in the fundamental representation of the flavour group, 
in analogy with the  case of quarks.  
From the point of view of effective Lagrangians, this case corresponds 
to $d=6$ effective Yukawa operators. 

In a first step we considered the $d=6$ operator contribution alone, such
that no $d=5$ operator is present.  In this context the general renormalizable scalar potential for scalar flavour fields 
in the fundamental representation was constructed, both for the case of two and three families of quarks, although conclusions
translate straightforwardly to leptons. By construction, this scenario 
results unavoidably in a strong hierarchy of masses: only one quark gets mass in each
sector: the top and bottom quark. Non-trivial mixing requires as expected a 
misalignment between the flavour fields associated to the up and down (neutrino and electron) left-handed quarks (leptons). 
In consequence, the minimal field 
content corresponds to four fields $\chi^L_{U\,(\nu)}$, $\chi^L_{D\,(E)}$, $\chi^R_{U\,(\nu)}$ and $\chi^R_{D\,(E)}$, and the physics of mixing lies 
in the interplay of the first two. In resume, for fundamental flavour fields it follows in a completely natural way: 
i) a strong mass hierarchy between quarks of the same charge, pointing to a distinctly heavier quark in each sector; 
ii) one non-vanishing mixing angle, which can be identified with the rotation in the $23$ sector for both quark
and leptons in the three generation case.

Finally, as a possible correction to the patterns above, we briefly explored the possibility of introducing simultaneously bi-fundamentals and fundamentals flavour fields. 
It is a very sensible possibility from the point of view of effective Lagrangians to consider both  $d=5$ and 
$d=6$ Yukawa operators when working to $\mathcal{O}(1/\Lambda_f^2)$. It suggests that $d=5$ operators, which bring in the 
bi-fundamentals, could give the dominant contributions, while the $d=6$ operator  - which brings in the fundamentals - 
should provide a correction inducing the masses of the two lighter families and non-zero angles. 
 
Overall, it is remarkable  that  the requirement of invariance under the flavour symmetry strongly constraints 
the scalar potential. Furthermore, one result of the analysis stands out among the rest.
In the minimum of the potential, at the renormalizable level the quark mixing angles 
vanish at leading order, whereas lepton mixing is found to be maximal.
The presence of the maximal angle in the lepton case is due to the $O(n_g)$ factor 
of the flavour group, which is in turn related of the Majorana nature of neutrinos. The explanation of the different mixing
patterns in quarks and leptons in this scheme is, utterly, the different fundamental nature of the two types of fermions: Dirac and Majorana.



\chapter{Appendix} 


\ifpdf
    \graphicspath{{8/figures/PNG/}{8/figures/PDF/}{8/figures/}}
\else
    \graphicspath{{8/figures/EPS/}{8/figures/}}
\fi


\section{$\mathcal{G}_\mathcal{F}^q$-invariant renormalizable potential formulae of the vevs of $\mathcal{Y}_{D,U}$}\label{App1}
In the following the complete vacua configuration of the bi-fundamental quark fields is
given in terms of the potential parameters. The only two assumptions are: i) a negative $g$ coefficient
($g<0$), since this yields the approximate observed alignment of up and down sectors, ii) both $I_{U}$ and
$I_D$ as defined in Eqs.~\ref{Invariants2F_BiFundAC1},\ref{InvBIF3F1} being non-zero at the minimum. The expression  
for $I_{U,D}$ at the minimum is:
\begin{equation}
\left(\begin{array}{c}
I_U\\
I_D
\end{array}\right)=\frac{1}{2}(\lambda+\lambda')^{-1}\cdot\mu^2\,,
\end{equation}
where $\lambda$ and $\lambda'$ are $2\times 2$ real symmetric matrices and $\mu^2$
is a real vector in 2 dimensions. Assumption ii) implies that the product $(\lambda+\lambda')^{-1}\cdot\mu^2$
is a positive 2-vector.  $\lambda$ and $\mu^2$ are defined in Eq.~\ref{Pot2FBi} and $\lambda'$ differs for each of the vacua
configurations detailed in Sec. \ref{potQ2f}. Each of the different vacua has a distinct breaking
pattern: $\mathcal{G}_\mathcal{F}^q\rightarrow \mathcal{H}^q$. The formulae for the different possibilities
are given next for $n_g$ families where $n_g=2,3$ although the results can presumably be extended to any $n_g$.
\begin{itemize}
\item[\bf I] The unbroken group is $\mathcal{H}^q=U(n_g-1)^3 \times U(1)$ and the solution for the field vevs:
\begin{equation}
\mathcal{Y}_D =\Lambda_f\mbox{Diag}\left(0\,,\,\cdots\,,\,0\,,\,y_b\right)\,,\qquad \mathcal{Y}_U =\Lambda_f
\mbox{Diag}\left( 0\,,\,\cdots\,,\,0\,,\,y_t\right)\,,
\end{equation}
\begin{equation}
\lambda'=\left(
\begin{array}{cc}
 h_U & \frac{g-|g|}{4} \\
 \frac{g-|g|}{4} & h_D
\end{array}
\right)\,.
\end{equation}
\item[\bf II] The unbroken group is $\mathcal{H}^q=U(n_g-1)^2 \times U(1)$ and the fields:
\begin{equation}
\mathcal{Y}_D =\Lambda_f\mbox{Diag}\left(y\,,\,\cdots\,,\,y\,,\,y_b\right)\,,\qquad \mathcal{Y}_U =\Lambda_f
\mbox{Diag}\left( 0\,,\,\cdots\,,\,0\,,\,y_t\right)\,,\end{equation}
\begin{equation}
\quad\frac{y_b^2-y^2}{y_b^2+(n_g-1)y^2}=\frac{-g}{2 h_D}\frac{I_{U}}{I_D}\,,\qquad \lambda'=\left(
\begin{array}{cc}
 h_U-\frac{n_g-1}{n_g}\frac{g^2}{4 h_D} & \frac{g}{2n_g} \\
 \frac{g}{2n_g} & \frac{h_D}{n_g}
\end{array}
\right)\,.\label{EqIIQ}
\end{equation} 
This solution requires the right hand side of the first equation in~\ref{EqIIQ} to lie between $-1/(n_g-1)$ and $1$,
if it reaches the upper value the minimum in on the edge of case {\bf I}, edge depicted by the horizontal line in
Fig.~\ref{VacuaMapQuarks}. The other limit of this solution is given by $g^2-4h_Uh_D=0$ beyond 
which the absolute minimum is the degenerate case; it is the line in between {\bf II} and {\bf IV} in 
Fig.~\ref{VacuaMapQuarks}.

\item[\bf III]The unbroken group is $\mathcal{H}^q=U(n_g-1)^2 \times U(1)$ and the fields:
\begin{equation} 
\mathcal{Y}_D =\Lambda_f\mbox{Diag}\left(0\,,\,\cdots\,,\,0\,,\,y_b\right)\,,\qquad \mathcal{Y}_U =\Lambda_f
\mbox{Diag}\left( y\,,\,\cdots\,,\,y\,,\,y_t\right)\,,\end{equation}
\begin{equation}
\frac{y_t^2-y^2}{y_t^2+(n_g-1)y^2}=\frac{-g}{2 h_U}\frac{I_D}{I_U}\,,\qquad \lambda'=\left(
\begin{array}{cc}
 \frac{h_U}{n_g} & \frac{g}{2n_g} \\
 \frac{g}{2n_g} & h_D-\frac{n_g-1}{n_g}\frac{g^2}{4 h_U}
\end{array}
\right)\,.
\end{equation}
The limit in which $-g/(2h_U)=I_U/I_D$ signals the end of validity of this
solution and the transition of the absolute minimum to case {\bf I}. This
limit is depicted as the vertical line of Fig.~\ref{VacuaMapQuarks}.

\item[\bf IV] The unbroken group is $\mathcal{H}^q=U(n_g)$ and the fields:
\begin{equation}
\mathcal{Y}_D =\Lambda_f\mbox{Diag}\left(y\,,\,\cdots\,,\,y\right)\,,\qquad \mathcal{Y}_U =\Lambda_f
\mbox{Diag}\left( y'\,,\,\cdots\,,\,y'\right)\,,\end{equation}
\begin{equation}
\lambda'=\left(
\begin{array}{cc}
 \frac{h_U}{n_g} & \frac{g}{2n_g} \\
 \frac{g}{2n_g} & \frac{h_D}{n_g}
\end{array}
\right)\,.
\end{equation}
This is the absolute minimum provided $g^2<4h_Uh_D$ with $h_U>0,h_D>0$, 
this condition in Fig.~\ref{VacuaMapQuarks} translates in the allowed region
above the curved line.
\end{itemize}

\section{$\mathcal{G}_\mathcal{F}^l$-invariant renormalizable potential formulae of the vevs of $\mathcal{Y}_{E,\nu}$}\label{App2}
The renormalizable potential allows for the configurations listed in Sec.~\ref{potL3g} for the vevs of the bi-fundamental fields
in the case of $h_{\nu}'>0$ whereas in the case of $h_{\nu}'<0$ the possibilities
are the same as in the quark case. The former case is examined in the following for three families, negative $g$ and the assumption
of both invariants $I_{E}, I_\nu$ taking non-zero vevs given by:
 \begin{equation}
\left(\begin{array}{c}
I_\nu\\
I_E
\end{array}\right)=\frac{1}{2}(\lambda+\lambda')^{-1}\mu^2\,,
\end{equation}
where $\lambda$ and $\mu^2$ as given in Eq. \ref{potential} are a $2\times 2$ real symmetric matrix
and a 2-vector respectively. $\lambda'$ is a $2\times 2$ real symmetric matrix different for each
vacuum alignment. Each vacuum configuration is in turn characterized by the remaining unbroken subgroup $\mathcal{H}^l$.
\begin{itemize}
\item[\bf I] The unbroken group is $\mathcal{H}^l=U(2)^2\times U(1)$ and the fields:
\begin{equation}
\mathcal{Y}_E =\Lambda_f \left( \begin{array}{ccc}
0&0&0\\
0&0&0\\
0&0&y_\tau \\
\end{array}\right),\qquad \mathcal{Y}_\nu =\Lambda_f
\left( \begin{array}{ccc}
0&0&0\\
0&0&0\\
y_\nu/\sqrt{2}&0&iy_\nu/\sqrt{2}\\
\end{array}\right),
\end{equation}
\begin{equation}
\lambda'=\left(
\begin{array}{cc}
 h_{\nu } & \frac{g}{2} \\
 \frac{g}{2} & h_E
\end{array}
\right)\,.
\end{equation}
\item[\bf II] The unbroken group is $\mathcal{H}^l=U(2)\times U(1)$ and the fields:
\begin{equation}
\mathcal{Y}_E =\Lambda_f \left( \begin{array}{ccc}
y&0&0\\
0&y&0\\
0&0&y_\tau \\
\end{array}\right),\qquad \mathcal{Y}_\nu =\Lambda_f
\left( \begin{array}{ccc}
0&0&0\\
0&0&0\\
y_\nu/\sqrt{2}&0&iy_\nu/\sqrt{2}\\
\end{array}\right),
\end{equation}
\begin{equation}
\frac{y_\tau^2-y^2}{y_\tau^2+2y^2}=\frac{-g}{2 h_E}\frac{I_\nu}{I_E}\,,\qquad\qquad
\lambda'=\left(
\begin{array}{cc}
 -\frac{g^2}{6 h_E}+h_{\nu } & \frac{g}{6} \\
 \frac{g}{6} & \frac{h_E}{3}
\end{array}
\right).
\end{equation}
For the negative region of $g$, this configuration turns into that of case {\bf I} for vanishing $y$,
which occurs for $-g/(2 h_E)=I_E /I_\nu$, a limit depicted as
the horizontal line of Fig.~\ref{VacuaMapLeptons}.

\item[\bf III] The unbroken group is $\mathcal{H}^l=U(2)\times U(1)$ and the fields:
\begin{equation}
\mathcal{Y}_E =\Lambda_f \left( \begin{array}{ccc}
0&0&0\\
0&0&0\\
0&0&y_\tau \\
\end{array}\right),\qquad \mathcal{Y}_\nu =\Lambda_f
\left( \begin{array}{ccc}
y_{\nu_1}/\sqrt{2}&0&-iy_{\nu_1}/\sqrt{2}\\
0&y_{\nu_2}&0\\
y_{\nu_3}/\sqrt{2}&0&iy_{\nu_3}/\sqrt{2}\\
\end{array}\right),
\end{equation}
\begin{equation}
\frac{y_{\nu_3}^2-y_{\nu_1}^2}{y_{\nu_3}^2+y_{\nu_2}^2+y_{\nu_1}^2}= \frac{-gI_E}{2 \left(h_{\nu }-h_\nu'\right)I_\nu}\,,\qquad
\frac{y_{\nu_2}^2-y_{\nu_1}^2}{y_{\nu_3}^2+y_{\nu_2}^2+y_{\nu_1}^2}= \frac{-g I_E h_\nu'}{2 \left(h_{\nu }^2-h_\nu'^2\right)I_\nu}\,,
\end{equation}
\begin{equation}
\lambda'=
\left(
\begin{array}{cc}
 \frac{1}{3} \left(h_{\nu }+h_\nu'\right) & \frac{g}{6} \\
 \frac{g}{6} & h_E-\frac{ g^2 (2 h_{\nu }+h_\nu')}
 {12 \left(h_{\nu }^2-\left(h_\nu'\right){}^2\right)}
\end{array}
\right)\,.
\end{equation}
This solution turns into case {\bf VII} when $y_{\nu_1}=0$, a limit
drawn as the vertical line to the right in Fig.~\ref{VacuaMapLeptons}.

\item[\bf IV] The unbroken group is $\mathcal{H}^l=SO(3)$ and the fields:
\begin{equation}
\mathcal{Y}_E =\Lambda_f \left( \begin{array}{ccc}
y&0&0\\
0&y&0\\
0&0&y \\
\end{array}\right),\qquad \mathcal{Y}_\nu =\Lambda_fy_\nu
\left( \begin{array}{ccc}
1/\sqrt{2}&0&-i/\sqrt{2}\\
0&1&0\\
1/\sqrt{2}&0&i/\sqrt{2}\\
\end{array}\right),
\end{equation}
\begin{equation}
\lambda'=\left(
\begin{array}{cc}
 \frac{1}{3} \left(h_{\nu }+h_\nu'\right) & \frac{g}{6} \\
 \frac{g}{6} & \frac{h_E}{3}
\end{array}
\right)\,.
\end{equation}
This case is the absolute minima provided $4(h_\nu-h_\nu')h_E>g^2$ and $h_E>0$.

\item[\bf V] The unbroken group is $\mathcal{H}^l=U(1)^2$ and the fields:
\begin{equation}
\mathcal{Y}_E =\Lambda_f \left( \begin{array}{ccc}
0&0&0\\
0&y_\mu&0\\
0&0&y_\tau \\
\end{array}\right),\qquad \mathcal{Y}_\nu =\Lambda_f
\left( \begin{array}{ccc}
y_{\nu_1}/\sqrt{2}&0&-i y_{\nu_1}/\sqrt{2}\\
0&y_{\nu_2}&0\\
y_{\nu_3}/\sqrt{2}&0&iy_{\nu_3}/\sqrt{2}\\
\end{array}\right)\,,
\end{equation}
\begin{gather}
\frac{y_{\nu_3}^2-y_{\nu_1}^2}{y_{\nu_3}^2+y_{\nu_2}^2+y_{\nu_1}^2}= \frac{-g I_E  \left(4 h_E \left(h_{\nu }+h_\nu'\right)-g^2\right)}{2 \left(8 h_E\left( h_{\nu }^2-\left(h_\nu'\right)^2\right)- g^2 \left(2 h_{\nu }-h_\nu'\right)\right)I_\nu}\,,\\
\frac{y_{\nu_2}^2-y_{\nu_1}^2}{y_{\nu_3}^2+y_{\nu_2}^2+y_{\nu_1}^2}= \frac{-g I_E \left(4 h_E h_{\nu }-g^2\right) }{2 \left(8 h_E \left(h_{\nu }^2-\left(h_\nu'\right){}^2\right)-g^2 \left(2 h_{\nu }-h_\nu'\right)\right)I_\nu}\,,\\
\frac{y_\tau^2-y_\mu^2}{y_\tau^2+y_\mu^2}=\frac{g^2  h_\nu'}{8 h_E \left(h_{\nu }^2-\left(h_\nu'\right){}^2\right)-g^2 \left(2 h_{\nu }-h_\nu'\right)}\,,
\end{gather}
\begin{equation}
\lambda'=\left(
\begin{array}{cc}
 \frac{1}{3} \left(h_{\nu }+h_\nu'\right) & \frac{g}{6} \\
 \frac{g}{6} & \frac{2 g^4+8 g^2 h_E \left(-4 h_{\nu }+h_\nu'\right)+96 h_E^2 \left(h_{\nu }^2-\left(h_\nu'\right){}^2\right)}{24 \left(g^2 \left(-2 h_{\nu }+h_\nu'\right)+8 h_E \left(h_{\nu }^2-\left(h_\nu'\right){}^2\right)\right)}
\end{array}
\right),
\end{equation}
This solution turns into case {\bf III} for $y_\mu=0$ and into case {\bf VIII} for $y_{\nu_1}=0$. This two
conditions translated into the potential parameters through the equations above allow to draw the lines in Fig.~\ref{VacuaMapLeptons} between the respective cases.

\item[\bf VI] The unbroken group is $\mathcal{H}^l=U(1)^2$ and the fields: 
\begin{equation}
\mathcal{Y}_E =\Lambda_f \left( \begin{array}{ccc}
y_e&0&0\\
0&y_\mu&0\\
0&0&y_\tau \\
\end{array}\right),\qquad \mathcal{Y}_\nu =\Lambda_f
\left( \begin{array}{ccc}
0&0&0\\
0&y_{\nu_2}&0\\
y_{\nu_3}/\sqrt{2}&0&iy_{\nu_3}/\sqrt{2}\\
\end{array}\right),
\end{equation}
\begin{gather}\label{ALH1}
\frac{y_{\tau}^2-y_{e}^2}{y_\tau^2+y_\mu^2+y_e^2}= \frac{-g I_{\nu } \left(4 h_E \left(h_{\nu }+h_\nu'\right)-g^2\right)}{4 h_E \left(2 h_E \left(2 h_{\nu }+h_\nu'\right)-g^2\right)I_E}\,,\\ \label{ALH2}
\frac{y_{\mu}^2-y_{e}^2}{y_\tau^2+y_\mu^2+y_e^2}= \frac{-gI_{\nu }\left(4  h_E h_{\nu }-g^2\right) }{4 h_E \left(2 h_E \left(2 h_{\nu }+h_\nu'\right)-g^2\right)I_E}\,,\\
\frac{y_{\nu_3}^2-y_{\nu_2}^2}{y_{\nu_3}^2+y_{\nu_2}^2}= \frac{2 h_E h_\nu'}{2 h_E \left(2 h_{\nu }+h_\nu'\right)-g^2}\,,
\end{gather}
\begin{equation}
\lambda'=\left(
\begin{array}{cc}
 -\frac{g^4+48 h_E^2 h_{\nu } \left(h_{\nu }+h_\nu'\right)-8 g^2 h_E \left(2 h_{\nu }+h_\nu'\right)}{24 h_E \left(g^2-2 h_E \left(2 h_{\nu }+h_\nu'\right)\right)} & \frac{g}{6} \\
 \frac{g}{6} & \frac{h_E}{3}
\end{array}
\right)\, .
\end{equation}
This solution in the limit $y_\mu=y_e$ becomes case {\bf II} and for $y_{e}=0$ it turns into
case {\bf VIII}. These limits are identified in Eqs.~\ref{ALH1},\ref{ALH2} and translated in Fig.~\ref{VacuaMapLeptons}
in the respective lines.

\item[\bf VII] The unbroken group is $\mathcal{H}^l=U(2)\times U(1)^2$ and the fields: 
\begin{equation}
\mathcal{Y}_E =\Lambda_f \left( \begin{array}{ccc}
0&0&0\\
0&0&0\\
0&0&y_\tau \\
\end{array}\right),\qquad \mathcal{Y}_\nu =\Lambda_f
\left( \begin{array}{ccc}
0&0&0\\
0&y_{\nu_2}&0\\
y_{\nu_3}/\sqrt{2}&0&iy_{\nu_3}/\sqrt{2}\\
\end{array}\right),
\end{equation}
\begin{gather}
\frac{y_{\nu_3}^2-y_{\nu_2}^2}{y_{\nu_3}^2+y_{\nu_2}^2}
= \frac{I_{\nu } h_\nu'-g I_E}{\left(2 h_{\nu }+h_\nu'\right)I_\nu}\,,
\end{gather}
\begin{equation}
\lambda'=\left(
\begin{array}{cc}
 \frac{h_{\nu } \left(h_{\nu }+h_\nu'\right)}{2 h_{\nu }+h_\nu'} & \frac{g \left(h_{\nu }+h_\nu'\right)}{2 \left(2 h_{\nu }+h_\nu'\right)} \\
 \frac{g \left(h_{\nu }+h_\nu'\right)}{2 \left(2 h_{\nu }+h_\nu'\right)} & \frac{-2 g^2+8 h_E \left(2 h_{\nu }+h_\nu'\right)}{8 \left(2 h_{\nu }+h_\nu'\right)}
\end{array}
\right)\,.
\end{equation}
This solution connects with the hierarchical case of {\bf I} for $y_{\nu_2}=0$, that is
at the vertical line on the left in Fig.~\ref{VacuaMapLeptons}.

\item[\bf VIII] The unbroken group is $\mathcal{H}^l=U(1)^3$ and the fields: 
\begin{equation}
\mathcal{Y}_E =\Lambda_f \left( \begin{array}{ccc}
0&0&0\\
0&y_{\mu}&0\\
0&0&y_\tau \\
\end{array}\right),\qquad \mathcal{Y}_\nu =\Lambda_f
\left( \begin{array}{ccc}
0&0&0\\
0&y_{\nu_2}&0\\
y_{\nu_3}/\sqrt{2}&0&iy_{\nu_3}/\sqrt{2}\\
\end{array}\right),
\end{equation}
\begin{gather}\label{ALH3}
\frac{y_\tau^2-y_\mu^2}{y_\tau^2+y_\mu^2}= \frac{-g h_\nu' I_{\nu }}{\left(2 h_E \left(2 h_{\nu }+h_\nu'\right)-g^2\right)I_E}\,,\\
\frac{y_{\nu_3}^2-y_{\nu_2}^2}{y_{\nu_3}^2+y_{\nu_2}^2}= \frac{2 h_E  h_\nu'}{2 h_E \left(2 h_{\nu }+h_\nu'\right)-g^2}\, ,
\end{gather}
\begin{equation}
\lambda'=\left(
\begin{array}{cc}
 \frac{-8 h_E h_{\nu }^2+g^2 h_\nu'+2 h_{\nu } \left(g^2-4 h_E h_\nu'\right)}{4 \left(g^2-2 h_E \left(2 h_{\nu }+h_\nu'\right)\right)} & \frac{g}{4} \\
 \frac{g}{4} & \frac{h_E}{2}
\end{array}
\right)\,.
\end{equation}
This solution becomes that of case {\bf VII} for $y_\mu=0$, or equivalently for the RHS of Eq.~\ref{ALH3}
equal to 1.
\end{itemize}

\section{Perturbations on a extremal degenerate neutrino matrix}\label{App3}
A possibility for an extremal or boundary configuration for the neutrino flavour field $\mathcal{Y}_\nu$  (see Sec. \ref{Jac3FL})
 yields the following neutrino matrix, where corrections to the pattern are implemented through $\epsilon_{ij}$,
\begin{equation}
\mathbf{m}_\nu=\frac{y^2v^2}{2M}
\left(
\begin{array}{ccc}
 1+\epsilon_{11} &\epsilon_{12} &\epsilon_{13} \\
\epsilon_{12} &\epsilon_{22} & 1 \\
\epsilon_{13} & 1 &\epsilon_{33}
\end{array}
\right)\,,
\end{equation}
such that masses, as defined in Eq.~\ref{numassdata}, read, 
\begin{gather}
m_{\nu_{1,2}}=\frac{y^2v^2}{2M}\left(1+\frac{\epsilon _{22}+2 \epsilon _{11}+\epsilon _{33}\mp \sqrt{D^2}}{4}\right) +\mathcal{O}(\epsilon^2)\, ,\\
m_{\nu_3}=\frac{y^2v^2}{2M}\left(1-\frac{\epsilon _{22}+\epsilon _{33}}{2}\right) +\mathcal{O}(\epsilon^2)\,,
\end{gather}
where $D^2=\left(\epsilon_{22}+\epsilon_{33}-2\epsilon_{11}\right)^2+8\left(\epsilon_{12}+\epsilon_{13}\right)^2$ and the mixing matrix,
\begin{gather}
U_{PMNS}=
\left(
\begin{array}{ccc}
  \cos(\omega) & \sin(\omega)&-i\epsilon'\cos(\phi)  \\
 -\frac{\sin(\omega)-\epsilon'\cos(\phi +\omega)}{\sqrt{2}} & \frac{\cos(\omega)+\epsilon'\sin(\phi +\omega)}{\sqrt{2}}&i \frac{1-\epsilon'\sin(\phi)}{\sqrt{2}} \\
 -\frac{\sin(\omega)+\epsilon'\cos(\phi +\omega)}{\sqrt{2}} & \frac{\cos(\omega)-\epsilon'\sin(\phi +\omega)}{\sqrt{2}}& -i\frac{1+\epsilon'\sin(\phi)}{\sqrt{2}} 
\end{array}
\right)
\,,\\[2mm]
\tan\omega=\frac{2\sqrt{2}\left(\epsilon_{12}+\epsilon_{13}\right)}{2\epsilon_{11}-\epsilon_{22}-\epsilon_{33}-\sqrt{D^2}}\,,\qquad 
\tan\phi=\frac{\epsilon_{22}-\epsilon_{33}}{\sqrt{2}(\epsilon_{12}-\epsilon_{13})}\,,\\[2mm]
\epsilon'=\frac{1}{4}\sqrt{2(\epsilon_{12}-\epsilon_{13})^2+(\epsilon_{22}-\epsilon_{33})^2}\,.
\end{gather}
to be compared to Eq.~\ref{pmns}.

\cleardoublepage



\begin{acknowledgements}      
The author acknowledges support from the (late) MICINN through the grant BES-2010-037869 during
the completion of this thesis and thanks the {\it Center for the Fundamental Laws of Nature} at Harvard
and the {\it Theory Divison} at CERN for kind hospitality.
 \\[1.5cm]

\begin{center}
{\Large \bf Agradecimientos}\\[1.5cm]
\end{center}
Por abrirme las puertas a la investigaci\'on y
ense\~narme el camino del trabajo riguroso y de calidad debo m\'as de lo que s\'e expresar a Bel\'en. 
Por otro lado, han contribuido directamente al trabajo aqu\'i presentado  Stefano, Dani, Gino, Luciano y
especialmente Luca, de quienes he aprendido gran parte de lo que s\'e, lo cual les agradezco junto con el escuchar a un insignificante
estudiante de doctorado.
Discusiones de F\'isica y el esparcimiento necesario deben ser agradecidos a Luis, David, Manos, Antonio, Eduardo,
Stefania, Juan y una larga lista. De un modo indirecto pero capital mi familia, en especial mi madre, ha ayudado a que
este manuscrito exista, junto con Josefina, Javier,  Saul, Fernando, Jaime, Linda
y el inimitable elenco de mi San Esteban natal.\\

Espero compensar el laconismo de estos agradecimientos al que urgencia obliga con activa muestra de aprecio en adelante a todas esas personas que me son importantes.
\cleardoublepage

\end{acknowledgements}







\begin{multicols}{2} 
\begin{tiny} 

\bibliographystyle{PhDbiblio-url2} 
\renewcommand{\bibname}{References} 

\bibliography{biblio} 

\begin{thebibliography}{100}

\bibitem{Aad:2012tfa}
{\sc Georges Aad et~al.}
\newblock {\bf {Observation of a new particle in the search for the Standard
  Model Higgs boson with the ATLAS detector at the LHC}}.
\newblock {\em Phys.Lett.}, {\bf B716}:1--29, 2012.

\bibitem{Chatrchyan:2012ufa}
{\sc Serguei Chatrchyan et~al.}
\newblock {\bf {Observation of a new boson at a mass of 125 GeV with the CMS
  experiment at the LHC}}.
\newblock {\em Phys.Lett.}, {\bf B716}:30--61, 2012.

\bibitem{Corbett:2012dm}
{\sc Tyler Corbett, O.J.P. Eboli, J.~Gonzalez-Fraile, and M.C.
  Gonzalez-Garcia}.
\newblock {\bf {Constraining anomalous Higgs interactions}}.
\newblock {\em Phys.Rev.}, {\bf D86}:075013, 2012.

\bibitem{Giardino:2013bma}
{\sc Pier~Paolo Giardino, Kristjan Kannike, Isabella Masina, Martti Raidal, and
  Alessandro Strumia}.
\newblock {\bf {The universal Higgs fit}}.
\newblock 2013.

\bibitem{Englert:1964et}
{\sc F.~Englert and R.~Brout}.
\newblock {\bf {Broken Symmetry and the Mass of Gauge Vector Mesons}}.
\newblock {\em Phys.Rev.Lett.}, {\bf 13}:321--323, 1964.

\bibitem{Higgs:1964ia}
{\sc Peter~W. Higgs}.
\newblock {\bf {Broken Symmetries, Massless Particles and Gauge Fields}}.
\newblock {\em Phys.Lett.}, {\bf 12}:132--133, 1964.

\bibitem{Higgs:1964pj}
{\sc Peter~W. Higgs}.
\newblock {\bf {Broken Symmetries and the Masses of Gauge Bosons}}.
\newblock {\em Phys.Rev.Lett.}, {\bf 13}:508--509, 1964.

\bibitem{Callaway:1988ya}
{\sc David~J.E. Callaway}.
\newblock {\bf {Triviality Pursuit: Can Elementary Scalar Particles Exist?}}
\newblock {\em Phys.Rept.}, {\bf 167}:241, 1988.

\bibitem{Degrassi:2012ry}
{\sc Giuseppe Degrassi, Stefano Di~Vita, Joan Elias-Miro, Jose~R. Espinosa,
  Gian~F. Giudice, et~al.}
\newblock {\bf {Higgs mass and vacuum stability in the Standard Model at
  NNLO}}.
\newblock {\em JHEP}, {\bf 1208}:098, 2012.

\bibitem{EliasMiro:2011aa}
{\sc Joan Elias-Miro, Jose~R. Espinosa, Gian~F. Giudice, Gino Isidori, Antonio
  Riotto, et~al.}
\newblock {\bf {Higgs mass implications on the stability of the electroweak
  vacuum}}.
\newblock {\em Phys.Lett.}, {\bf B709}:222--228, 2012.

\bibitem{Alonso:2012jc}
{\sc R.~Alonso, M.B. Gavela, L.~Merlo, S.~Rigolin, and J.~Yepes}.
\newblock {\bf {Minimal Flavour Violation with Strong Higgs Dynamics}}.
\newblock {\em JHEP}, {\bf 1206}:076, 2012.

\bibitem{Alonso:2012pz}
{\sc R.~Alonso, M.B. Gavela, L.~Merlo, S.~Rigolin, and J.~Yepes}.
\newblock {\bf {Flavor with a light dynamical "Higgs particle"}}.
\newblock {\em Phys.Rev.}, {\bf D87}:055019, 2013.

\bibitem{Alonso:2012px}
{\sc R.~Alonso, M.B. Gavela, L.~Merlo, S.~Rigolin, and J.~Yepes}.
\newblock {\bf {The Effective Chiral Lagrangian for a Light Dynamical
  'Higgs'}}.
\newblock 2012.

\bibitem{Alonso:2012ji}
{\sc R.~Alonso, M.~Dhen, M.B. Gavela, and T.~Hambye}.
\newblock {\bf {Muon conversion to electron in nuclei in type-I seesaw
  models}}.
\newblock {\em JHEP}, {\bf 1301}:118, 2013.

\bibitem{Alonso:2011yg}
{\sc R.~Alonso, M.~B. Gavela, L.~Merlo, and S.~Rigolin}.
\newblock {\bf {On the Scalar Potential of Minimal Flavour Violation}}.
\newblock {\em JHEP}, {\bf 07}:012, 2011.

\bibitem{Alonso:2012fy}
{\sc R.~Alonso, M.B. Gavela, D.~Hernandez, and L.~Merlo}.
\newblock {\bf {On the Potential of Leptonic Minimal Flavour Violation}}.
\newblock {\em Phys.Lett.}, {\bf B715}:194--198, 2012.

\bibitem{onpreparation}
{\sc R.~Alonso, M.B. Gavela, G.~Isidori, and L.~Maiani}.
\newblock {\bf {Neutrino Mixing and Masses from a Minimum Principle}}.
\newblock 2013.

\bibitem{onpreparation2}
{\sc R.~Alonso, M.B. Gavela, D.~Hernandez, L.~Merlo, and S.~Rigolin}.
\newblock {\bf {Leptonic Dynamical Yukawa Couplings}}.
\newblock 2013.

\bibitem{Maiani:2011zz}
{\sc N.~Cabibbo and L.~Maiani}.
\newblock {\bf {Weak interactions and the breaking of hadron symmetries}}.
\newblock {\em Evolution of particle physics}, pages 50--80, 1970.

\bibitem{Froggatt:1978nt}
{\sc C.~D. Froggatt and Holger~Bech Nielsen}.
\newblock {\bf {Hierarchy of Quark Masses, Cabibbo Angles and CP Violation}}.
\newblock {\em Nucl. Phys.}, {\bf B147}:277, 1979.

\bibitem{Georgi:1979md}
{\sc Howard Georgi}.
\newblock {\bf {Towards a Grand Unified Theory of Flavor}}.
\newblock {\em Nucl.Phys.}, {\bf B156}:126, 1979.

\bibitem{Berezhiani:1983hm}
{\sc Z.G. Berezhiani}.
\newblock {\bf {The Weak Mixing Angles in Gauge Models with Horizontal
  Symmetry: A New Approach to Quark and Lepton Masses}}.
\newblock {\em Phys.Lett.}, {\bf B129}:99--102, 1983.

\bibitem{Chivukula:1987py}
{\sc R.~Sekhar Chivukula and Howard Georgi}.
\newblock {\bf {Composite Technicolor Standard Model}}.
\newblock {\em Phys. Lett.}, {\bf B188}:99, 1987.

\bibitem{Barbieri:1995uv}
{\sc Riccardo Barbieri, G.R. Dvali, and Lawrence~J. Hall}.
\newblock {\bf {Predictions from a U(2) flavor symmetry in supersymmetric
  theories}}.
\newblock {\em Phys.Lett.}, {\bf B377}:76--82, 1996.

\bibitem{Berezhiani:2001mh}
{\sc Zurab Berezhiani and Anna Rossi}.
\newblock {\bf {Flavor structure, flavor symmetry and supersymmetry}}.
\newblock {\em Nucl.Phys.Proc.Suppl.}, {\bf 101}:410--420, 2001.

\bibitem{D'Ambrosio:2002ex}
{\sc G.~D'Ambrosio, G.~F. Giudice, G.~Isidori, and A.~Strumia}.
\newblock {\bf {Minimal Flavour Violation: an Effective Field Theory
  Approach}}.
\newblock {\em Nucl. Phys.}, {\bf B645}:155--187, 2002.

\bibitem{Cirigliano:2005ck}
{\sc Vincenzo Cirigliano, Benjamin Grinstein, Gino Isidori, and Mark~B. Wise}.
\newblock {\bf {Minimal flavor violation in the lepton sector}}.
\newblock {\em Nucl. Phys.}, {\bf B728}:121--134, 2005.

\bibitem{Davidson:2006bd}
{\sc Sacha Davidson and Federica Palorini}.
\newblock {\bf {Various Definitions of Minimal Flavour Violation for Leptons}}.
\newblock {\em Phys. Lett.}, {\bf B642}:72--80, 2006.

\bibitem{Alonso:2011jd}
{\sc Rodrigo Alonso, Gino Isidori, Luca Merlo, Luis~Alfredo Munoz, and Enrico
  Nardi}.
\newblock {\bf {Minimal Flavour Violation Extensions of the Seesaw}}.
\newblock {\em JHEP}, {\bf 06}:037, 2011.

\bibitem{Barbieri:2011ci}
{\sc Riccardo Barbieri, Gino Isidori, Joel Jones-Perez, Paolo Lodone, and
  David~M. Straub}.
\newblock {\bf {U(2) and Minimal Flavour Violation in Supersymmetry}}.
\newblock {\em Eur.Phys.J.}, {\bf C71}:1725, 2011.

\bibitem{Joshipura:2009gi}
{\sc Anjan~S. Joshipura, Ketan~M. Patel, and Sudhir~K. Vempati}.
\newblock {\bf {Type I seesaw mechanism for quasi degenerate neutrinos}}.
\newblock {\em Phys.Lett.}, {\bf B690}:289--295, 2010.

\bibitem{Susskind:1978ms}
{\sc Leonard Susskind}.
\newblock {\bf {Dynamics of Spontaneous Symmetry Breaking in the Weinberg-
  Salam Theory}}.
\newblock {\em Phys. Rev.}, {\bf D20}:2619--2625, 1979.

\bibitem{Dimopoulos:1979es}
{\sc Savas Dimopoulos and Leonard Susskind}.
\newblock {\bf {Mass without Scalars}}.
\newblock {\em Nucl. Phys.}, {\bf B155}:237--252, 1979.

\bibitem{Dimopoulos:1981xc}
{\sc Savas Dimopoulos and John Preskill}.
\newblock {\bf {Massless Composites with Massive Constituents}}.
\newblock {\em Nucl.Phys.}, {\bf B199}:206, 1982.

\bibitem{Kaplan:1983fs}
{\sc David~B. Kaplan and Howard Georgi}.
\newblock {\bf {$SU(2)\times U(1)$ Breaking by Vacuum Misalignment}}.
\newblock {\em Phys.Lett.}, {\bf B136}:183, 1984.

\bibitem{Kaplan:1983sm}
{\sc David~B. Kaplan, Howard Georgi, and Savas Dimopoulos}.
\newblock {\bf {Composite Higgs Scalars}}.
\newblock {\em Phys. Lett.}, {\bf B136}:187, 1984.

\bibitem{Agashe:2004rs}
{\sc Kaustubh Agashe, Roberto Contino, and Alex Pomarol}.
\newblock {\bf {The Minimal Composite Higgs Model}}.
\newblock {\em Nucl.Phys.}, {\bf B719}:165--187, 2005.

\bibitem{Agashe:2006at}
{\sc Kaustubh Agashe, Roberto Contino, Leandro Da~Rold, and Alex Pomarol}.
\newblock {\bf {A Custodial Symmetry for Z B Anti-B}}.
\newblock {\em Phys. Lett.}, {\bf B641}:62--66, 2006.

\bibitem{Gripaios:2009pe}
{\sc Ben Gripaios, Alex Pomarol, Francesco Riva, and Javi Serra}.
\newblock {\bf {Beyond the Minimal Composite Higgs Model}}.
\newblock {\em JHEP}, {\bf 0904}:070, 2009.

\bibitem{Manohar:1983md}
{\sc Aneesh Manohar and Howard Georgi}.
\newblock {\bf {Chiral Quarks and the Nonrelativistic Quark Model}}.
\newblock {\em Nucl.Phys.}, {\bf B234}:189, 1984.

\bibitem{GeorgiWeakInt}
{\sc H.~Georgi}.
\newblock {\bf {Weak Interactions}}.
\newblock {\em http://www.people.fas.harvard.edu/~hgeorgi/}.

\bibitem{Weinberg:1979sa}
{\sc Steven Weinberg}.
\newblock {\bf {Baryon and Lepton Nonconserving Processes}}.
\newblock {\em Phys.Rev.Lett.}, {\bf 43}:1566--1570, 1979.

\bibitem{Minkowski:1977sc}
{\sc Peter Minkowski}.
\newblock {\bf {mu --> e gamma at a Rate of One Out of 1-Billion Muon Decays?}}
\newblock {\em Phys.Lett.}, {\bf B67}:421, 1977.

\bibitem{GellMann:1980vs}
{\sc Murray Gell-Mann, Pierre Ramond, and Richard Slansky}.
\newblock {\bf {COMPLEX SPINORS AND UNIFIED THEORIES}}.
\newblock {\em Conf.Proc.}, {\bf C790927}:315--321, 1979.

\bibitem{Mohapatra:1979ia}
{\sc Rabindra~N. Mohapatra and Goran Senjanovic}.
\newblock {\bf {Neutrino Mass and Spontaneous Parity Violation}}.
\newblock {\em Phys.Rev.Lett.}, {\bf 44}:912, 1980.

\bibitem{Magg:1980ut}
{\sc M.~Magg and C.~Wetterich}.
\newblock {\bf {NEUTRINO MASS PROBLEM AND GAUGE HIERARCHY}}.
\newblock {\em Phys.Lett.}, {\bf B94}:61, 1980.

\bibitem{Schechter:1980gr}
{\sc J.~Schechter and J.W.F. Valle}.
\newblock {\bf {Neutrino Masses in SU(2) x U(1) Theories}}.
\newblock {\em Phys.Rev.}, {\bf D22}:2227, 1980.

\bibitem{Wetterich:1981bx}
{\sc C.~Wetterich}.
\newblock {\bf {Neutrino Masses and the Scale of B-L Violation}}.
\newblock {\em Nucl.Phys.}, {\bf B187}:343, 1981.

\bibitem{Lazarides:1980nt}
{\sc George Lazarides, Q.~Shafi, and C.~Wetterich}.
\newblock {\bf {Proton Lifetime and Fermion Masses in an SO(10) Model}}.
\newblock {\em Nucl.Phys.}, {\bf B181}:287, 1981.

\bibitem{Mohapatra:1980yp}
{\sc Rabindra~N. Mohapatra and Goran Senjanovic}.
\newblock {\bf {Neutrino Masses and Mixings in Gauge Models with Spontaneous
  Parity Violation}}.
\newblock {\em Phys.Rev.}, {\bf D23}:165, 1981.

\bibitem{Foot:1988aq}
{\sc Robert Foot, H.~Lew, X.G. He, and Girish~C. Joshi}.
\newblock {\bf {SEESAW NEUTRINO MASSES INDUCED BY A TRIPLET OF LEPTONS}}.
\newblock {\em Z.Phys.}, {\bf C44}:441, 1989.

\bibitem{Ma:1998dn}
{\sc Ernest Ma}.
\newblock {\bf {Pathways to naturally small neutrino masses}}.
\newblock {\em Phys.Rev.Lett.}, {\bf 81}:1171--1174, 1998.

\bibitem{Beringer:1900zz}
{\sc J.~Beringer et~al.}
\newblock {\bf {Review of Particle Physics (RPP)}}.
\newblock {\em Phys.Rev.}, {\bf D86}:010001, 2012.

\bibitem{GonzalezGarcia:2010er}
{\sc M.~C. Gonzalez-Garcia, Michele Maltoni, and Jordi Salvado}.
\newblock {\bf {Updated Global Fit to Three Neutrino Mixing: Status of the
  Hints of $\theta_{13} > 0$}}.
\newblock {\em JHEP}, {\bf 04}:056, 2010.

\bibitem{PhysRevLett.105.031301}
{\sc Shaun~A. Thomas, Filipe~B. Abdalla, and Ofer Lahav}.
\newblock \href{http://link.aps.org/doi/10.1103/PhysRevLett.105.031301}{{\bf
  Upper Bound of 0.28~eV on Neutrino Masses from the Largest Photometric
  Redshift Survey}}.
\newblock {\em Phys. Rev. Lett.}, {\bf 105}:031301, Jul 2010.

\bibitem{Georgi:1974sy}
{\sc H.~Georgi and S.L. Glashow}.
\newblock {\bf {Unity of All Elementary Particle Forces}}.
\newblock {\em Phys.Rev.Lett.}, {\bf 32}:438--441, 1974.

\bibitem{Glashow:1970gm}
{\sc S.L. Glashow, J.~Iliopoulos, and L.~Maiani}.
\newblock {\bf {Weak Interactions with Lepton-Hadron Symmetry}}.
\newblock {\em Phys.Rev.}, {\bf D2}:1285--1292, 1970.

\bibitem{Pakvasa:1977in}
{\sc Sandip Pakvasa and Hirotaka Sugawara}.
\newblock {\bf {Discrete Symmetry and Cabibbo Angle}}.
\newblock {\em Phys.Lett.}, {\bf B73}:61, 1978.

\bibitem{Harrison:2002er}
{\sc P.~F. Harrison, D.~H. Perkins, and W.~G. Scott}.
\newblock {\bf {Tri-Bimaximal Mixing and the Neutrino Oscillation Data}}.
\newblock {\em Phys. Lett.}, {\bf B530}:167, 2002.

\bibitem{Altarelli:2005yp}
{\sc Guido Altarelli and Ferruccio Feruglio}.
\newblock {\bf {Tri-Bimaximal Neutrino Mixing from Discrete Symmetry in Extra
  Dimensions}}.
\newblock {\em Nucl. Phys.}, {\bf B720}:64--88, 2005.

\bibitem{Altarelli:2005yx}
{\sc Guido Altarelli and Ferruccio Feruglio}.
\newblock {\bf {Tri-Bimaximal Neutrino Mixing, $A_4$ and the Modular
  Symmetry}}.
\newblock {\em Nucl. Phys.}, {\bf B741}:215--235, 2006.

\bibitem{Altarelli:2006kg}
{\sc Guido Altarelli, Ferruccio Feruglio, and Yin Lin}.
\newblock {\bf {Tri-Bimaximal Neutrino Mixing from Orbifolding}}.
\newblock {\em Nucl. Phys.}, {\bf B775}:31--44, 2007.

\bibitem{Bazzocchi:2009pv}
{\sc Federica Bazzocchi, Luca Merlo, and Stefano Morisi}.
\newblock {\bf {Fermion Masses and Mixings in a $S_4$-Based Model}}.
\newblock {\em Nucl. Phys.}, {\bf B816}:204--226, 2009.

\bibitem{Altarelli:2009gn}
{\sc Guido Altarelli, Ferruccio Feruglio, and Luca Merlo}.
\newblock {\bf {Revisiting Bimaximal Neutrino Mixing in a Model with $S_4$
  Discrete Symmetry}}.
\newblock {\em JHEP}, {\bf 05}:020, 2009.

\bibitem{Kubo:2003iw}
{\sc J.~Kubo, A.~Mondragon, M.~Mondragon, and E.~Rodriguez-Jauregui}.
\newblock {\bf {The Flavor symmetry}}.
\newblock {\em Prog.Theor.Phys.}, {\bf 109}:795--807, 2003.

\bibitem{Feruglio:2007uu}
{\sc Ferruccio Feruglio, Claudia Hagedorn, Yin Lin, and Luca Merlo}.
\newblock {\bf {Tri-Bimaximal Neutrino Mixing and Quark Masses from a Discrete
  Flavour Symmetry}}.
\newblock {\em Nucl. Phys.}, {\bf B775}:120--142, 2007.

\bibitem{Feruglio:2013hia}
{\sc Ferruccio Feruglio, Claudia Hagedorn, and Robert Ziegler}.
\newblock {\bf {S4 and CP in a SUSY Model}}.
\newblock 2013.

\bibitem{Holthausen:2012dk}
{\sc Martin Holthausen, Manfred Lindner, and Michael~A. Schmidt}.
\newblock {\bf {CP and Discrete Flavour Symmetries}}.
\newblock {\em JHEP}, {\bf 1304}:122, 2013.

\bibitem{Randall:1999ee}
{\sc Lisa Randall and Raman Sundrum}.
\newblock {\bf {A Large mass hierarchy from a small extra dimension}}.
\newblock {\em Phys.Rev.Lett.}, {\bf 83}:3370--3373, 1999.

\bibitem{Randall:1999vf}
{\sc Lisa Randall and Raman Sundrum}.
\newblock {\bf {An Alternative to compactification}}.
\newblock {\em Phys.Rev.Lett.}, {\bf 83}:4690--4693, 1999.

\bibitem{Grossman:1999ra}
{\sc Yuval Grossman and Matthias Neubert}.
\newblock {\bf {Neutrino masses and mixings in nonfactorizable geometry}}.
\newblock {\em Phys.Lett.}, {\bf B474}:361--371, 2000.

\bibitem{Gherghetta:2000kr}
{\sc Tony Gherghetta and Alex Pomarol}.
\newblock {\bf {A Warped supersymmetric standard model}}.
\newblock {\em Nucl.Phys.}, {\bf B602}:3--22, 2001.

\bibitem{ArkaniHamed:1998rs}
{\sc Nima Arkani-Hamed, Savas Dimopoulos, and G.R. Dvali}.
\newblock {\bf {The Hierarchy problem and new dimensions at a millimeter}}.
\newblock {\em Phys.Lett.}, {\bf B429}:263--272, 1998.

\bibitem{ArkaniHamed:1999dc}
{\sc Nima Arkani-Hamed and Martin Schmaltz}.
\newblock {\bf {Hierarchies without symmetries from extra dimensions}}.
\newblock {\em Phys.Rev.}, {\bf D61}:033005, 2000.

\bibitem{Mirabelli:1999ks}
{\sc Eugene~A. Mirabelli and Martin Schmaltz}.
\newblock {\bf {Yukawa hierarchies from split fermions in extra dimensions}}.
\newblock {\em Phys.Rev.}, {\bf D61}:113011, 2000.

\bibitem{Branco:2000rb}
{\sc G.C. Branco, Andre de~Gouvea, and M.N. Rebelo}.
\newblock {\bf {Split fermions in extra dimensions and CP violation}}.
\newblock {\em Phys.Lett.}, {\bf B506}:115--122, 2001.

\bibitem{Hall:1999sn}
{\sc Lawrence~J. Hall, Hitoshi Murayama, and Neal Weiner}.
\newblock {\bf {Neutrino mass anarchy}}.
\newblock {\em Phys.Rev.Lett.}, {\bf 84}:2572--2575, 2000.

\bibitem{deGouvea:2003xe}
{\sc Andre de~Gouvea and Hitoshi Murayama}.
\newblock {\bf {Statistical test of anarchy}}.
\newblock {\em Phys.Lett.}, {\bf B573}:94--100, 2003.

\bibitem{deGouvea:2012ac}
{\sc Andre de~Gouvea and Hitoshi Murayama}.
\newblock {\bf {Neutrino Mixing Anarchy: Alive and Kicking}}.
\newblock 2012.

\bibitem{Buchmuller:1985jz}
{\sc W.~Buchmuller and D.~Wyler}.
\newblock {\bf {Effective Lagrangian Analysis of New Interactions and Flavor
  Conservation}}.
\newblock {\em Nucl.Phys.}, {\bf B268}:621, 1986.

\bibitem{Grzadkowski:2010es}
{\sc B.~Grzadkowski, M.~Iskrzynski, M.~Misiak, and J.~Rosiek}.
\newblock {\bf {Dimension-Six Terms in the Standard Model Lagrangian}}.
\newblock {\em JHEP}, {\bf 1010}:085, 2010.

\bibitem{Wyler:1982dd}
{\sc D.~Wyler and L.~Wolfenstein}.
\newblock {\bf {Massless Neutrinos in Left-Right Symmetric Models}}.
\newblock {\em Nucl.Phys.}, {\bf B218}:205, 1983.

\bibitem{Mohapatra:1986bd}
{\sc R.N. Mohapatra and J.W.F. Valle}.
\newblock {\bf {Neutrino Mass and Baryon Number Nonconservation in Superstring
  Models}}.
\newblock {\em Phys.Rev.}, {\bf D34}:1642, 1986.

\bibitem{Branco:1988ex}
{\sc G.C. Branco, W.~Grimus, and L.~Lavoura}.
\newblock {\bf {THE SEESAW MECHANISM IN THE PRESENCE OF A CONSERVED LEPTON
  NUMBER}}.
\newblock {\em Nucl.Phys.}, {\bf B312}:492, 1989.

\bibitem{Raidal:2004vt}
{\sc Martti Raidal, Alessandro Strumia, and Krzysztof Turzynski}.
\newblock {\bf {Low-scale standard supersymmetric leptogenesis}}.
\newblock {\em Phys.Lett.}, {\bf B609}:351--359, 2005.

\bibitem{Gavela:2009cd}
{\sc M.B. Gavela, T.~Hambye, D.~Hernandez, and P.~Hernandez}.
\newblock {\bf {Minimal Flavour Seesaw Models}}.
\newblock {\em JHEP}, {\bf 0909}:038, 2009.

\bibitem{Broncano:2002rw}
{\sc A.~Broncano, M.B. Gavela, and Elizabeth~Ellen Jenkins}.
\newblock {\bf {The Effective Lagrangian for the seesaw model of neutrino mass
  and leptogenesis}}.
\newblock {\em Phys.Lett.}, {\bf B552}:177--184, 2003.

\bibitem{Antusch:2006vwa}
{\sc S.~Antusch, C.~Biggio, E.~Fernandez-Martinez, M.B. Gavela, and
  J.~Lopez-Pavon}.
\newblock {\bf {Unitarity of the Leptonic Mixing Matrix}}.
\newblock {\em JHEP}, {\bf 0610}:084, 2006.

\bibitem{Abazov:2010hv}
{\sc Victor~Mukhamedovich Abazov et~al.}
\newblock {\bf {Evidence for an anomalous like-sign dimuon charge asymmetry}}.
\newblock {\em Phys.Rev.}, {\bf D82}:032001, 2010.

\bibitem{Aaij:2011in}
{\sc R.~Aaij et~al.}
\newblock {\bf {Evidence for CP violation in time-integrated $D^0 \to h^-h^+$
  decay rates}}.
\newblock {\em Phys.Rev.Lett.}, {\bf 108}:111602, 2012.

\bibitem{Aguilar:2001ty}
{\sc A.~Aguilar-Arevalo et~al.}
\newblock {\bf {Evidence for neutrino oscillations from the observation of
  anti-neutrino(electron) appearance in a anti-neutrino(muon) beam}}.
\newblock {\em Phys.Rev.}, {\bf D64}:112007, 2001.

\bibitem{Isidori:2013ez}
{\sc Gino Isidori}.
\newblock {\bf {Flavor physics and CP violation}}.
\newblock 2013.

\bibitem{Adam:2013mnn}
{\sc J.~Adam et~al.}
\newblock {\bf {New constraint on the existence of the mu+$\rightarrow$ e+
  gamma decay}}.
\newblock 2013.

\bibitem{Dohmen:1993mp}
{\sc C.~Dohmen et~al.}
\newblock {\bf {Test of lepton flavor conservation in mu --$\rightarrow$ e
  conversion on titanium}}.
\newblock {\em Phys.Lett.}, {\bf B317}:631--636, 1993.

\bibitem{Isidori:2012ts}
{\sc Gino Isidori and David~M. Straub}.
\newblock {\bf {Minimal Flavour Violation and Beyond}}.
\newblock {\em Eur.Phys.J.}, {\bf C72}:2103, 2012.

\bibitem{Michel:1970mua}
{\sc L.~Michel and L.~A. Radicati}.
\newblock {\bf {Breaking of the $SU_{3}xSU_{3}$ symmetry in hadronic physics}}.
\newblock {\em Evolution of particle physics}, pages 191--203, 1970.

\bibitem{Michel:1971th}
{\sc L.~Michel and L.A. Radicati}.
\newblock {\bf {Properties of the breaking of hadronic internal symmetry}}.
\newblock {\em Annals Phys.}, {\bf 66}:758--783, 1971.

\bibitem{Berezhiani:2005tp}
{\sc Z.~Berezhiani and F.~Nesti}.
\newblock {\bf {Supersymmetric SO(10) for fermion masses and mixings: Rank-1
  structures of flavor}}.
\newblock {\em JHEP}, {\bf 0603}:041, 2006.

\bibitem{Jenkins:2009dy}
{\sc Elizabeth~Ellen Jenkins and Aneesh~V. Manohar}.
\newblock {\bf {Algebraic Structure of Lepton and Quark Flavor Invariants and
  CP Violation}}.
\newblock {\em JHEP}, {\bf 0910}:094, 2009.

\bibitem{Hanany:2010vu}
{\sc Amihay Hanany, Elizabeth~E. Jenkins, Aneesh~V. Manohar, and Giuseppe
  Torri}.
\newblock {\bf {Hilbert Series for Flavor Invariants of the Standard Model}}.
\newblock {\em JHEP}, {\bf 1103}:096, 2011.

\bibitem{Blankenburg:2012nx}
{\sc Gianluca Blankenburg, Gino Isidori, and Joel Jones-Perez}.
\newblock {\bf {Neutrino Masses and LFV from Minimal Breaking of $U(3)^5$ and
  $U(2)^5$ flavor Symmetries}}.
\newblock {\em Eur.Phys.J.}, {\bf C72}:2126, 2012.

\bibitem{Feldmann:2009dc}
{\sc Thorsten Feldmann, Martin Jung, and Thomas Mannel}.
\newblock {\bf {Sequential Flavour Symmetry Breaking}}.
\newblock {\em Phys. Rev.}, {\bf D80}:033003, 2009.

\bibitem{Nardi:2011st}
{\sc Enrico Nardi}.
\newblock {\bf {Naturally large Yukawa hierarchies}}.
\newblock {\em Phys.Rev.}, {\bf D84}:036008, 2011.

\bibitem{Espinosa:2012uu}
{\sc Jose~R. Espinosa, Chee~Sheng Fong, and Enrico Nardi}.
\newblock {\bf {Yukawa hierarchies from spontaneous breaking of the
  $SU(3)_L\times SU(3)_R$ flavour symmetry?}}
\newblock {\em JHEP}, {\bf 1302}:137, 2013.

\bibitem{Grinstein:2010ve}
{\sc Benjamin Grinstein, Michele Redi, and Giovanni Villadoro}.
\newblock {\bf {Low Scale Flavor Gauge Symmetries}}.
\newblock {\em JHEP}, {\bf 1011}:067, 2010.

\bibitem{Feldmann:2010yp}
{\sc Thorsten Feldmann}.
\newblock {\bf {See-Saw Masses for Quarks and Leptons in $SU(5)$}}.
\newblock {\em JHEP}, {\bf 04}:043, 2011.

\bibitem{Guadagnoli:2011id}
{\sc Diego Guadagnoli, Rabindra~N. Mohapatra, and Ilmo Sung}.
\newblock {\bf {Gauged Flavor Group with Left-Right Symmetry}}.
\newblock {\em JHEP}, {\bf 1104}:093, 2011.

\bibitem{Lopez-Honorez:2013wla}
{\sc Laura Lopez-Honorez and Luca Merlo}.
\newblock {\bf {Dark matter within the minimal flavour violation ansatz}}.
\newblock {\em Phys.Lett.}, {\bf B722}:135--143, 2013.

\bibitem{Buras:2011wi}
{\sc Andrzej~J. Buras, Maria~Valentina Carlucci, Luca Merlo, and Emmanuel
  Stamou}.
\newblock {\bf {Phenomenology of a Gauged $SU(3)^3$ Flavour Model}}.
\newblock {\em JHEP}, {\bf 1203}:088, 2012.

\bibitem{Maiani:2013lma}
{\sc Luciano Maiani}.
\newblock {\bf {Universality of the Weak Interactions, Cabibbo theory and where
  they led us}}.
\newblock {\em Rivista del Nuovo Cimento, 34,}, {\bf 679}, 2011.

\bibitem{Kagan:2009bn}
{\sc Alexander~L. Kagan, Gilad Perez, Tomer Volansky, and Jure Zupan}.
\newblock {\bf {General Minimal Flavor Violation}}.
\newblock {\em Phys. Rev.}, {\bf D80}:076002, 2009.

\bibitem{An:2012eh}
{\sc F.~P. An et~al.}
\newblock {\bf {Observation of Electron-Antineutrino Disappearance at Daya
  Bay}}.
\newblock 2012.

\bibitem{Ahn:2012nd}
{\sc J.K. Ahn et~al.}
\newblock {\bf {Observation of Reactor Electron Antineutrino Disappearance in
  the RENO Experiment}}.
\newblock {\em Phys.Rev.Lett.}, {\bf 108}:191802, 2012.

\bibitem{Pascoli:2013wca}
{\sc Silvia Pascoli and Thomas Schwetz}.
\newblock {\bf {Prospects for neutrino oscillation physics}}.
\newblock {\em Adv.High Energy Phys.}, {\bf 2013}:503401, 2013.

\bibitem{DeGerone:2012iya}
{\sc M.~De~Gerone}.
\newblock {\bf {Latest results from the MEG experiment}}.
\newblock {\em Nucl.Phys.Proc.Suppl.}, {\bf 233}:231--236, 2012.

\bibitem{Blondel:2013ia}
{\sc A.~Blondel, A.~Bravar, M.~Pohl, S.~Bachmann, N.~Berger, et~al.}
\newblock {\bf {Research Proposal for an Experiment to Search for the Decay
  ${\mu}$ $\rightarrow$ eee}}.
\newblock 2013.

\bibitem{Hungerford:2009zz}
{\sc Ed~V. Hungerford}.
\newblock {\bf COMET/PRISM muon to electron conversion at J-PARC}.
\newblock {\em AIP Conf.Proc.}, {\bf 1182}:694--697, 2009.

\bibitem{Carey:2008zz}
{\sc R.~M.~Carey {\it et al.}~[Mu2e~Collaboration]}.
\newblock {\bf Proposal to search for mu- N -> e- N with a single event
  sensitivity below 10 -16}.
\newblock {\em FERMILAB-PROPOSAL-0973}.

\bibitem{Ade:2013zuv}
{\sc P.A.R. Ade et~al.}
\newblock {\bf {Planck 2013 results. XVI. Cosmological parameters}}.
\newblock 2013.

\bibitem{GomezCadenas:2011it}
{\sc J.J. Gomez-Cadenas, J.~Martin-Albo, M.~Mezzetto, F.~Monrabal, and
  M.~Sorel}.
\newblock {\bf {The Search for neutrinoless double beta decay}}.
\newblock {\em Riv.Nuovo Cim.}, {\bf 35}:29--98, 2012.

\bibitem{Alonso:2010wu}
{\sc Rodrigo Alonso, Stefan Antusch, Mattias Blennow, Pilar Coloma, Andre
  de~Gouvea, et~al.}
\newblock {\bf {Summary report of MINSIS workshop in Madrid}}.
\newblock 2010.

\bibitem{Eboli:2011ia}
{\sc O.J.P. Eboli, J.~Gonzalez-Fraile, and M.C. Gonzalez-Garcia}.
\newblock {\bf {Neutrino Masses at LHC: Minimal Lepton Flavour Violation in
  Type-III See-saw}}.
\newblock {\em JHEP}, {\bf 1112}:009, 2011.

\bibitem{Dinh:2012bp}
{\sc D.N. Dinh, A.~Ibarra, E.~Molinaro, and S.T. Petcov}.
\newblock {\bf {The mu - e Conversion in Nuclei, mu -$\rightarrow$ e gamma, mu
  -$\rightarrow$ 3e Decays and TeV Scale See-Saw Scenarios of Neutrino Mass
  Generation}}.
\newblock {\em JHEP}, {\bf 1208}:125, 2012.

\bibitem{Sierra:2012yy}
{\sc D.~Aristizabal~Sierra, A.~Degee, and J.F. Kamenik}.
\newblock {\bf {Minimal Lepton Flavor Violating Realizations of Minimal Seesaw
  Models}}.
\newblock {\em JHEP}, {\bf 1207}:135, 2012.

\bibitem{Casas:2001sr}
{\sc J.A. Casas and A.~Ibarra}.
\newblock {\bf {Oscillating neutrinos and muon --$\rightarrow$ e, gamma}}.
\newblock {\em Nucl.Phys.}, {\bf B618}:171--204, 2001.

\bibitem{Altarelli:2004za}
{\sc Guido Altarelli and Ferruccio Feruglio}.
\newblock {\bf {Models of neutrino masses and mixings}}.
\newblock {\em New J.Phys.}, {\bf 6}:106, 2004.

\end{thebibliography}

\end{tiny}
\end{multicols}








\end{document}